\RequirePackage[english=usenglishmax]{hyphsubst}
\documentclass[oneside,openright,titlepage,numbers=noenddot,headinclude,footinclude=true,cleardoublepage=empty,listof=totoc,paper=a4,fontsize=11pt,australian,DIV=calc]{scrreprt}

\newcommand{\myTitle}{Improving the understanding of the dynamics of open quantum systems}
\newcommand{\mySubmissionYear}{2023}
\newcommand{\mySubmissionMonth}{June}
\newcommand{\mySubmissionDay}{12}

\newcommand{\myFirstName}{Ali Raza}
\newcommand{\myLastName}{Mirza}
\newcommand{\myWebsite}{https://www.surrey.ac.uk/people/ali-raza-mirza}
\newcommand{\myBirthYear}{1990}
\newcommand{\myBirthMonth}{January}
\newcommand{\myBirthDay}{06}
\newcommand{\myBirthPlace}{Barnali}

\newcommand{\myProfTitle}{Dr.}
\newcommand{\myProfFirstName}{Adam Zaman}
\newcommand{\myProfLastName}{Chaudhry}
\newcommand{\myProfWebsite}{https://lums.edu.pk/lums_employee/4056}

\newcommand{\myOtherProfTitle}{Dr.}
\newcommand{\myOtherProfFirstName}{Salman Khan}
\newcommand{\myOtherProfLastName}{Safi}
\newcommand{\myOtherProfWebsite}{http://ww2.comsats.edu.pk/faculty/FacultyDetails.aspx?Uid=2203}

\newcommand{\intAFirstName}{Muhammad Sabieh}
\newcommand{\intALastName}{Anwar}
\newcommand{\intAWebsite}{https://lums.edu.pk/lums_employee/1617}

\newcommand{\intBFirstName}{Ata ul}
\newcommand{\intBLastName}{Haq}
\newcommand{\intBWebsite}{https://lums.edu.pk/lums_employee/3312}

\newcommand{\intCFirstName}{Muhammad Imran}
\newcommand{\intCLastName}{Cheema}
\newcommand{\intCWebsite}{https://lums.edu.pk/lums_employee/4623}

\newcommand{\myDepartment}{Department of Physics}
\newcommand{\mySchool}{Syed Babar Ali School of Science and Engineering}
\newcommand{\myUni}{Lahore University of Management Sciences}
\newcommand{\myFaculty}{}

\input{classicthesis-config}
\usepackage[a4paper,top=25.4mm,bottom=25.4mm,left=25mm, right=25mm,bindingoffset=8mm]{geometry}

\begin{document}

  \frenchspacing
  \raggedbottom
  \selectlanguage{australian}
  
  \pagestyle{plain}
  \pagenumbering{roman}
  
\singlespacing
\begin{titlepage}
  \doublespacing
  \large
  \hfill
  \vfill
  \vspace*{0.5cm}
  \begin{center}
    \doublespacing
    \textcolor{Maroon}{\huge\textbf{\myTitle}}
  \end{center}
  \vspace{1.25cm}
  \hrule
  \vspace{1.5cm}
  \onehalfspacing
  \begin{center}

    \includegraphics[width=10cm]{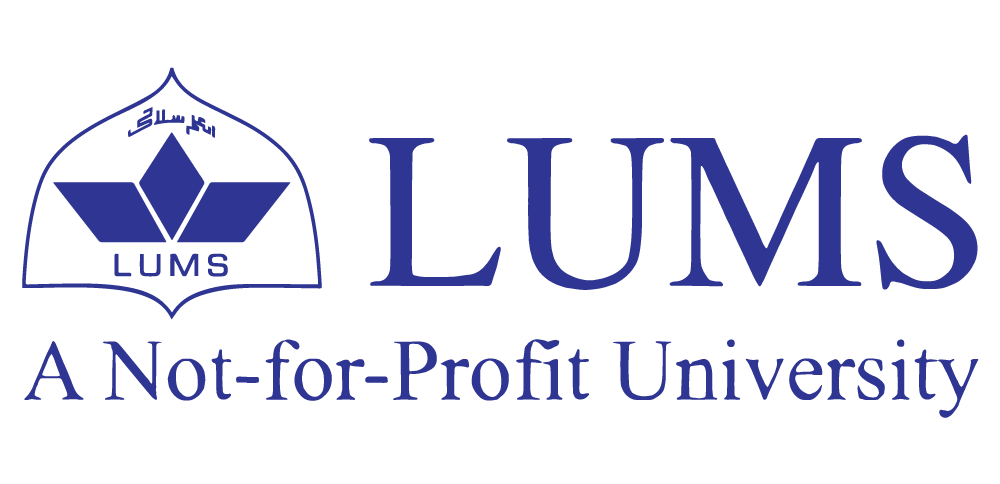}

\hfill
  \vfill
\vspace*{0.5cm}

    \href{\myWebsite}{{\myFirstName} \textsc{\myLastName}}

    \myDepartment\\
    \mySchool\\
    \myFaculty
    \href{https://lums.edu.pk}{\myUni}\\
     Pakistan\\

\hfill
  \vfill
  \vspace*{0.5cm}

A thesis submitted in fulfillment \\
    of the requirements for the degree of\\
    {Doctor of Philosophy}\\[1cm]
     
    {\mySubmissionMonth} {\mySubmissionYear}
  \end{center}
\end{titlepage}
\hfill
\vfill

\noindent \textit{\myTitle,} {\textcopyright} {\mySubmissionMonth} {\mySubmissionYear}

\bigskip

\noindent Author:\\
\href{\myWebsite}{{\myFirstName} \textsc{\myLastName}}

\medskip

\noindent Supervisor:\\
\href{\myProfWebsite}{{\myProfTitle} {\myProfFirstName} \textsc{\myProfLastName}}

\medskip

\noindent Institute:\\
\href{https://lums.edu.pk}{\myUni}, Pakistan

\chapter*{Dissertation Approval}
\addcontentsline{toc}{chapter}{Desertation Approval}
The members of the Committee approve the dissertation entitled \emph{\myTitle}, defended on 26/06/2023. It is recommended that this dissertation be used in partial fulfillment of the requirements for the degree of Doctor of Philosophy from Department of Physics in Syed Babar Ali School of Science and Engineering.

\bigskip
\bigskip
\bigskip
\bigskip
\bigskip
\bigskip
\bigskip
\bigskip
\bigskip
\bigskip
\bigskip
\bigskip

\noindent Supervisor:\\
\href{\myProfWebsite}{{\myProfTitle} {\myProfFirstName} \textsc{\myProfLastName}}

\bigskip
\bigskip
\bigskip
\bigskip

\noindent External Examiner:\\
\href{\myOtherProfWebsite}{{\myOtherProfTitle} {\myOtherProfFirstName} \textsc{\myOtherProfLastName}}

\bigskip
\bigskip
\bigskip
\bigskip

\noindent FDC member:\\
\href{\intAWebsite}{{\myProfTitle} {\intAFirstName} \textsc{\intALastName}}

\bigskip
\bigskip
\bigskip
\bigskip

\noindent FDC member:\\
\href{\intBWebsite}{{\myProfTitle} {\intBFirstName} \textsc{\intBLastName}}

\bigskip
\bigskip
\bigskip
\bigskip

\noindent FDC member:\\
\href{\intCWebsite}{{\myProfTitle} {\intCFirstName} \textsc{\intCLastName}}

\bigskip
\bigskip
\bigskip
\bigskip

\noindent Institute:\\
\href{https://lums.edu.pk}{\myUni}, Pakistan   

\chapter*{Declaration}
\addcontentsline{toc}{chapter}{Declaration}
\noindent I, {\myFirstName} {\myLastName}, born in {\myBirthMonth} {\myBirthDay}, {\myBirthYear} at {\myBirthPlace} (Pakistan), declare that this thesis titled \emph{\myTitle} and its content is solely my own work. I also hereby confirm that all this work has been done as a Ph. D. student at Lahore University of Management Sciences (LUMS), Lahore. I have acknowledged wherever I have used someone else's work the in the main text of my thesis. This work has not been submitted for the award of any degree in any other university around the globe previously.
\vspace{5em}

\noindent\hspace{0.5em}\includegraphics[width=15em]{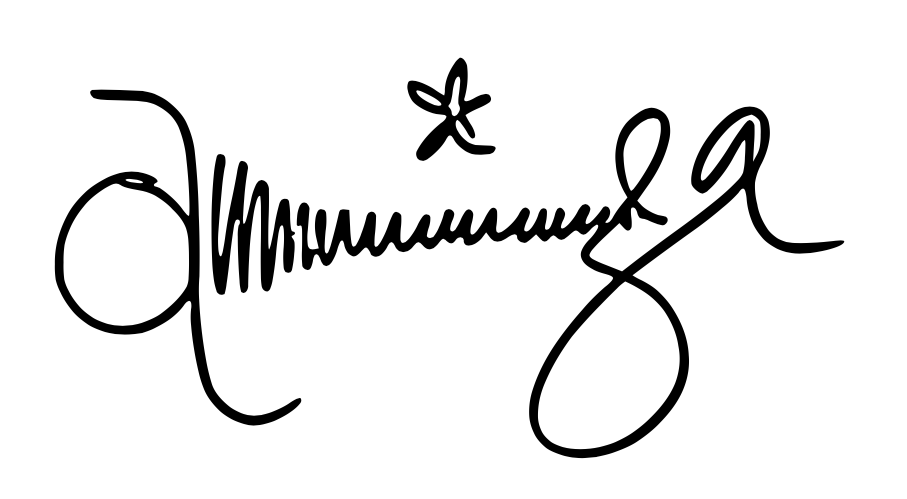}\\
\noindent\rule[1em]{16.5em}{0.5pt}

\vspace{-1.5em}
\noindent {\myFirstName} {\myLastName} \hspace{12em} {\mySubmissionMonth} {\mySubmissionDay}, {\mySubmissionYear}

\chapter*{Certificate}
\addcontentsline{toc}{chapter}{Certificate}
This is to certify that the work presented by Ali Raza Mirza on the thesis titled "\emph{\myTitle}" is based on the results of the research study conducted by the candidate under my supervision. No portion of this work has been formerly offered for higher degree in this university or any other institute of learning and to the best of the author's knowledge, no material has been presented in this thesis which is not his own work, except where due acknowledgements have been made. He has fulfilled all the requirements and is qualified to submit this thesis in the partial fulfilment for the degree of Doctor of Philosophy in Physics.

\vspace{5em}

\noindent Supervisor:\\
\href{\myProfWebsite}{{\myProfTitle} {\myProfFirstName} \textsc{\myProfLastName}}\\
Associate Professor \\
Department of Physics\\
Lahore University of Management Sciences.  
\chapter*{Dedication}

\hspace{0pt}
\vfill
\hfil \emph{This thesis is dedicated to my \st{late} father \textbf{Muhammad Aslam Mirza}}
\\
\newline Father! When I was a child, you left me and went to ALLAH (SWT), the \emph{world's creator and possessor}. Everyone has to leave but you went away so early. You were not forgotten by me. You always seem so close to me. In fact, \emph{I miss you frantically} every second of my life. May ALLAH continue to bless and favor you.
\vfill
\hspace{0pt}

\chapter*{Acknowledgments}
\addcontentsline{toc}{chapter}{Acknowledgments}

My Ph.D. thesis would not have been possible without all the assistance and positive reinforcement from others. Throughout my stay here at {\myUni}, I had the honor of associating with  a lot of competent people, numerous of whom got to be my comrades. To all of you, I thusly say \emph{Thank You very much! This work would not have been conceivable without you!} Nevertheless, there are a few people who deserve special mention.

Firstly, I would like to commend my advisor \textbf{\myProfTitle} \textbf{\myProfFirstName} \textbf{\myProfLastName} for providing me with immense support and guidance throughout my Ph.D. There are few people in this world who are selfless, my supervisor is one of them. He always trusted in me and never got harsh on me. His door always remained open for me. He directed me not only in academics but also in my private matters. I never saw him in a rage but in a polite and affectionate mood.

I also like to say thanks to \href{https://research.manchester.ac.uk/en/persons/ahsan.nazir}{\textbf{Dr. Ahsan Nazir}} who hosted me at the University of Manchester, England. He is a man of extraordinarily high calibre. He introduced me \emph{Counting Statistics} which was a new field for me. However, it went well solely due to his marvelous cooperation. This visit was sponsored by the Higher Education Commission (HEC) of Pakistan. Indeed, HEC deserves a lot of appreciation for supporting this foreign research project.

Furthermore, I sincerely acknowledge \textbf{Mr. Muhammad Zia} and  \textbf{Miss Mehwish Majeed} who assisted me in coding. Special credit goes to \textbf{Prof. Bilal Ali} for helping me refine my English grammar and \textbf{Dr. Ata ul Haq}, \textbf{Dr. Muhammad Faryad} and \textbf{Dr. Atif Shahbaz} for endorsing my IRSIP (International Research Support Initiative Program) and Commonwealth Scholarship applications.

Additionally, I want to express my deepest gratitude for my courageous mother \textbf{Mrs. Khursheed Begum} whose prayers and tireless hard work empowered me to achieve all goals, and my elder brothers \textbf{Mr. Zaheer Aslam} and \textbf{Mr. Ihtasham ul Haq} for their moral and financial support, and indeed my elder sister \textbf{Mrs. Sobia Aslam} for her prayers and affection. And of course last but not the least, I duly acknowledge my wife \textbf{Huma Ali Raza Mirza}. Indeed she is a lovely and friend-like wife. She has proven herself a loyal companion through her priceless prayers and taking care of me in every aspect.  
\chapter*{Publications}
\addcontentsline{toc}{chapter}{Publications}

\begin{itemize}

  \item Ali Raza Mirza, Muhammad Zia, and Adam Zaman Chaudhry ``\textbf{Master equation incorporating the system-environment correlations present in the joint equilibrium state}'' published in \emph{Physical Review A}, \\DOI: \href{https://doi.org/10.1103/PhysRevA.104.042205}{10.1103/PhysRevA.104.042205}.

  \item Ali Raza Mirza, Mah Noor Jamil, and Adam Zaman Chaudhry ``\textbf{The role of initial system-environment correlations with a spin environment}'', \href{https://doi.org/10.48550/arXiv.2301.07332}{arXiv.2301.07332}.

  \item Ali Raza Mirza and Adam Zaman Chaudhry ``\textbf{Improving the estimation of the environment parameters via a two-qubit scheme}'',  \href{https://doi.org/10.48550/arXiv.2305.12278}{arXiv.2305.12278}.

  \item Ali Raza Mirza and Ahsan Nazir ``\textbf{Work counting statistics in periodically driven strongly coulped quantum systems}'' (in progress).
  
\end{itemize}

\chapter*{Useful Identities}
\addcontentsline{toc}{chapter}{Useful Identities}

\begin{itemize}
    
\item   \textbf{Magnus Expansion}. This is useful for us in calculating the time-evolution operator corresponding to the time-dependent Hamiltonian. 
\begin{align*}
    U\left(t\right)
    = \text{exp}\left\{\sum_{i=1}^\infty A_i\left(t\right)\right\}, 
\end{align*}
where
\begin{align*}
    A_1&=-i \int_0^t H_I \left(t_1\right) \, dt_1,\nonumber
\\
    A_2&=-\frac{1}{2}\int _0^t dt_1\int_0^{t_1} \left[H_I \left(t_1\right),H_I \left(t_2\right)\right] \, dt_{2}.\nonumber
\end{align*}
The higher-order terms are given by further commutators. 
    \item     \textbf{ Hadamard's lemma}. This tells us that, for operators $A$ and $B$, 
 \begin{align*}
    e^{\theta A} B e^{-\theta A} 
    = B +  \theta\left[A,B\right] + \frac{\theta^2}{2!} \left[A,\left[A,B\right]\right] \cdots.
 \end{align*}   
    \item     \textbf{Kubo Identity}. For any two operators $A$ and $B$, we have
 \begin{align*}
     e^{\beta(A+B)}
    =e^{\beta A}\left\{1+\int_{0}^{\beta} e^{-\lambda A}B e^{\lambda B}d\lambda\right\}.
 \end{align*}   
    \item     \textbf{Bloch identity}. If $C$ is a linear combination of the harmonic oscillator raising and lowering operators, then \label{blochID}
 \begin{align*}
     \av{e^C} 
     = e^{\av{C^2}/2}.
 \end{align*}   
   \item     \textbf{Weyl identity}. If operator $A$ and operator $B$ commute with their commutator $[A,B]$, then
\begin{align*}
    e^{A+B} 
    = e^{A} e^{B} e^{-\left[A,B\right]/2}.
\end{align*}    
    \item \textbf{Exponential of a Pauli matrix}. For a `unit' vector $\hat{n}$, we have that
\begin{align*}
    e^{i a \left( \hat{n} . \vec{\sigma} \right)} 
    = \mathds{1} \cos a + i  \left( \hat{n} . \vec{\sigma} \right) \sin a.
\end{align*}
\end{itemize}

\chapter*{}
\chapter*{Abstract}
\addcontentsline{toc}{chapter}{Abstract}

This thesis presents studies performed on open quantum systems, that is, quantum systems interacting with their surrounding environment. Such systems are important not only in understanding the quantum-to-classical transition but also for the practical implementation of modern quantum technologies. In studies of open quantum systems performed to date, a very common assumption is that the system and the environment are in separated initial states to begin with. One primary objective of this thesis is to critically analyze this assumption. As such, the core of this thesis incorporates the effect of the initial system-environment $(\mathcal{SE})$ correlations that are present in the joint thermal equilibrium state of the system and the environment on the subsequent system dynamics. In this regard, we follow two different approaches to investigate the dynamics. First, we solve an exactly solvable spin-spin model where a central spin system interacts with a collection of quantum spins. We analyze exactly the central spin dynamics, starting from both initially correlated and uncorrelated $\mathcal{SE}$ states, and look at the dynamical differences due to the different starting states. Second, we consider an arbitrary system interacting with an arbitrary environment and derive a master equation that describes the system dynamics and also incorporates the effect of the initial $\mathcal{SE}$ correlations. This effect of initial correlations is captured by an extra term appearing in the master equation. The master equation is subsequently applied to the paradigmatic $\mathcal{SE}$ models such as the spin-boson model and the spin-spin model. We demonstrate that the role played by initial correlations can be noticeable even if the $\mathcal{SE}$ coupling strength is kept smaller.

The next part of the thesis deals with estimating the parameters characterizing the environment of a quantum system. After all, in order to predict the dynamics of a quantum system, one needs to know, for example, the cutoff frequency of the environment as well as its temperature. Recently, the use of a single qubit system to infer the characteristics of an environment has attracted considerable interest. We show that the use of two two-level systems can greatly enhance the estimation of the environment parameters. The reason is simple - two two-level systems also get correlated with each other due to their interactions with the environment, and information about the environment is imprinted onto these correlations. We quantitatively demonstrate this by calculating the quantum Fisher information for a two-qubit probe. Finally, in the last part of this thesis, we study the \emph{work} counting statistics via a Markovian master equation for a periodically driven spin system weakly coupled to its environment of harmonic oscillators.

  \onehalfspacing
  %----------------------------------------------------------------------------------------
%  Table of Contents
%----------------------------------------------------------------------------------------
\refstepcounter{dummy}
\pdfbookmark[1]{\contentsname}{tableofcontents} % Bookmark name visible in a PDF viewer
\setcounter{tocdepth}{2}
\setcounter{secnumdepth}{3}
\manualmark
\markboth{\spacedlowsmallcaps{\contentsname}}{\spacedlowsmallcaps{\contentsname}}
\tableofcontents
\automark[section]{chapter}
\renewcommand{\chaptermark}[1]{\markboth{\spacedlowsmallcaps{#1}}{\spacedlowsmallcaps{#1}}}
\renewcommand{\sectionmark}[1]{\markright{\thesection\enspace\spacedlowsmallcaps{#1}}}

\clearpage

\begingroup 
\let\clearpage\relax
\let\cleardoublepage\relax
\let\cleardoublepage\relax

%----------------------------------------------------------------------------------------
%  List of Figures
%----------------------------------------------------------------------------------------
\refstepcounter{dummy}
\listoffigures

\endgroup
  
  \pagestyle{scrheadings}
  \pagenumbering{arabic}
   
  \chapter{Introduction}\label{c:intro}
  Quantum mechanics, one of the pillars of modern physics, has led to many technological developments such as lasers, Global Positioning System (GPS) and the transistors. Currently, one of the primary goals is to control precisely individual small quantum systems (such as single electrons or photons), and to make them interact in a precise manner. The ability to do so opens the door to many novel technologies. For example, we can achieve unbreakable encryption in communication, the uncertainty in measurements can be greatly reduced, and we can build far more powerful computers. Such emerging quantum technologies depend on the quantum properties of physical systems since such technologies rely on harnessing the `quantumness' of these systems. What we mean by quantum over here are properties that cannot be explained by using the laws of classical physics. For example, considering the famous double slit experiment, how is it that a particle can be found here and there at the same time? Objects that we observe in our everyday life around us - moving boxes, cars, human beings - do not show such behavior, meaning that they have lost their quantumness. The reason is that physical systems interact with their environment.

The surrounding environment has two main effects on the quantum system. The first one is a classical phenomenon, namely that the system can exchange energy with its environment. This is called \emph{dissipation}, analogous to how a cup of tea exchanges energy with the surrounding air. The second is purely quantum mechanical whereby the system undergoes \emph{decoherence} - the relative phase within the quantum state gets scrambled due to the interaction with the surrounding environment. Alternatively, decoherence is the decay of the off-diagonal entries of the reduced density matrix describing the system dynamics. The study of decoherence enables us to understand how the classical world transpires out of the quantum world - why superposition states are not observable in the classical world? Actually, most quantum states decohere very quickly due to the system-environment $(\mathcal{SE})$ interaction. Only special states survive. Such states, which are more robust to the effect of the environment, are dubbed \emph{pointer states}. Generally, the larger the quantum system, the more quickly this decoherence process occurs. Moreover, decoherence can happen without dissipation, but the converse is not true.

In order to make use of unique quantum properties in modern technologies (such as quantum communications and quantum computing), decoherence must be properly understood, and, generally, we must try to minimize its effect. Indeed, decoherence is one of the main hurdles in the enlargement of such technologies. We need to have some idea of how quickly a given physical system decoheres, and, if possible, lengthen this timescale. Therefore the ultimate objective of this thesis is to make a small contribution to the understanding of the open systems dynamics.

\section{A brief overview}

In the theoretical study of open quantum system dynamics, we generally need to consider various assumptions and approximations in order to make the problem manageable. The most commonly used assumptions are \cite{de2017dynamics, breuer2016colloquium}:

\begin{itemize}
 	\item Weak $\mathcal{SE}$ coupling (Born approximation),
    \item Memoryless environment (Markov approximation),
    \item The initial $\mathcal{SE}$ state is a factorized state (a product state), that is, initial $\mathcal{SE}$ correlations are ignored.
\end{itemize}

In many realistic situations, such as transfer of energy in photosynthetic complexes \cite{engel2007evidence} and superconducting qubits \cite{buluta2011natural}, where the system is strongly interacting with its environment, these assumptions are simply not valid. Therefore, one main focus of our thesis is to develop better strategies to investigate open system dynamics. In particular, in the first part of this thesis, we aim to include the effect of initial $\mathcal{SE}$ correlations. The study of these correlations has become a topic of great interest because this effect can play a great role whenever strong $\mathcal{SE}$ coupling is involved. Many efforts have been made to better analyze the effect of these initial correlations \cite{HakimPRA1985, HaakePRA1985, Grabert1988, SmithPRA1990, GrabertPRE1997, PazPRA1997, LutzPRA2003, BanerjeePRE2003, vanKampen2004, BanPRA2009, HanggiPRL2009, UchiyamaPRA2010, TanimuraPRL2010, SmirnePRA2010, DajkaPRA2010, ZhangPRA2010,TanPRA2011, CKLeePRE2012,MorozovPRA2012, SeminPRA2012,  ChaudhryPRA2013a,ChaudhryPRA2013b,ChaudhryCJC2013,FanchiniSciRep2014,FanSciRep2015,ChenPRA2016,VegaRMP2017,VegaPRA2017,ShibataJPhysA2017,CaoPRA2017}. Since the role of the initial correlations is expected to be most remarkable in the strong $\mathcal{SE}$ coupling regime, we usually cannot simply apply perturbation theory. One possible solution is to use exactly solvable models \cite{MorozovPRA2012, chaudhry2013amplification}. However, these studies have limitations, namely, the pure dephasing models used do not consider the effect of dissipation. Another study of these correlations found in the literature uses the Jaynes-Cummings model \cite{SmirnePRA2010}. A Spin-Spin $(\mathcal{SS})$ model has also been considered, with the initial system state prepared by performing a suitable projective measurement on the quantum system \cite{majeed2019effect}. As the first problem that we tackle in this thesis, we extend this study to consider the system state preparation via a suitable unitary operation instead. 

The most commonly used method to study open quantum system dynamics is to use a master equation \cite{BPbook}. This approach considers the system and the environment together as a \emph{closed} quantum system whose dynamics are governed by the so-called Schrodinger equation. Thereafter, the environment is `removed' by taking the partial trace over the environment. The leftover first-order differential equation that describes only the system dynamics is known as the \emph{master equation}. Finding and solving a master equation are, unfortunately, not easy tasks. Usually, we assume that the $\mathcal{SE}$ interaction strength is weak, and thereby the $\mathcal{SE}$ time-evolution operator is found perturbatively \cite{Weissbook, breuer2002theory}. To include the effect of the initial $\mathcal{SE}$ correlations in the master equation, we assume that the system and its environment are in a joint thermal equilibrium state, and thereafter, a unitary operation is performed to prepare the desired initial system state, with the system Hamiltonian possibly changing thereafter as well. We first consider an arbitrary system which is interacting with an arbitrary environment and derive a master equation perturbatively. The role of the initial $\mathcal{SE}$ correlations is captured by an extra term appearing in our master equation. To scrutinize the role of initial correlations quantitatively, we apply our master equation to the paradigmatic Spin-Boson $(\mathcal{SB})$ model as well as a $\mathcal{SS}$ model. We demonstrate that, in general, the initial $\mathcal{SE}$ correlations need to be accounted for in order to accurately obtain the system dynamics especially when the number of central spins is large and the temperature is low.

We next note that in order to predict the role of the environment on a given quantum system, one must know, as precisely as possible, various parameters such as the environment's cutoff frequency, the temperature, and the $\mathcal{SE}$ coupling strength. One useful method is to consider a quantum probe - a small and controllable quantum system which is interacting with its environment \cite{brunelli2011qubit, brunelli2012qubit, neumann2013high, benedetti2014quantum, correa2015individual,elliott2016nondestructive,norris2016qubit,tamascelli2016characterization,streif2016measuring,benedetti2018quantum,cosco2017momentum,sone2017exact,salari2019quantum,razavian2019quantum,gebbia2020two,wu2020quantum,tamascelli2020quantum,gianani2020discrimination}. By studying the dynamics of the probe, one can estimate the various parameters of interest. To estimate these parameters as precisely as possible, it is important to obtain as large a quantum Fisher information (QFI) as possible  \cite{helstrom1976quantum,fujiwara2001quantum,monras2006optimal,paris2009quantum,monras2007optimal,genoni2011optical,spagnolo2012phase,pinel2013quantum,chaudhry2014utilizing,chaudhry2015detecting}. Our goal in this thesis is to show how using two two-level systems (rather than a single two-level system) can drastically increase the QFI and consequently the accuracy of our estimates. 

Finally, we consider some thermodynamic aspects in open quantum systems by looking at work statistics. Development in experimental methods enables us to explore the dissipative dynamics of quantum systems \cite{collin2005verification,liphardt2001reversible,douarche2005experimental,pekola2013calorimetric}. Considerable attention has been paid aiming at the derivation of the quantum kinds of fluctuation relations for open systems \cite{campisi2009fluctuation,silaev2014lindblad,campisi2011erratum,esposito2009nonequilibrium,crooks1999entropy,crooks2008jarzynski,mukamel2003quantum,de2004quantum}. Sub-Poissonian statistics for photon counts have been investigated indicating the nonclassical states of an electromagnetic field in quantum optics \cite{mandel1995optical}. Counting statistics of heat transfer and charge also been scrutinized mainly in nonequilibrium mesoscopic systems previously \cite{bagrets2003full,silaev2014lindblad,kindermann2004statistics,esposito2009nonequilibrium}. Of particular interest, full work statistics via the Lindblad master equation approach have been presented, where the environment is supposed to be Markovian \cite{silaev2014lindblad}. In this thesis, generating functions are derived that determine the counting statistics of \emph{work} in the presence of a driving field. Our derivation is based on the $\mathcal{SB}$ model where a single two-level system is interacting with a harmonic oscillator environment. We follow a two-point measurement scheme to construct the characteristic function. Our goal is to study the trade of energy between our system and the environment in terms of bosons under the action of the driving field. 
\newpage
\section{Thesis Outline}
\label{s:Outline}

\begin{description}

  \item[Chapter \ref{c:formulism}] provides the basic concepts employed in the theory of open systems such as density matrix formalism, Bloch vector representation, and entanglement. We also derive a general Markovian master equation and present detailed calculations for an exactly solvable pure dephasing model.
  
  \item[Chapter \ref{c:SpinEnv}] explores the role of initial $\mathcal{SE}$ correlations in an exactly solvable $\mathcal{SS}$ model.
  
  \item[Chapter \ref{c:MasterEq}] details a different formalism aiming at elucidating the importance of the initial $\mathcal{SE}$ correlations. We derive a generalized non-Markovian master equation that incorporates the effect of the initial correlations. 
  
  \item[Chapter \ref{c:fisher}] considers the estimation of environment parameters by using two-qubit system interaction with a common environment, thereby showing how the estimation can be greatly improved compared to the use of a single two-level system.  
  
\item[Chapter \ref{c:workstat}] contains our calculations for the work statistics in a periodically driven quantum system. 

\item[Chapter \ref{c:Conclusions}] concludes the thesis. After some concluding remarks, we briefly highlight the possible extension of the work performed in this thesis.
\end{description}
   
  \chapter{Preliminaries}\label{c:formulism}
  In this chapter, we briefly review the fundamental `tools of the trade', most notably density matrices and master equations. We also review the pure dephasing model that will be used often in this thesis \cite{chaudhry2013understanding}.

\section{Density matrices}
    In an open quantum system, we use density matrices to denote the system state. Density matrices supply a general representation of a quantum state since they can be used to represent pure as well as mixed quantum states. 
    
\subsection{Pure quantum states}
    In closed quantum systems, we denote a quantum state by a `ket' vector $\ket{\phi}$ which encapsulates all the information about our physical system. Given this vector, we can define a density matrix $\varrho$ (also known as the density operator) corresponding to the pure state $\ket{\phi}$ as
\begin{align}
    \varrho \equiv \ket{\phi} \bra{\phi},
\end{align}    
    which is in fact a projection operator onto the quantum state $\ket{\phi}$. If we write $\ket{\phi}$ in terms of a set of basis states $\ket{\phi_i}$ as
\begin{align}
    \ket{\phi}
    = \sum_i b_i \ket{\phi_i},
\end{align}
    the corresponding density matrix can be written as  
\begin{align}
    \varrho
    = \sum_{uv} b_i b^*_j \ket{\phi_i}\bra{\phi_j}. \label{eq3.1}
\end{align}
The terms for which $i \neq j$ represent the off-diagonal entries of the matrix $\varrho$. They are also known as \emph{interference terms}, or the terms that tell us about quantum coherence between the basis states. Now we define the \emph{trace operation} performed on some operator $A$ as
\begin{align}
    \text{Tr}\left\{A\right\}
    = \sum_i \bra{\phi_i} A \ket{\phi_i},
\end{align}
where the ${\ket{\phi_i}}$ form any orthonormal basis for the system's Hilbert space. Now, if we choose $A = \varrho B$, with $\emph{B}$ being a Hermitian operator having eigenvalues $\emph{b}_i$ and eigenstates $\ket{b_i}$, we can write
\begin{align}
    \text{Tr}\left\{A\right\}
    &= \sum_i \bra{b_i} \varrho B \ket{b_i},\nonumber
\\
    &= \sum_i \bra{b_i} \left(\ket{\phi}\bra{\phi}\right) B \ket{b_i},\nonumber
\\
    &= \sum_i b_i \abs{\ip{b_i}{\phi}}^2. \nonumber
\end{align}
    Since $\abs{\ip{b_i}{\phi}}^2$ is the usual Born probability for the measurement result $\emph{b}_i$, we have that 
\begin{align}
    \av{B}
    =\text{Tr}\left\{\varrho B\right\}.
\end{align}    
    If we set $B = \mathds{1}$, then
\begin{align}
    \text{Tr}\left\{\varrho\right\} = 1.
\end{align}    
       
\subsection{Mixed quantum states}\label{mixstate}

    Besides pure states, we can also have \emph{mixed states} \cite{BPbook}. We illustrate this by an example of a spin-$\frac{1}{2}$ particle. If the state of this particle is $\ket{\phi_1}$ with probability $\frac{1}{2}$ and $\ket{\phi_2}$ with probability $\frac{1}{2}$, then the state of this particle can be written as $\varrho = \frac{1}{2}\ket{\phi_1}\bra{\phi_1} + \frac{1}{2}\ket{\phi_2}\bra{\phi_2}$. Note that this should not be confused with a coherent superposition state. Generalizing further, we can write 
\begin{align}
    \varrho
    = \sum_i \mathds{P}_i \ket{\phi_i} \bra{\phi_i},
\end{align}
    where $\mathds{P}_i$ is the probability associated with state $\ket{\phi_i}$. Once again, we have $\text{Tr}(\varrho) =1$, and the expectation value of operator $A$ is given as
\begin{align}
    \langle A \rangle = \text{Tr} \left\{\varrho A\right\}.
\end{align}
    
\subsection{Quantifying the purity of a state}
    
    Since a pure state is a projection operator onto the quantum state $\ket{\phi}$, we immediately have that $\varrho^2 = \varrho$, and hence $\text{Tr}\left\{\varrho^2\right\}  = \text{Tr} \left\{\varrho\right\} = 1$. However, for the case of a mixed state $\varrho^2 \neq \varrho$, which leads to $\text{Tr}\left\{\varrho^2\right\} < 1$. Consequently, we can come up with a definition of the `purity' of the quantum state as \cite{van2009mixed}
\begin{align}
    \xi = \text{Tr} \left\{\varrho^2\right\},
\end{align}    
    where $\xi \leq 1$. It is straightforward to prove that the maximally mixed state is $\varrho = \frac{1}{D} \mathds{1}$ where $D$ is the dimension of the associated Hilbert space. One can also use the \emph{von-Neumann entropy} defined as 
\begin{align}
    S(\varrho) \equiv - \text{Tr} \left\{\varrho \, \text{log}_2 \varrho\right\} \equiv -\sum_i \lambda_i \text{log}_2  \lambda_i,
\end{align}    
    where the $\lambda_i$ are the eigenvalues of $\varrho$. For a pure quantum state $S(\varrho) = 0$, while $S(\varrho) > 0$ for a mixed state. 

\subsection{Dynamics}

   By dynamics we mean, If an initial state $\varrho$ is given, what is the state at later time $t$? For a closed system, this is essentially done by applying the unitary operator $U(t, t_0) = e^{-\frac{i}{\hbar} H(t-t_0)}$ which connects the initial state $\varrho(t_0)$ to the final state $\varrho(t)$. We have      
\begin{align}
    \varrho(t)
    = U(t, t_0) \varrho(t_0) U^{\dagger}(t, t_0). \label{ch3:eq2}
\end{align}
    Using this relation along with the cyclic invariance feature of the trace operation, we can prove that the purity $\xi$ is time-invariant for a closed system. From equation \eqref{ch3:eq2}, we can also derive the first-order differential equation (called \emph{von-Neumann equation})
\begin{align}
    i \frac{\partial}{\partial t} \varrho(t)
    = \comm{H}{\varrho (t)},
\end{align}
    to determine the time evolution of the density matrix. For convenience, we have taken $\hbar = 1$ throughout this thesis.
    
\subsection{Entanglement}

Consider we have a \emph{bipartite} quantum system - a system made of the system $A$ and system $B$ - represented by a ket vector $\ket{\phi_{AB}}$, that is, the bipartite system state is a pure state. This state is said to be an entangled state if it cannot be cast as a tensor product of the form $\ket{\phi_{AB}} = \ket{\phi_A} \otimes \ket{\phi_B}$, where $\ket{\phi_A}$ and $\ket{\phi_B}$ are the states belonging to system $A$ and system $B$ respectively. An example is given by the spin singlet state. However, if the bipartite system state is a mixed state (described in subsection \autoref{mixstate}), we have to be more careful. A bipartite mixed state is said to be an entangled state if it is not separable, that is if it cannot be written in the following form
\begin{align}
    \varrho=\sum_k \mathds{P}_k \varrho^k_A \otimes \varrho^k_B,   
\end{align}
where the $\mathds{P}_k$ are probabilities with $\sum_k \mathds{P}_k = 1$, and $\varrho^k_A$ and $\varrho^k_B$ are states for system $A$ and $B$ respectively. 
    
\subsection{The reduced density matrix}
    
    The reduced density matrix formalism plays an important role in the description of open quantum systems. Suppose that a quantum system $S$ is interacting with another quantum system $E$. The combined state, which, in general, will be an entangled state, is given by $\varrho_{\text{SE}}$. If we are interested in system $S$ alone, we can take a partial trace over $E$. This is written as 
\begin{align}
    \varrho_S
    = \text{Tr}_E \left\{\varrho_{\text{SE}}\right\}.
\end{align}
    Let us look at the action of the partial trace more explicitly. Let $\varrho_S = \ket{s}\bra{s}$ and $\varrho_E = \ket{e}\bra{e}$. The combined state is also pure then. Now  
\begin{align}
    \text{Tr}_E \left\{\varrho_{\text{SE}}\right\}
    = \text{Tr}_E \underbrace{\left\{\ket{s}\bra{s} \otimes \ket{e}\bra{e} \right\}}_{\emph{operator in } \mathcal{H}_S \otimes \mathcal{H}_E}
    = \ket{s}\bra{s} \text{Tr}_E \left\{\ket{e}\bra{e}\right\}
    = \underbrace{\ket{s}\bra{s}}_{\emph{operator in } \mathcal{H}_S},
\end{align}
as should be the case. Note that the partial trace is linear. 

\subsection{Bloch sphere representation}

\begin{figure}
    \centering
    \includegraphics[scale = 0.4]{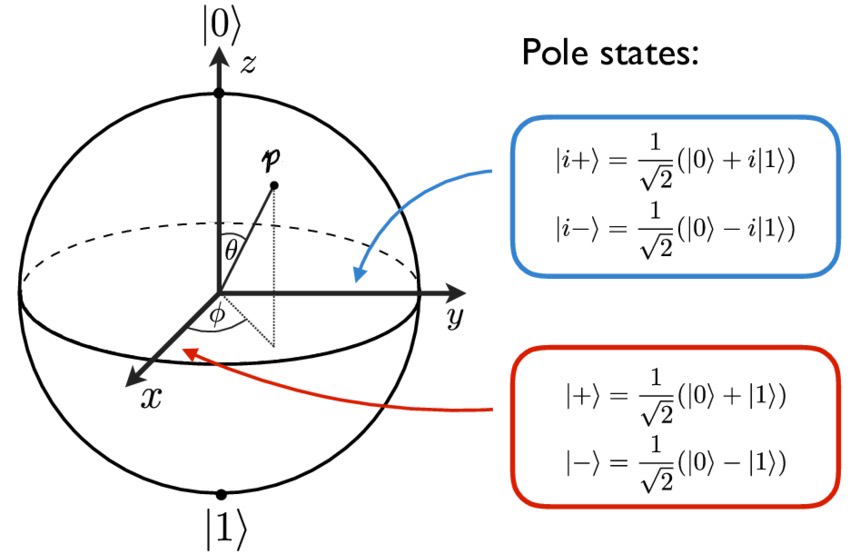}
    \caption{The Bloch sphere. The diagram has been taken from \href{https://nlbao.page/quantum/}{NL Bao's blog} with due permission.}
    \label{ch3:bloch}
\end{figure}

The \href{https://en.wikipedia.org/wiki/Bloch_sphere}{Bloch sphere} provides a nice visualization tool for the state of a two-level system. It is a sphere of unit radius. The points lying on the surface of the Bloch sphere describe a \emph{pure state} whereas the points lying inside it represent a \emph{mixed state}. As shown in Fig.~\ref{ch3:bloch}, the poles correspond to the states $\ket{0}$ and $\ket{1}$, which are eigenstates of $\sigma_z$ and the corresponding eigenvalues are $+1$ and $-1$ respectively. Any arbitrary state vector $\ket{\psi}$ can be written as a linear combination of $\ket{0}$ and $\ket{1}$ as 
\begin{align}
    \ket{\psi}
    = \cos{\left(\frac{\theta}{2}\right)}\ket{0} + e^{i\phi} \sin{\left(\frac{\theta}{2}\right)}\ket{1},
\end{align}
where $0 \leq \phi < 2\pi$ and $0\leq \theta \leq \pi$ are the spherical polar angles. Every state of a two-level quantum system can be mapped to vector lying either on the surface of this Bloch sphere (for pure states) or within it (for mixed states), and the dynamics of the quantum state can be translated to the dynamics of this Bloch vector instead. Note that the density matrix for a two-level system can be written in terms of Bloch vector components and Pauli matrices as
\begin{align*}
    \varrho
    =\frac{1}{2} \left(\mathds{1} + \bm{p}.\bm{\sigma} \right)
    = \frac{1}{2} \left( \begin{array}{cc}
        1+ p_y &  p_x + ip_y\\
        p_x - ip_y & 1 - p_y
    \end{array}
    \right),
\end{align*}
where $\bm{p} = \hat{x}p_x + \hat{y}p_y + \hat{z}p_z$ gives the Bloch vector. These numbers can be found from $p_i = \text{Tr}\left\{\varrho \sigma_i\right\}$. 

\section{derivation of master equation}

Now we quickly review the usual Born-Markov master equation \cite{BPbook}. We consider the system $S$ interacting with its surrounding environment $E$. The total System-Environment $(\mathcal{SE})$ state is denoted by $\varrho_{\text{SE}}$. The system's reduced density matrix can found by taking a partial trace over the environment, that is
\begin{align}
	\varrho_S(t)
	&= \text{Tr}_{E}\left\{\varrho_{\text{SE}}(t) \right\} 
	= \text{Tr}_{E}\left\{U(t)\varrho_{\text{SE}}(0) U^{\dagger}(t)\right\}  \label {eq2.1},
\end{align}
	here $U(t)$ is the unitary time-evolution operator for the composite system. Thus, in order to perceive the system dynamics, we first need to understand the dynamics of a composite system, which is generally an impossible task. Nevertheless, with some approximations and assumptions as detailed below, we can obtain a first-order differential equation for the system density matrix, that is, the master equation. The general form of this master equation is 
\begin{align}	
	\frac{d\varrho_S}{dt} = -i \left[H'_S, \varrho_{S} (t) \right] + \mathcal{D} [\varrho_{S} (t)]
\end{align}
	The first term corresponds to the system's `free' coherent evolution. Note that, due to the environment, a restructuring of the system's energy levels can take place, thereby changing the system Hamiltonian from $H_S$ to $H_S'$. The second term, namely $\mathcal{D} [\varrho_{S} (t)]$, called the dissipator, includes the effect of dissipation and decoherence. 
	
	Before presenting the derivation of suchlike master equation, let us highlight the usual assumptions and approximations made. These are: 
	\begin{enumerate}
	\item  \emph{Born approximation}. This states that the interaction strength between the system and its environment is significantly weak such that perturbation theory can be applied. 
	\item  \emph{Markov approximation}. This ignores `memory' effects, and hence is for `memoryless' environments. What we mean by a `memoryless' environment here is that the environment correlation time is very short - the environment forgets any information of the system very rapidly \cite{schlosshauer2007decoherence}.   
	\item \emph{Ignore initial $\mathcal{SE}$ correlations}. The initial state of the composite system is taken as a simple product state with the environment in a thermal equilibrium state. 
	\end{enumerate}
Now we succinctly outline the derivation of the master equation with these approximations. First we write the total Hamiltonian of the system and the environment as 
	$$ H = H_S + H_E + H_{\text{SE}},$$
	where $H_{\text{SE}}$ denotes the $\mathcal{SE}$ interaction Hamiltonian. It is convenient to set down the `free' Hamiltonian as $H_0= H_S + H_E$, so that $H=H_0 + H_{\text{SE}}$. We then have 
\begin{align}
    	\Dot{\varrho}
    	&= -i\left[H_0 , \varrho (t)\right] - i\left[H_{\text{SE}} , \varrho (t)\right],
\end{align}
	where $\varrho(t)$ is the density matrix for the composite system and $\Dot{\varrho}$ represents its total time derivative.
	For convenience, we now transform to the interaction picture via
\begin{align}
	\widetilde{\varrho}(t)
	&=e^{iH_{0}t} \varrho(t) e^{-iH_{0}t},\label{ch2-7}
\\
	\widetilde{H}_{\text{SE}}(t)
	&=e^{iH_{0}t} H_{\text{SE}}e^{-iH_{0}t}, \label{ch2-8}
\end{align}
	where the `tildes' denote operators written in the interaction picture. Now taking the time derivative on both sides of Eq.~\eqref{ch2-7},
\begin{align}
	\Dot{\widetilde{\varrho}}(t)
	&=iH_{0}e^{iH_{0}t} \varrho(t) e^{-iH_{0}t} + e^{iH_{0}t} \Dot{\varrho}(t) e^{-iH_{0}t} - ie^{iH_{0}t} \varrho(t) e^{-iH_{0}t}H_{0},\nonumber
\\
	&=i\left[H_{0} ,\widetilde{\varrho}(t)\right]  - i\left[H_0 , \widetilde{\varrho}(t)\right] - i\left[\widetilde{H}_{\text{SE}} (t), \widetilde{\varrho}(t)\right],\nonumber
\\
	&= -i\left[\widetilde{H}_{\text{SE}} (t), \widetilde{\varrho}(t)\right]. \label{ch2-4}
\end{align}
Integrating both sides of Eq.~\eqref{ch2-4}, we obtain
\begin{align}
    	\widetilde{\varrho}(t)
    	=\varrho(0) -i\int_{0}^{t} ds\left[ \widetilde{H}_{\text{SE}} (s) , \widetilde{\varrho}( s)\right].\nonumber
\end{align}
	Putting this back in Eq.~\eqref{ch2-4}, we have
\begin{align}
    	\Dot{\widetilde{\varrho}}(t)
    	=&-i\left[ \widetilde{H}_{\text{SE}} (t) , \varrho(0)\right] - \int_{0}^{t} ds \left[\widetilde{H}_{\text{SE}} (t) , \left[ \widetilde{H}_{\text{SE}} (s) , \widetilde{\varrho}( s)\right]\right].\nonumber
\end{align}
	We can now acquire the required system density operator by taking a partial trace over the environment, that is,
\begin{align}	
	\Dot{\widetilde{\varrho}}_{S}(t)
    	=&-i \text{Tr}_{E}\left[ \widetilde{H}_{\text{SE}} (t) , \varrho(0)\right] - \int_{0}^{t} ds \text{Tr}_{E}\left[\widetilde{H}_{\text{SE}} (t) , \left[ \widetilde{H}_{\text{SE}} (s) , \widetilde{\varrho}( s)\right]\right].\label{ch2-5}
\end{align}
	We now write the interaction Hamiltonian as $\widetilde{H}_{\text{SE}}=\sum _{\alpha}\widetilde{S}_{\alpha} \otimes \widetilde{E}_{\alpha}$ where $\widetilde{S}_\alpha$ are operators live in the system Hilbert space and $\widetilde{E}_\alpha$ are operators belonging to the environment Hilbert space. Doing so, it is quit explicit that the first term in Eq.~\eqref{ch2-5} is proportional to $\av{E_\alpha}_{E}$. It is straightforward to prove this term to be zero for most $\mathcal{SE}$ models. We are then left with
\begin{align}	
	\Dot{\widetilde{\varrho}}_{S}(t)
    	=- \sum _{\alpha\beta} \int_{0}^{t} ds \text{Tr}_{E}\left\{\widetilde{S}_{\alpha} (t)\otimes \widetilde{E}_{\alpha}(t) , \left[\widetilde{S}_{\beta}(s) \otimes \widetilde{E}_{\beta} (s), \widetilde{\varrho}(s)\right]\right\}.\nonumber
\end{align}
Until now, no approximations have been made. However, it should be recognized that the evaluation of the right-hand side necessitates knowing the dynamics of the system density matrix, but that is precisely what we are trying to calculate. Therefore, to actually achieve a practically useful differential equation, we generally need to make a number of approximations and assumptions. First, we presume that  
there are no correlations present between the system and its environment initially, and the environment is in a thermal equilibrium state. The total $\mathcal{SE}$ initial state is then put down as  
\begin{align}
	\varrho(0)
	=\varrho_{S}(0)\otimes\varrho_{E}(0),
\end{align}
with an equivalent relation $\widetilde{\varrho}(0)=\widetilde{\varrho}_{S}(0)\otimes\varrho_{E}(0)$ in the interaction picture \cite{brito2015knob}. We remind ourselves that this supposition is questionable if the $\mathcal{SE}$ interaction is even weak. Similarly, again assuming that the $\mathcal{SE}$ coupling is weak (the Born approximation), it is reasonable to suppose that we can replace $\widetilde{\varrho}(s)$ by $\widetilde{\varrho}_S(s) \otimes \varrho_E$. The master equation then takes the following form
\begin{align}	
	\Dot{\widetilde{\varrho}}_{S}(t)
    	=- \sum _{\alpha\beta} \int_{0}^{t} ds \text{Tr}_{E}\left\{\left[\widetilde{S}_{\alpha} (t)\otimes \widetilde{E}_{\alpha}(t) , \left[ \widetilde{S}_{\beta}(s) \otimes \widetilde{E}_{\beta} (s), \widetilde{\varrho}_{S}(s)\otimes \varrho_{E}\right]\right]\right\}.\label{ch2-6}
\end{align}
Notice now that the trace over the environment degree of freedom leads us to the emergence the environment correlation functions  
\begin{align}
	E_{\text{corr}}^{\alpha\beta} (t,s)
	\equiv \text{Tr}_{E} \left\{ \varrho_{E} \widetilde{E}_{\alpha}(t) \widetilde{E}_{\beta}(s) \right\} 
	=\av{\widetilde{E}_{\alpha}(t)\widetilde{E}_{\beta}(s)}_E.
\end{align}
Now using the cyclic invariance feature of the trace operation, we find   
\begin{align}
    \av{\widetilde{E}_{\alpha}(t)\widetilde{E}_{\beta}(s)}_E
    &=\text{Tr}_E\left\{e^{i H_E t}E_{\alpha}e^{-i H_E t}e^{i H_E s}E_{\beta}e^{-i H_E s}\varrho_E\right\},\nonumber
\\
    &=\text{Tr}_E\left\{e^{i(t-s) H_E }E_{\alpha}e^{-i(t-s) H_E }E_{\beta}\varrho_E\right\},\nonumber
\\
    &=\text{Tr}_E\left\{\widetilde{E}_{\alpha}(t-s)E_{\beta}\varrho_E\right\},\nonumber
\\
    &=\av{\widetilde{E}_{\alpha}(t-s)E_{\beta}}_E,\nonumber.
\end{align}
Consequently, we are allowed to rewrite the environment correlation functions as 
\begin{align}
	E_{\text{corr}}^{\alpha\beta} (t,s)
	= \av{\widetilde{E}_{\alpha}(t-s)E_{\beta}} 
	\equiv E_{\text{corr}}^{\alpha\beta} (t-s).\label{ch2-17}
\end{align}
Writing the double commutator in Eq.~\eqref{ch2-6} explicitly along with the environment correlation function
\begin{align}
	\Dot{\widetilde{\varrho}}_{S}(t)
    	= &- \sum _{\alpha\beta} \int_{0}^{t} ds \Big\{E_{\text{corr}}^{\alpha\beta} (t-s) \left[\widetilde{S}_{\alpha} (t) \widetilde{S}_{\beta}(s) \widetilde{\varrho}_{S}(s) - \widetilde{S}_{\beta} (s) \widetilde{\varrho}_{S}(s)\widetilde{S}_{\alpha}(t) \right]\nonumber
\\	
	& + E_{\text{corr}}^{\beta\alpha} (s-t) \left[\widetilde{\varrho}_{S}(s)\widetilde{S}_ {\beta}(s) \widetilde{S}_{\alpha}(t) - 		\widetilde{S}_{\alpha} (t) \widetilde{\varrho}_{S}(s)\widetilde{S}_{\beta}(s) \right]\Big\}.
\end{align}
In deriving the above equation we have once again used the cyclic invariance property of the trace.

At this point, the Markov approximation is usually made. This assumes that the environment is `memoryless'. More rigorously, the environment correlation functions decay as a function of time. If the environment correlation time is very small in comparison with the characteristic time scale in which the system state changes noticeably, then we are in the Markovian regime. Particularly, the environment correlation function $E_{\text{corr}}^{\alpha\beta}(t-s)$ is sharply peaked about $t-s=0$ and quickly decays. By changing variable to $\tau = t - s$, our master equation can then be put in the following form
\begin{align}
	\Dot{\widetilde{\varrho}}_{S}(t)
    	= - \sum _{\alpha\beta} \int_{0}^{\infty} &d\tau \Big\{E_{\text{corr}}^{\alpha\beta} (\tau) \left[\widetilde{S}_{\alpha} (t) \widetilde{S}_{\beta}(t-\tau) \widetilde{\varrho}_{S}(t) - \widetilde{S}_{\beta} (t-\tau) \widetilde{\varrho}_{S}(t) \widetilde{S}_{\alpha}(t) \right],\nonumber
\\	
	& + E_{\text{corr}}^{\beta\alpha} (-\tau) \left[\widetilde{\varrho}_{S}(t) \widetilde{S}_ {\beta}(t-\tau) \widetilde{S}_{\alpha}(t) - 	\widetilde{S}_{\alpha} (t) \widetilde{\varrho}_{S}(t) \widetilde{S}_{\beta}(t-\tau) \right]\Big\}. \label{ch2-9}
\end{align}
Finally, we switch back to the Schrodinger picture to obtain
\begin{align*}
	\Dot{\varrho_S}
    	= -i \left[H_{S} , \varrho_{S}(t) \right] - \sum _{\alpha\beta} \int_{0}^{\infty} & d\tau \left\{E_{\text{corr}}^{\alpha\beta} (\tau) \left[S_{\alpha} , S_{\beta}(-\tau) \varrho_{S}(t) \right] 
	+ E_{\text{corr}}^{\beta\alpha} (-\tau) \left[\varrho_{S}(t)S_ {\beta}(-\tau) , S_{\alpha} \right]\right\}.
\end{align*}
Making use of the integrals 
\begin{align}
P_{\alpha}
	&\equiv\int^{\infty}_{0} d\tau \sum _{\beta} E_{\text{corr}}^{\alpha\beta}(\tau) \widetilde{S}_{\beta}(-\tau),
\\	
    Q_{\alpha}
	&\equiv\int^{\infty}_{0} d\tau \sum _{\beta} E_{\text{corr}}^{\beta\alpha}(-\tau) \widetilde{S}_{\beta}(-\tau),
\end{align}
	where $\widetilde{S}_{\beta}$ is in fact a system operator $S_{\beta}$ but now written in the interaction picture, the Born-Markov master equation can be put in the more compact and final form 
\begin{equation}
    \mathcolorbox{Apricot}{\varrho_{S}(t)
    	= -i \left[H_{S} , \varrho_{S}(t) \right] - \sum _{\alpha}  \Big\{ \left[S_{\alpha} , P_{\alpha} \varrho_{S}(t) \right] + \left[\varrho_{S}(t)Q_ {\alpha} , S_{\alpha} \right]\Big\}.}
\end{equation}

\section{The pure dephasing model} 
\label{pure}

We now present a example of an open quantum system where the system dynamics can be found exactly. We consider $N$ identical two-level systems interacting with a common environment of harmonic oscillators, with the assumption that we can ignore dissipation and only consider decoherence \cite{chaudhry2013understanding}. The dynamics of such a system undergoing pure dephasing, can be described by the Hamiltonian
\begin{align}
    H_\text{tot}
    =H_S+H_E+H_{\text{SE}}, \nonumber
\end{align}
with,
\begin{align*}
    H_S
    &= \varepsilon J_z,
\\
    H_E
    &= \sum_k \omega_k b_{k}^{\dagger} b_k,
\\
    H_{\text{SE}}
    &=2J_z \sum_{k} \left(g_{k}^{*}b_k + g_k b_k^{\dagger} \right).
\end{align*}
Here $J_{x,y,z}$ are the relevant effective spin operators with $J_i = \sum_{k = 1}^N \frac{\sigma_i^{(k)}}{2}$, $\varepsilon$ denotes energy bias, $H_E$ is the Hamiltonian for the environment composed of collection of harmonic oscillators, while $H_{\text{SE}}$ corresponds to the $\mathcal{SE}$ interaction. Since the system Hamiltonian commutes with the interaction Hamiltonian, the system does not undergo dissipation. To work out the dynamics of the effective central large spin, we first switch the interaction Hamiltonian to the interaction picture, that is, 
\begin{align}
    H_{\text{SE}} \left(t\right)
    &=e^{i \left(H_S+H_E\right) t} H_{\text{SE}}  e^{-i  \left(H_S+H_E\right)t},\nonumber
\\ 
    &=2J_z \sum_{k} \left(g_{k}^{*}b_k e^{-i\omega_k t} + g_k b_k^{\dagger} e^{i\omega_k t}\right).
\end{align}
To find the unitary operator corresponding to the this interaction Hamiltonian, we use the Magnus expansion \cite{magnus1954exponential}, namely
\begin{align}
    U_{\text{SE}}\left(t\right)= \text{exp}\left\{\sum_{i=1}^\infty A_i\left(t\right)\right\}, \label{unitary}
\end{align}
where the first two terms in the exponent are
\begin{align*}
    A_1&=-i \int_0^t H_{\text{SE}} \left(t_1\right)  dt_1,
\\
    A_2&=-\frac{1}{2}\int _0^t dt_1\int_0^{t_1} \left[H_{\text{SE}} \left(t_1\right),H_{\text{SE}} \left(t_2\right)\right]  dt_{2}.
\end{align*}
A simple calculation leads to 
\begin{align}
    A_1 = J_z\sum_k\left[ \alpha_k \left(t\right)b_k^{\dagger} -  \alpha_k^* \left(t\right)b_k\right],
\end{align}
with
\begin{align}
    \alpha_k\left( t \right)=  \frac{2g_k}{\omega_k}\left(1-e^{i\omega_k t} \right).
\end{align}
In order to calculate the second term in the exponential of the Magnus expansion ($A_2$), we note that 
\begin{align*}
    \left[H_{\text{SE}} \left(t_1\right),H_{\text{SE}} \left(t_2\right)\right]
    =-8iJ_z^2 \sum_k \abs{g_k}^2 \sin\left[\omega_k (t_2-t_1)\right].
\end{align*}
It then follows that 
\begin{align}
    A_2
    = -iJ_z^2 \Delta \left( t \right),
\end{align}
with
\begin{align}
    \Delta \left(t\right) 
    = \sum_k \frac{4\abs{g_k}^2}{\omega_k^2}\left[\sin(\omega_k t) - \omega_k t\right].
\end{align}
Since this is a c-number, also $J_z^2$ is proportional to identity, It means that the higher order terms in the Magnus expansion are all zero. To sum up, the total unitary operator (in the Schrodinger picture) can be written as
\begin{align}
U\left(t\right) 
    &= U_S(t) U_E \left( t \right)U_{\text{SE}}\left( t \right), \nonumber 
\\
    &= e^{-i\varepsilon J_z t} e^{-i H_E t} \text{exp} \left\{ J_z \sum_k \left[ \alpha_k \left(t\right)b_k^{\dagger} -  \alpha_k^* \left(t\right)b_k \right] -iJ_z^2   \Delta \left( t \right)  \right\}.
\end{align}
We use this exact time-evolution operator to find the reduced dynamics of the system. We first allow our system and environment to evolve together under the time evolution operator. After that, we take trace over the environment to obtain
\begin{align}
    \varrho_S \left(t\right) 
    = \text{Tr}_E \left\{ U(t) \varrho(0)U^{\dagger}(t) \right\},
\end{align}
here $\varrho (0)$ is the joint initial state of the system and its environment. It is useful to write system density operator in matrix form using the $J_z$ eigenbasis, namely
\begin{align}
    \left[\varrho_S \left(t\right)\right]_{uv} 
    = \text{Tr}_{\text{S,E}} \left\{ U\left(t\right) \varrho \left(0\right)U^{\dagger}\left(t\right) P_{uv} \right\}, \label{ch2-10}
\end{align}
where we have introduced an operator $P_{uv} = \ket{v}\bra{u}$ belonging to the system Hilbert space and $\ket{v}$ and $\ket{u}$ are eigenstates of $J_z$ with $J_z\ket{u}=u\ket{u}$. This enables us to write 
\begin{align}
    [\varrho_S(t)]_{uv} 
    = \text{Tr}_{\text{S,E}} \left\{ \varrho\left(0\right) P_{uv}\left(t\right) \right\},\label{ch2-12}
\end{align}
where $P_{uv}\left(t\right) = U^{\dagger} \left(t\right) P_{uv} U\left(t\right)$. Using our found time-evolution operator, it is easy to check that 
\begin{align}
    P_{uv}\left(t\right) 
    &= e^{-i\varepsilon \left( u-v \right) t}  e^{-i \left( u^2 - v^2 \right) \Delta \left( t \right)} e^{-R_{uv}\left(t\right)} P_{uv} ,\label{ch2-11}
\end{align}
with
\begin{align}
    R_{uv}\left(t\right) = \left(u-v\right) \sum_k \left[ \alpha_k \left(t\right)b_k^{\dagger} -  \alpha_k^* \left(t\right)b_k \right].
\end{align}
Inserting Eq.~\eqref{ch2-11} into Eq.~\eqref{ch2-12}, we obtain 
\begin{align}
    \mathcolorbox{Apricot}{\left[\varrho_S \left(t\right)\right]_{uv} 
    = e^{-i\varepsilon \left( u-v \right) t} e^{-i\Delta \left( t \right) \left( u^2 - v^2 \right) t}\text{Tr}_{\text{S,E}} \left\{ \varrho \left(0\right) e^{-R_{uv}\left(t\right)} P_{uv} \right\}.}\label{ch2-13} 
\end{align}
This is a general expression for the reduced system dynamics which does not assume a particular form of the initial $\mathcal{SE}$ state. In other words, it is valid for both correlated and uncorrelated initial states. In the following two subsections, we will work out $\left[\varrho_S \left(t\right)\right]_{uv} $ for correlated and uncorrelated cases separately.

\subsection{Dynamics for uncorrelated initial state}

We proceed by considering the usual initial product state
\begin{align}
    \varrho\left(0\right) 
    = \varrho_S\left(0\right) \otimes \varrho_E,
\end{align}
where $\varrho_S(0) = \frac{e^{-\beta H_S}}{Z_S}$ and $\varrho_E = \frac{e^{-\beta H_E}}{Z_E}$ with  $Z_E = \text{Tr}_E \left\{ e^{-\beta H_E} \right\}$ and $Z_S = \text{Tr}_S \left\{ e^{-\beta H_S} \right\}$ being their partition functions respectively. $\beta$, of course, denotes the inverse of temperature with $k_B=1$. Eq.~\eqref{ch2-13} can now be written as
\begin{align}
    \left[\varrho_S \left(t\right)\right]_{uv} 
    &=\left[\varrho_S \left(0\right)\right]_{uv} e^{-i\varepsilon \left( u-v \right) t} e^{-i\Delta \left( t \right) \left( u^2 - v^2 \right) } \text{Tr}_{E} \left\{ \varrho_Ee^{-R_{uv}\left(t\right)}\right\}.
    \label{ref4}
\end{align}
Our aim now is to simplify $\text{Tr}_{E} \left\{ \varrho_Ee^{-R_{uv}\left(t\right)}\right\}=\av{e^{-R_{uv}\left(t\right)}}$. Since the environment modes of the harmonic oscillators are independent of each other, 
\begin{align}
    \text{Tr}_{E} \left\{ \varrho_E e^{-R_{uv}\left(t\right)}\right\} 
    &= \prod_k \av{e^{ -\left( v -u\right) \left[\alpha_k \left(t\right)b_k^{\dagger} -  \alpha_k^* \left(t\right)b_k\left(t\right)\right] } }.
\end{align}
Since $R_{uv}\left(t\right)$ is a linear combination of $b_k$ and $b_k^{\dagger}$, we can write
\begin{align}
    \av{e^{-R_{uv}\left(t\right)}}
    &= e^ {\frac{1}{2}\av{-R^2_{vu}\left(t\right)}} \nonumber 
\\
    &= \prod_k  \text{exp} \left\{-\frac{1}{2}\left( v -u\right)^2 \abs{\alpha_k\left(t\right)}^2 \av{2n_k + 1 }\right\},
\end{align}
where we have defined $n_k = \av{b_k^{\dagger} b_k}$. Since the environment is in a thermal equilibrium state, therefore $n_k$ is simply the Bose-Einstein distribution, that is, $n_k =  \frac{1}{e^{{\beta\omega_k}} -1}=\frac{1}{2}\left\{\coth\left(\frac{\beta \omega_k}{2}\right)-1\right\}$. Using $\abs{\alpha_k \left(t\right)}^2=\frac{8\abs{g_k}^2}{\omega_k^2}\left\{1-\cos{\left(\omega_kt\right)}\right\}$, we obtain
\begin{align}
    \text{Tr}_{E} \left\{ \varrho_Ee^{-R_{uv}\left(t\right)}\right\}
    = \text{exp}\left\{ -\sum_k (u-v)^2  \frac{4 \abs{g_k}^2}{\omega_k^2} \left[1-\cos\left(\omega_k t \right)\right] \coth{\left( \frac{\beta \omega_k}{2}\right)}\right\}. \nonumber
\end{align}
Using this, Eq.~\eqref{ref4} becomes 
\begin{align}
    \mathcolorbox{Apricot}{\left[\varrho_S \left(t\right)\right]_{uv} 
    = \left[\varrho_S \left(0\right)\right]_{uv}e^{-i\varepsilon \left( u-v \right) t} e^{-i\Delta \left( t \right) \left( u^2 - v^2 \right) }e^{-\Gamma\left(t\right)\left( u-v\right)^2},}
\end{align}
with
\begin{align}
    \Gamma\left(t\right)
    =\sum_k   \frac{4 \abs{g_k}^2}{\omega_k^2} \left[1-\cos\left(\omega_k t \right)\right]\coth\left(\frac{\beta \omega_k}{2} \right).
\end{align}
The factor $e^{-i\Delta \left( t \right) \left( u^2 - v^2 \right) }$ encapsulates the indirect inter-spin interactions due to the common environment while $e^{-\Gamma\left(t\right)\left( u-v\right)^2}$ captures the effect of decoherence.

\subsection{Dynamics for correlated initial state}
\label{corrStatPrep}
For the case of the correlated initial $\mathcal{SE}$ state, we consider \cite{chaudhry2013understanding}
\begin{align}
    \varrho \left(0\right) 
    = \frac{\Theta e^{-\beta H_{\text{tot}}} \Theta^{\dagger}}{Z}, \label{ref5}
\end{align}
where $Z= \text{Tr}_{\text{S,E}}\left\{\Theta e^{-\beta H_{\text{tot}}} \Theta^{\dagger}\right\}$ is the total partition function. Since we have not specified $\Theta$ yet, so  $\Theta$ can be any operator, acting on the system only, to prepare the desired initial state. Subsequently, we will specify this operator and study its effect separately. For the time being, we derive a general expression for the reduced dynamics. To proceed, we first simplify the partition function by inserting a completeness relation over the $J_z$ eigenstates
\begin{align}
    Z
    &=  \text{Tr}_{\text{S,E}}\left\{\Theta e^{-\beta H} \Theta^\dagger\right\}, \nonumber 
\\
    &= \text{Tr}_{\text{S,E}}\left\{\Theta^\dagger\Theta e^{-\beta H_S}e^{-\beta( H_{E}+ H_{\text{SE}})} \right\}, \nonumber 
\\
    &= \sum_{l}\text{Tr}_{\text{S,E}}\left\{\Theta^\dagger\Theta e^{-\beta\varepsilon J_z  }   e^{-\beta \left\{H_{E} + 2 J_z \sum_{k} \left(g_{k}^{*}b_{k}  + g_{k} b_{k}^\dagger \right)\right\}}\ket{l}\bra{l} \right\}, \nonumber
\\
    &=\sum_{l} e^{-\beta\varepsilon l } \bra{l} \Theta^\dagger\Theta \ket{l} \text{Tr}_{E}\left\{ e^{-\beta \left\{H_{E} + 2 J_z \sum_{k} \left(g_{k}^{*}b_{k}  + g_{k} b_{k}^\dagger \right)\right\}} \right\}, \nonumber
\\
    &=  \sum_{l} e^{-\beta\varepsilon l } \bra{l} \Theta^\dagger\Theta \ket{l}  \text{Tr}_{E}\left\{e^{-\beta H_{E}^{\left(l\right)}} \right\}, \label{ch2-14}
\end{align}
where $H_E^{\left(l\right)} 
    = H_E + 2l \sum_k \left( g_{k}^{*}b_k + g_k b_k^{\dagger} \right)$ is a `shifted' Hamiltonian of the environment. To further simplify the trace over the environment, we introduce displaced harmonic oscillator modes defined as
\begin{align}
    B_{k,l}
    = b_k + \frac{2lg_k}{\omega_k},\label{ch2-1}
\\
    B_{k,l}^{\dagger}= b_k^{\dagger} + \frac{2lg_k^*}{\omega_k}.\label{ch2-2}
\end{align}
The trace over the environment then simplifies to 
\begin{align}
    \text{Tr}_{E}\left\{ e^{-\beta H_{E}^{\left(l\right)}} \right\}
    &=\text{Tr}_{E}\left\{ e^{-\beta\sum_{k}  \omega_{k} b_{k}^{\dagger} b_{k} - 2 \beta  l \sum_{k} \left( g_{k}^{*}b_{k} + g_{k} b_{k}^\dagger \right)} \right\},\nonumber
\\
    &= \text{Tr}_{E}\left\{ e^{-\beta\sum_{k}  \left[\omega_{k} \left(B_{k,l}^\dagger-\frac{ 2l g_{k}^*}{\omega_{k}}\right) \left(B_{k,l}-\frac{ 2l g_{k}}{\omega_{k}}\right) + 2 l  \left\{ g_{k}^{*}\left(B_{k,l}-\frac{ 2l g_{k}}{\omega_{k}}\right) + g_{k} \left(B_{k,l}^\dagger-\frac{ 2l g_{k}^*}{\omega_{k}}\right) \right\}\right]} \right\},\nonumber
\\
    &= e^{\beta l^2 \sum_{k} \frac{4\abs{g_{k}}^2}{\omega_{k}}}\text{Tr}_{E}\left\{ e^{-\beta\sum_{k}  \left\{\omega_{k} B_{k,l}^\dagger B_{k,l}
    \right\}} \right\},\nonumber
\\
    &=e^{\beta l^2 \mathcal{C} }Z_E,    
\end{align}
where $\mathcal{C} = \sum_k \frac{4 \abs{g_k}^2}{\omega_k}$, and $Z_E=\text{Tr}_{E}\left\{ e^{-\beta\sum_k  \omega_k B_{k,l}^{\dagger} B_{k,l}
    } \right\}$. In short
\begin{align}
    Z
    &=\sum_l e^{-\beta \varepsilon l}\bra{l}\Theta^{\dagger} \Theta \ket{l}e^{\beta l^2\mathcal{C}
    }Z_E.
\end{align}
Before proceeding, it is useful to write $R_{uv}\left(t\right)$ [see Eq.~\eqref{ch2-13}] in terms of the displaced harmonic modes. We get
\begin{align}
    R_{uv}\left(t\right)
    &= \left(u-v\right) \sum_{k} \left[  \alpha _{k}\left(t\right)b_{k}^\dagger -   \alpha ^*_{k} \left(t\right)b_{k} \right],\nonumber
\\
    &=\left(u-v\right) \sum_{k} \left[  \alpha _{k}\left(t\right)\left(B_{k,l}^\dagger-\frac{ 2l g_{k}^*}{\omega_{k}}\right) -   \alpha ^*_{k} \left(t\right)\left(B_{k,l}-\frac{ 2l g_{k}}{\omega_{k}}\right)\right],\nonumber
\\
     &= \left(u-v\right) \sum_{k} \left[  \alpha _{k}\left(t\right)B_{k,l}^\dagger-  \alpha ^*_{k} \left(t\right) B_{k,l}+ 2il \left(\frac{e^{i\omega_{k} t}-e^{-i\omega_{k} t} }{2i}\right)\frac{4 \abs{g_{k}}^2}{\omega_{k}^2} \right],\nonumber
\\
     &=\left(u-v\right) \sum_{k} \left[  \alpha _{k}\left(t\right)B_{k,l}^\dagger-  \alpha ^*_{k} \left(t\right) B_{k,l} \right]+2 i l \left(u-v\right)\sum_{k} \sin{\left(\omega_{k} t\right)}\frac{4\abs{g_{k}}^2}{\omega_{k}^2},\nonumber
\\
     &= \left(u-v\right) \sum_{k} \left[  \alpha _{k}\left(t\right)B_{k,l}^\dagger - \alpha ^*_{k} \left(t\right)B_{k,l} \right] + i \Phi^{(l)}_{uv}\left(t\right),
\end{align}
where
\begin{align}
    \Phi_{vu}^{\left(l\right)}\left(t\right)
    &= 2l\left(u-v\right) \phi\left(t\right),
\\
    \phi\left(t\right) 
&= \sum_k \frac{4\abs{g_k}^2}{\omega_k^2}\sin\left(\omega_k t\right).
\end{align}
We then simplify the trace in Eq.~\eqref{ch2-13}. A straightforward calculation leads to 
\begin{align}
    &\text{Tr}_{\text{S,E}} \left\{ \rho \left(0\right) e^{-R_{uv}\left(t\right)}P_{uv} \right\}
    =\frac{1}{Z}\text{Tr}_{\text{S,E}} \left\{ {\Theta e^{-\beta H_S}e^{-\beta( H_{E}+ H_{\text{SE}})} \Theta^\dagger} e^{-R_{uv}\left(t\right)}P_{uv} \right\},\nonumber
\\
    &=\frac{1}{Z}\sum_{l}\text{Tr}_{\text{S,E}}\left\{\Theta e^{-\beta\varepsilon J_z }   e^{-\beta \left\{H_{E}+ 2 J_z \sum_{k} \left(g_{k}^{*}b_{k}  + g_{k} b_{k}^\dagger \right)\right\}}\ket{l}\bra{l}\Theta^\dagger e^{-R_{uv}\left(t\right)}P_{uv}\right\},\nonumber
\\
    &=\frac{1}{Z}\sum_{l}\text{Tr}_{\text{S,E}}\left\{\Theta e^{-\beta\varepsilon l}   e^{-\beta \left\{H_{E}+ 2 l \sum_{k} \left(g_{k}^{*}b_{k}  + g_{k} b_{k}^\dagger \right)\right\}}\ket{l}\bra{l} \Theta^\dagger e^{-R_{uv}\left(t\right)}P_{uv}\right\},\nonumber
\\
    &=\frac{1}{Z}\sum_{l}e^{-\beta\varepsilon l}\bra{l}\Theta^\dagger  P_{uv}\Theta\ket{l}\text{Tr}_{E}\left\{{e^{-\beta H_{E}^{\left(l\right)}}}e^{-\left(u-v\right)\sum_{k} \left[  \alpha _{k} \left(t\right)B_{k,l}^\dagger- \alpha _{k}^*\left(t\right)B_{k,l} \right]}e^{-i \Phi^{(l)}_{uv}\left(t\right)} \right\},\nonumber
\\
    &=\frac{1}{Z}\sum_{l}e^{-\beta\varepsilon l}\bra{l}\Theta^\dagger  P_{uv}\Theta\ket{l}e^{-i\Phi^{(l)}_{uv}\left(t\right)}\text{Tr}_{E}\left\{ e^{-\beta\sum_{k}  \left\{\omega_{k} B_{k,l}^\dagger B_{k,l}- \frac{4 l^2 \abs{g_{k}}^2 }{\omega_{k}}
    \right\}} e^{- \left(u-v\right) \sum_{k} \left[  \alpha _{k} \left(t\right)B_{k,l}^\dagger- \alpha _{k}^*\left(t\right)B_{k,l} \right]} \right\},\nonumber
\\
    &=\frac{1}{Z}\sum_{l}e^{-\beta\varepsilon l}\bra{l}\Theta^\dagger  P_{uv}\Theta\ket{l}e^{-i  \Phi^{(l)}_{uv}\left(t\right)}e^{\beta  l ^2 \sum_{k}\frac{4 \abs{g_{k}}^2}{\omega_{k}}}\text{Tr}_{E}\left\{e^{-\beta\sum_{k}  \omega_{k} B_{k,l}^\dagger B_{k,l}
    } e^{-\left(u-v\right)\sum_{k} \left[  \alpha _{k} \left(t\right)B_{k,l}^\dagger- \alpha _{k}^*\left(t\right)B_{k,l} \right]} \right\},\nonumber
\\
    &=\frac{1}{Z}\sum_{l}e^{-\beta\varepsilon l}\bra{l}\Theta^\dagger  P_{uv}\Theta\ket{l}e^{-i  \Phi^{(l)}_{uv}\left(t\right)}e^{\beta  l ^2 \sum_{k}\frac{4 \abs{g_{k}}^2}{\omega_{k}}} Z_E \text{Tr}_{E}\left\{\varrho_E e^{-\left(u-v\right)\sum_{k} \left[  \alpha _{k} \left(t\right)B_{k,l}^\dagger- \alpha _{k}^*\left(t\right)B_{k,l} \right]} \right\},\nonumber    
\\
    &=\frac{1}{Z}\sum_{l}e^{-\beta\varepsilon l}\bra{l}\Theta^\dagger  P_{uv}\Theta\ket{l}e^{-i  \Phi^{(l)}_{uv}\left(t\right)}e^{\beta  l ^2 \mathcal{C}} Z_E \prod_k \expval{e^{-\left(u-v\right) \left[  \alpha _{k} \left(t\right)B_{k,l}^\dagger- \alpha _{k}^*\left(t\right)B_{k,l} \right]}} ,\nonumber
\\
    &=\frac{1}{Z}\sum_{l}e^{-\beta\varepsilon l}\bra{l}\Theta^\dagger  P_{uv}\Theta\ket{l}e^{-i \Phi^{(l)}_{uv}\left(t\right)}e^{\beta  l ^2 \mathcal{C}}Z_E e^{-\left(u-v\right)^2\Gamma\left(t\right)},
\end{align}
leading to
\begin{align}
    \left[\varrho_S\left(t\right)\right]_{uv} 
    &= e^{-i\varepsilon \left( u-v \right) t} e^{-i\Delta \left(t\right) \left(u^2 - v^2 \right) }e^{-\Gamma\left(t\right)\left( u-v\right)^2 }   \nonumber 
    \frac{\sum_l \bra{l}\Theta^{\dagger} P_{uv}\Theta \ket{l} e^{-\beta \varepsilon l}e^{\beta l^2 \mathcal{C}}e^{-i\Phi_{vu}^{\left(l\right)}\left(t\right)}}{\sum_l \bra{l}\Theta^{\dagger}\Theta \ket{l} e^{-\beta \varepsilon l}e^{\beta l^2 \mathcal{C}}}.
\end{align}
Note that at $t=0$, the above expression reduces to
\begin{align}
    \left[\varrho_S \left(0\right)\right]_{uv} = \frac{\sum_l \bra{l}\Theta^{\dagger} P_{uv}\Theta \ket{l} e^{-\beta \varepsilon l}e^{\beta l^2 \mathcal{C}}}{\sum_l \bra{l}\Theta^{\dagger} \Theta \ket{l} e^{-\beta \varepsilon l}e^{\beta l^2 \mathcal{C}}},
\end{align}
we can then write
\begin{align}
    \mathcolorbox{Apricot}{\left[\varrho_S\left(t\right)\right]_{uv} 
    =\left[\varrho_S\left(0\right)\right]_{uv} e^{-i\varepsilon \left( u-v \right) t} e^{-i\Delta \left( t \right) \left( u^2 - v^2 \right) }e^{-\Gamma\left(t\right)\left( u-v\right)^2 }\sum_l \mu^{(l)}_{uv} e^{-i\Phi_{vu}^{\left(l\right)}\left(t\right)},} \label{ch2-15}
\end{align}
with
\begin{align}
    \mu^{(l)}_{uv}
    =\frac{ \bra{l}\Theta^{\dagger} P_{uv}\Theta \ket{l} e^{-\beta \varepsilon l}e^{\beta l^2 \mathcal{C}}}{\sum_l \bra{l}\Theta^{\dagger} P_{uv}\Theta \ket{l} e^{-\beta \varepsilon l}e^{\beta l^2 \mathcal{C}}}.
\end{align}
    
    Let us now specify the operator $\Theta$. If we envisage a projective measurement to prepare the desired initial system state, we should consider $\Theta = \sum_i \mathds{P}_i \ket{\psi_i} \bra{\psi_i}$, that is, a sum of projection operators. In this case, we have
\begin{align}
        \mathcolorbox{Apricot}{\left[\varrho_S\left(t\right)\right]_{uv} 
        =\left[\varrho_S\left(0\right)\right]_{uv} e^{-i\varepsilon \left( u-v \right) t} e^{-i\Delta \left( t \right) \left( u^2 - v^2 \right) }e^{-\Gamma\left(t\right)\left( u-v\right)^2 }
        \sum_{l,i} \left[ \mathds{P}_i \mu^{(l)}_{uv} e^{-i\Phi_{vu}^{\left(l\right)}\left(t\right)}\right],}
\end{align}
with the initial state  
\begin{align}
    \left[\varrho_S\left(0\right)\right]_{uv}
    =\frac{\sum_{l,i} \mathds{P}_i \abs{\ip{l}{\psi_i}}^2\bra{\psi_i}P_{uv}\ket{\psi_i}   e^{-\beta \varepsilon l}e^{\beta l^2 \mathcal{C}}}{\sum_{l,i} \mathds{P}_i \abs{\ip{l}{\psi_i}}^2 e^{-\beta \varepsilon l}e^{\beta l^2 \mathcal{C}} }.
\end{align}
On the other hand, if we apply a unitary operator to the system to prepare the desired initial state, then $\Theta = R$ with $R^\dagger R = \mathds{1}$. In this case, we have 
\begin{align}
        \mathcolorbox{Apricot}{\left[\varrho_S\left(t\right)\right]_{uv} 
        = \left[\varrho_S\left(0\right)\right]_{uv} e^{-i\varepsilon \left( u-v \right) t} e^{-i\Delta \left( t \right) \left( u^2 - v^2 \right) } e^{-\Gamma\left(t\right)\left( u-v\right)^2 } \sum_{l} \left[  \mu^{(l)}_{uv} e^{-i\Phi_{vu}^{\left(l\right)}\left(t\right)}\right],}
\end{align}
with 
\begin{align}
    \mu^{(l)}_{uv}
    &=\frac{ \bra{l}R^{\dagger}\ket{v}\bra{u}R\ket{l} e^{-\beta \varepsilon l}e^{\beta l^2 \mathcal{C}}  }{\sum_l \bra{l}R^{\dagger}\ket{v}\bra{u}R\ket{l} e^{-\beta \varepsilon l}e^{\beta l^2 \mathcal{C}}  },
\\
    \left[\varrho_S\left(0\right)\right]_{uv}
    &=\frac{\sum_l \bra{l}R^{\dagger}\ket{v}\bra{u}R\ket{l} e^{-\beta \varepsilon l}e^{\beta l^2 \mathcal{C}}}{\sum_l e^{-\beta \varepsilon l}e^{\beta l^2 \mathcal{C}}}.
 \end{align}

\section{Summary}
\label{ch2-Sum}

In this chapter, we established the fundamental terminology needed to understand open quantum systems. We derived a standard master equation under the Born-Markov approximation. In the end, we presented a detailed derivation of the system density operator for a pure dephasing model. In the upcoming chapters, these results will be used. For example, we will derive a master equation without specifying that the initial $\mathcal{SE}$ state is a product state. This master equation will then be applied to a collection of two-level systems coupled to the harmonic oscillator environment.

  \chapter{Initial correlations in spin environment}\label{c:SpinEnv}
  In this chapter, to gain insights into the role of initial correlations, we solve an exactly solvable model of a single qubit (or two-level system) interacting with a collection of qubits (spin environment), both with and without considering the initial System-Environment $(\mathcal{SE})$ correlations. We are able to obtain the dynamics of the central qubit exactly. We show that the effect of the  initial correlations can be important. We then extend our study to investigate the dynamics of the entanglement between two qubits interacting with a common spin environment. Once again, we demonstrate that the effect of the initial correlations can be very significant.

This chapter is organized as follows. In section \autoref{spinspin}, we present a Spin-Spin $(\mathcal{SS})$ model with a single central spin. We prepare the desired initial system spin state via a unitary operator, and we analyze the subsequent dynamics to show the impact of the initial $\mathcal{SE}$ correlations. In the next section \autoref{2qubits}, we shall consider two central spins coupled to the common environment. We look at the entanglement dynamics and quantify the role of the initial correlations. In the last section \autoref{ch3-sum}, we present a summary of this chapter. We put some detailed mathematical derivations in Appendix \autoref{app:A}.

\section{The Spin-spin Model}

\label{spinspin}
We first consider a single spin-half system (a qubit) interacting with $N$ spin-half systems (the spin environment). We write the $\mathcal{SE}$ Hamiltonian as 
\begin{align}
    H_{\text{tot}} =
    \begin{cases}
      H_{\text{S0}} + H_E + H_{\text{SE}} & t\le 0,\\
      H_{S} + H_E + H_{\text{SE}} & t > 0.\\
    \end{cases}   
\end{align}
Here $H_E$ is the Hamiltonian of the spin environment alone, and $H_{\text{SE}}$ is the $\mathcal{SE}$ interaction. We prepare a desired initial state at time $t = 0$; the system Hamiltonian after this state preparation process can be different as compared to the system Hamiltonian before the state preparation process. As such, $H_S$ denotes the system Hamiltonian corresponding to the coherent evolution of the system only after the initial time $t = 0$ at which the system state is prepared. $H_{\text{S0}}$ is the system Hamiltonian before the system state preparation, with the parameters in $H_{\text{S0}}$ chosen so as to aid the state preparation process. Note that $H_{\text{S0}}$ is similar to $H_S$ in the sense that both operators live in the same Hilbert space, but they may have different parameters. For the $\mathcal{SS}$ model that we are discussing, we have 
\begin{align}
    H_{\text{S0}}
    &= \frac{\varepsilon_0}{2} \sigma_z +\frac{\Delta_0}{2} \sigma_x,
\\
 H_{S}
    &= \frac{\varepsilon}{2} \sigma_z +\frac{\Delta_0}{2} \sigma_x,
\\
    H_E
    &=\sum_{i=1}^{N} \frac{\varepsilon_i}{2} \sigma^{(i)}_z + \sum_{i=1}^{N} \alpha_i \sigma^{(i)}_z \sigma^{(i+1)}_z ,
\\
    H_{\text{SE}}
    &= \frac{1}{2} \sigma_z \otimes \sum_{i=1}^{N} g_i\sigma^{(i)}_z .
\end{align}
Here $\sigma_{x, y, z}$ represent the standard Pauli spin matrices, $\varepsilon_0$ and $\varepsilon$ denote the energy biasing of the central spin system before and after the state preparation respectively, $\Delta_0$ is the tunneling amplitude, and $\varepsilon_{i}$ denotes the energy bias for the $i^{\text{th}}$ spin of the environment. 
Environment spins interact with each other via $\sum_{i=1}^{N} \alpha_i \sigma^{(i)}_z \sigma^{(i+1)}_z$, where $\alpha_i$ denotes the nearest neighbor coupling strength between the environment spins. The central spin system interacts with the environmental spins via interaction Hamiltonian $H_{\text{SE}}$, where $g_i$ denotes the coupling strength between the central qubit and the $i^{\text{th}}$ spin of the environment. It is clear that our system Hamiltonian $H_S$ does not commute with the total Hamiltonian, meaning that the system energy is not conserved. 

Our primary goal is to find the dynamics of our central qubit system. To do that, we first obtain the total unitary time evolution operator $U(t)$ for the system and its environment as a whole. We write $H_{\text{SE}}=S\otimes E$, where $S$ is a system operator and $E$ is an environment operator. Now, the states $\ket{n}=\ket{n_1}\ket{n_2}\ket{n_3}...\ket{n_N} $ are the eigenstates of $E$ with $n_i=0$ signifying the spin-up along $z$ state while $n_i = 1$ is the spin-down state. We then have
\begin{align}
    E\ket{n}
    = e_n\ket{n},
\end{align}
with $e_n= \sum^N_{i=1} (-1)^{n_i} g_i$. We also have
\begin{align}
    \sum_{i=1}^{N} \varepsilon_i \sigma^{(i)}_z\ket{n}
    = \epsilon_n \ket{n},\label{eq1}
\\
    \sum_{i=1}^{N} \alpha_i\sigma^{(i)}_z\sigma^{(i+1)}_z  \ket{n}
    = \lambda_n \ket{n},\label{eq2}
\end{align}
where $\epsilon_n= \sum^N_{i=1} (-1)^{n_i} \varepsilon_i$ and $\lambda_n= \sum^N_{i=1}\alpha_i (-1)^{n_i}(-1)^{n_{i+1}}$ are the eigenvalues of the first and second terms of the environment Hamiltonian respectively. 

\subsection{\label{sec:level2A}Initial state preparation without correlations}
We now discuss the preparation of the initial system state. Ignoring the $\mathcal{SE}$ correlations, we can write the $\mathcal{SE}$ equilibrium state as a product state, namely 
\begin{align}
    \varrho 
    = \varrho_{\text{S0}} \otimes \varrho_{E}.\label{eq6}
\end{align}
Here $\varrho_{\text{S0}} =e^{-\beta H_{\text{S0}}}/Z_{\text{S0}}$ and $ \varrho_{E} = e^{-\beta H_E}/Z_E$ with the partition functions $Z_{\text{S0}} = \text{Tr}_S \left\{e^{-\beta H_{\text{S0}}}\right\}$ and $Z_E = \text{Tr}_E \left\{e^{-\beta H_E}\right\}$. Note that writing the state in this form is only justified if we can ignore the $\mathcal{SE}$ coupling $H_{\text{SE}}$ (or, in other words, we are in the weak coupling regime) since $H_{\text{SE}}$ does not commute with the system Hamiltonian. Now, a relatively large value of $\varepsilon_0$ and a small value of $\Delta_0$, that is, $\beta \varepsilon_0 \gg 1$, will correspond to the system state being approximately `down' along the $z\text{-}$axis. At time $t = 0$, we apply a unitary operator to prepare the desired initial state. For example, if the desired initial state is `spin up' along the $x\text{-}$axis, then the unitary operator $R= e^{i \frac{\pi}{4} \sigma_y}$, realized by the application of a suitable control pulse, is applied to the system only. During the pulse operation, we assume that the pulse duration is much smaller than the cutoff frequency of the environment $\omega_c$ and the effective Rabi frequency  $\sqrt{\varepsilon^2 + \Delta^2}$. Once the pulse has been applied, the total $\mathcal{SE}$ initial state is
\begin{equation}
    \varrho^R_{\text{tot}}
     = \varrho^R_{\text{S0}} \otimes \varrho_{E}, \label{eq3}
\end{equation}
with $ \varrho^R_{\text{S0}} =  e^{-\beta H^R_{\text{S0}}} /Z_{\text{S0}} $ and $H^R_{\text{S0}} = R H_{\text{S0}} R^{\dagger}$. Once we have prepared our system's initial state, we can change the system Hamiltonian parameters as needed. For example, we can change the energy bias to $\varepsilon$ so that the contribution of the tunneling term ($\frac{\Delta}{2} \sigma_x$) becomes more significant. Once again, we assume that this change occurs in a very short time duration. We now write the initial system state as (the superscript `woc' stands for `without correlations' since we are ignoring the $\mathcal{SE}$ interaction when preparing the initial system state) 
\begin{align}
    \mathcolorbox{Apricot}{\varrho_{\text{S0}}^{\text{woc}}
    =\frac{1}{Z_{\text{S0}}} \left\{\mathds{1} \cosh	\left(\beta\widetilde{\Delta}_0\right) - \frac{\sinh\left(\beta\widetilde{\Delta}_0\right)}{\widetilde{\Delta}_0}H^{R}_{\text{S0}}\right\},}\nonumber
\end{align}
with $\widetilde{\Delta}_0 = \frac{1}{2} \sqrt{\varepsilon_0^2 + \Delta_0^2} $. It is convenient to find the Bloch vector components using $ p^{\text{woc}}_i = \text{Tr}_S \left\{\sigma_i  \varrho_{\text{S0}}^{\text{wc}} \right\} $ (with $i= x,y,z$) and cast them into the column vector
\begin{equation}
    \left( \begin{array}{c}
   p^{\text{woc}}_x\\
   p^{\text{woc}}_y\\
   p^{\text{woc}}_z
  \end{array} \right)
  	= \frac{\sinh\left(\beta\widetilde{\Delta}_0\right)}{Z_{\text{S0}}\widetilde{\Delta}_0}
	\left( {\begin{array}{c}
  	\varepsilon_0\\
   	0\\
   -\Delta_0\\
  \end{array} } \right). \label{eq10}
\end{equation}

\subsection{\label{sec:level2A}Initial state preparation with correlations}

We now consider the initial $\mathcal{SE}$ state that includes the effect of the initial $\mathcal{SE}$ correlations. We imagine that the spin system has been interacting with its surrounding environment for very a long time before coming to a joint thermal equilibrium state with the environment; the $\mathcal{SE}$ state is then the standard canonical Gibbs state $\varrho_{\text{th}}= e^{-\beta H}/Z_{\text{tot}}$. In general, we can not write this state as a product state since the $\mathcal{SE}$ interaction does not commute with the system Hamiltonian. However, it is quite clear that if the $\mathcal{SE}$ coupling is weak, this state would approximate the product state given in Eq.~\eqref{eq6}. Now, at time $t=0$, as before, we apply a suitable pulse to prepare the initial system state. Consequently, the correlated $\mathcal{SE}$ state becomes (the superscript `wc' stands for `with correlations')
\begin{align}
\varrho^{\text{wc}}_{\text{tot}}
&= \frac{1}{Z_{\text{tot}}} e^{-\beta \left(H^R_{\text{S0}} + H_E + H^R_{\text{SE}}\right)},\label{eq5}
\end{align}
where $Z_{\text{tot}} = \text{Tr}_{\text{SE}} \left\{ e^{-\beta \left(H^R_{\text{S0}} + H_E + H^R_{\text{SE}}\right)} \right\}$ is the combined partition function for the system and the environment as a whole. Looking at Equations \eqref{eq1} and \eqref{eq2}, we can write $e^{-\beta H_E} \ket{n}
    = k_n \ket{n}$ with $k_n = e^{-\beta (\frac{\epsilon_n}{2} + \lambda_n)}$. Also
\begin{align}
    \left(H^R_{\text{S0}} + H^R_{\text{SE}}\right)\ket{n}
    &= \left(\frac{\varepsilon^n_{0}}{2} \sigma_z  -\frac{\Delta_0}{2} \sigma_x\right)\ket{n} \equiv H_{\text{S0},n}\ket{n},\nonumber
\end{align}
where $H_{\text{S0},n}$ is a `shifted' system Hamiltonian due to the $\mathcal{SE}$ interaction with the new parameter $\varepsilon_{0,n} = e_n + \varepsilon_0$. Following the same steps as in the previous section, we can eventually write
\begin{equation}
\mathcolorbox{Apricot}{
    \left( \begin{array}{c}
   p^{\text{wc}}_x\\
   p^{\text{wc}}_y\\
   p^{\text{wc}}_z
  \end{array} \right)
  	= \sum_n \frac{k_n\sinh\left(\beta\widetilde{\Delta}^n_0\right)}{Z_{\text{tot}}\widetilde{\Delta}^n_0}
	\left( {\begin{array}{c}
  	\varepsilon^n_0\\
   	0\\
   -\Delta_0\\
  \end{array} } \right),}    
\end{equation}
where we now have $\widetilde{\Delta}^n_0 = \frac{1}{2}\sqrt{(\varepsilon_0^{n})^2 + \Delta_0^2} $.

\subsection{\label{sec:level2A}System dynamics without initial correlations}

To find the dynamics, we construct the total time-evolution unitary operator. For this purpose, we insert the completeness relation over the environment states $\ket{n}$ over all the possible environment spin orientations. This gives us
\begin{align}
    U(t)
    &=\sum_{n} e^{-i\frac{\epsilon_n}2t}e^{-i\lambda_nt}e^{-iH_{S,n}t}\ket{n}\bra{n},\nonumber
\\
    &=\sum^{2^{N}-1}_{n=0} U_n(t)\ket{n}\bra{n},\label{eq4}
\end{align}
where $H_{S,n}$ is similar to $H_{\text{S0},n}$, the only difference being that the latter contains the energy bias $\varepsilon_{0,n}$ and the former $\varepsilon_n = e_n + \varepsilon$. Now, we can write
\begin{align}
    U_n(t)
    &=e^{-i\frac{\epsilon_n}2t}e^{-i\lambda_nt}\left\{\mathds{1}\cos\left(\widetilde{\Delta}_nt\right)-\frac{i\sin\left(\widetilde{\Delta}_nt\right)}{\widetilde{\Delta}_n} H_{S,n} \right\},
\end{align}
which is the effective unitary operator that only acts in the system's Hilbert space, with $\widetilde{\Delta}_n = \sqrt{\varepsilon^2 + \Delta^2} $. The reduced density matrix for the system at time $t$ can then be obtained via $\varrho^{\text{woc}}_S(t)
    =\text{Tr}_E\left\{U(t)\varrho^{R}_{\text{tot}}U^\dagger(t)\right\}$. Upon taking $U(t)$ from \eqref{eq4}, and the simple product state $\varrho^{R}_{\text{tot}}$ from \eqref{eq3}, we obtain, after some algebra (see Appendix \autoref{app:A} for details)
\begin{align}
    \mathcolorbox{Apricot}{\varrho^{\text{R}}_S(t)
    =\frac{1}{Z_E}\sum^{2^{N}-1}_{n=0}k_n{U_n(t)\varrho_{\text{S0}}^{\text{woc}}U^\dagger_n(t)}.}
\end{align}
Here $Z_{E}=\sum_n k_n$ which is sensible because every environment spin configuration $\ket{n}$ occurs with probability $k_n/Z_E$. $U_n(t)$ generates dynamics for each configuration, meaning that, to obtain the total reduced density matrix for the system, we need to take into account all the possible environment spin configurations. 

It is useful to find the Bloch vector components for the time-evolved density matrix. We can determine the Bloch vector $\mathbf{p}(t)$ at time $t$ via $\mathbf{p}^{\text{woc}}(t) = \frac{1}{Z_E}\mathbf{M}^{\text{woc}}(t)\mathbf{p}^{\text{woc}}$. Written out explicitly, this is  
\begin{equation}
\left( {\begin{array}{c}
   p^{\text{woc}}_x(t)\\
   p^{\text{woc}}_y(t)\\
   p^{\text{woc}}_z(t)
  \end{array} } \right) 
  = \frac{1}{Z_E} \left( {\begin{array}{ccc}
   M^{\text{woc}}_{11} & M^{\text{woc}}_{12} & M^{\text{woc}}_{13} \\
   M^{\text{woc}}_{21} & M^{\text{woc}}_{22} & M^{\text{woc}}_{23} \\
   M^{\text{woc}}_{31} & M^{\text{woc}}_{32} & M^{\text{woc}}_{33}
  \end{array} } \right) \left( {\begin{array}{c}
   p^{\text{woc}}_x\\
   p^{\text{woc}}_y\\
   p^{\text{woc}}_z
  \end{array} } \right),
\end{equation}
with 
\begin{align}
M^{\text{woc}}_{11}(t) 
&= \sum_{n} \frac{k_n}{4\widetilde{\Delta}_n^2} \left[\Delta^2 + \varepsilon_n^2 \cos(2\widetilde{\Delta}_n t) \right], \notag 
\\
M^{\text{woc}}_{12}(t) 
&= -\sum_{n} \frac{k_n \varepsilon_n}{2\widetilde{\Delta}_n}  \sin(2\widetilde{\Delta}_n t), \notag
\\
M^{\text{woc}}_{13}(t) 
&= \sum_{n} \frac{k_n \Delta \varepsilon_n}{2\widetilde{\Delta}_n^2} \sin^2(\widetilde{\Delta}_n t),\notag
\\
M^{\text{woc}}_{21}(t) 
&= \sum_{n} \frac{k_n \varepsilon_n}{2\widetilde{\Delta}_n}  \sin(2\widetilde{\Delta}_n t), \notag
\\
M^{\text{woc}}_{22}(t) 
&= \sum_{n} k_n \cos(2\widetilde{\Delta}_n t), \notag
\\
M^{\text{woc}}_{23}(t) 
&= -\sum_{n} \frac{k_n \Delta}{2\widetilde{\Delta}_n} \sin(2\widetilde{\Delta}_n t),\notag
\\
M^{\text{woc}}_{31}(t) 
&= \sum_{n} \frac{k_n\Delta \varepsilon_n}{2\widetilde{\Delta}_n^2}  \sin^2(\widetilde{\Delta}_n t),\notag 
\\
M^{\text{woc}}_{32}(t) 
&= \sum_{n} \frac{k_n \Delta}{2\widetilde{\Delta}_n} \sin(2\widetilde{\Delta}_n t),\notag 
\\
M^{\text{woc}}_{33}(t) 
&= \sum_{n} \frac{k_n}{4\widetilde{\Delta}_n^2} \left[\varepsilon_n^2 + \Delta^2 \cos(2\widetilde{\Delta}_n t)\right].
\end{align}
To find these matrix elements, we need to compute sums over all possible $2^N$ environment configurations. We emphasize that this solution is exact which means that it is also valid even for large coupling strengths $g_i$. Furthermore, it is understood that, in general, both the diagonal and the off-diagonal entries of the system density matrix evolve with time.

\begin{figure}[t]\begin{framed}
 		\includegraphics[scale = 1]{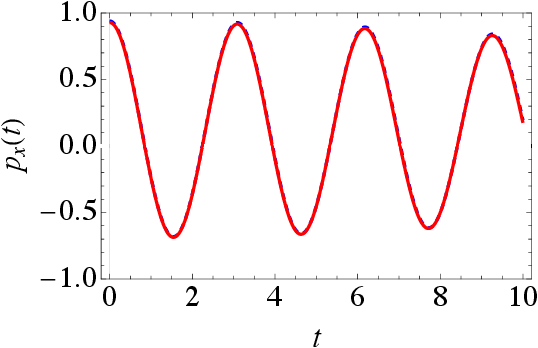}
 		\centering
		\caption{ Dynamics of $p_x(t)$ for relatively weak $\mathcal{SE}$ coupling without initial correlations (dashed, blue line) and with initial correlations (solid, red line). We work in dimensionless units throughout and we have set $\Delta_0 = 1$. Other $\mathcal{SE}$ parameters are $g_i = 0.01$, $\varepsilon_0 = 4, \varepsilon = 2$, $\varepsilon_i = 1$, $\beta = 1, \kappa=0$ and $N = 50$.}
		\label{weakcouplingspin}
\end{framed}\end{figure}
\begin{figure}[t]\begin{framed}
				\includegraphics[scale = 1]{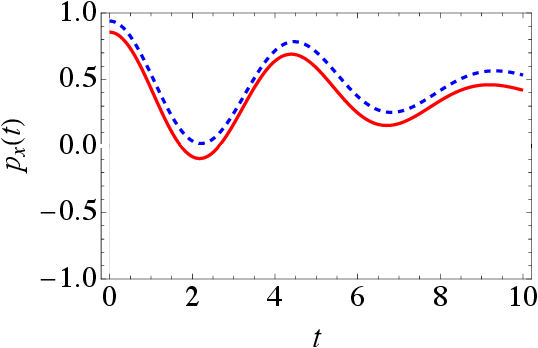}
 				\centering	
				\caption{ Same as Fig.~\ref{weakcouplingspin}, but now  $g_i = 0.05$.}
				\label{midcoupling}
\end{framed}\end{figure}
\begin{figure}[t]\begin{framed}
				\includegraphics[scale = 1]{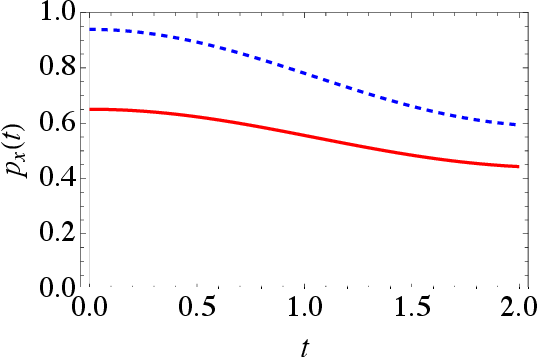}
 				\centering	
				\caption{ Same as Fig.~\ref{weakcouplingspin}, but now  $g_i = 0.1$.}
				\label{comparison}
\end{framed}\end{figure}
\begin{figure}[t]\begin{framed}
				\includegraphics[scale = 1]{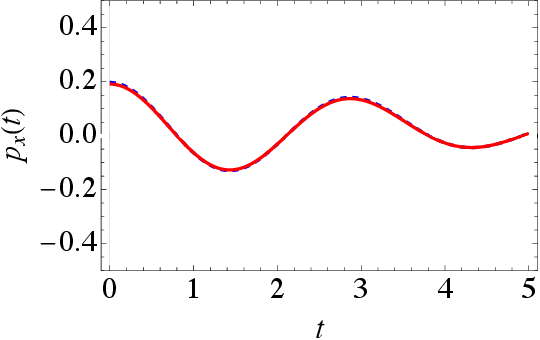}
 				\centering	
				\caption{ Same as Fig.~\ref{midcoupling}, but now $\beta = 0.1$.}
				\label{hightemp}
\end{framed}\end{figure}

\subsection{Dynamics with the correlated initial state}

We now study the dynamics while incorporating the initial $\mathcal{SE}$ correlations. We take the initial state given in Eq.~\eqref{eq5}, let it evolve under the unitary operator given in Eq. \eqref{eq4}, thereby taking a trace over the environment to obtain the following reduced system density matrix   
\begin{align}
    \mathcolorbox{Apricot}{\varrho^{\text{wc}}_S (t)
    =\frac{1}{Z_{\text{tot}}}\sum_{n}A_n k_n {U_n(t)\varrho_{S}^{\text{wc}}U^\dagger_n(t)},}
\end{align}
where now $Z_{\text{tot}}=\sum_n{A_n k_n}$ with $A_n = \text{Tr}\left\{e^{-\beta \left(\frac{\varepsilon^n_0}{2}\sigma_z + \frac{\Delta_0}{2}\sigma_x \right)}\right\} = 2\cosh\left(\beta \widetilde{\Delta}^n_0\right)$. The Bloch vector $\mathbf{p}(t)$ at time $t$ is now given by $\mathbf{p^{\text{wc}}}(t) = \frac{1}{Z_{\text{tot}}}\mathbf{M}^{\text{wc}}(t)\mathbf{p}^{\text{wc}}$, with 
\begin{align}
M^{\text{wc}}_{11}(t) 
&= \sum_{n} \frac{A_nk_n}{4\widetilde{\Delta}_n^2} \left[\Delta^2 + \varepsilon_n^2 \cos(2\widetilde{\Delta}_n t) \right], \notag 
\\
M^{\text{wc}}_{12}(t) 
&= -\sum_{n} \frac{A_nk_n \varepsilon_n}{2\widetilde{\Delta}_n}  \sin(2\widetilde{\Delta}_n t), \notag
\\
M^{\text{wc}}_{13}(t) 
&= \sum_{n} \frac{A_nk_n \Delta \varepsilon_n}{2\widetilde{\Delta}_n^2} \sin^2(\widetilde{\Delta}_n t),\notag
\\
M^{\text{wc}}_{21}(t) 
&= \sum_{n} \frac{A_nk_n \varepsilon_n}{2\widetilde{\Delta}_n}  \sin(2\widetilde{\Delta}_n t), \notag
\\
M^{\text{wc}}_{22}(t) 
&= \sum_{n} A_nk_n \cos(2\widetilde{\Delta}_n t), \notag
\\
M^{\text{wc}}_{23}(t) 
&= -\sum_{n} \frac{A_nk_n \Delta}{2\widetilde{\Delta}_n} \sin(2\widetilde{\Delta}_n t),\notag
\\
M^{\text{wc}}_{31}(t) 
&= \sum_{n} \frac{A_nk_n\Delta \varepsilon_n}{2\widetilde{\Delta}_n^2}  \sin^2(\widetilde{\Delta}_n t),\notag 
\\
M^{\text{wc}}_{32}(t) 
&= \sum_{n} \frac{A_nk_n \Delta}{2\widetilde{\Delta}_n} \sin(2\widetilde{\Delta}_n t),\notag 
\\
M^{\text{wc}}_{33}(t) 
&= \sum_{n} \frac{A_nk_n}{4\widetilde{\Delta}_n^2} \left[\varepsilon_n^2 + \Delta^2 \cos(2\widetilde{\Delta}_n t)\right].
\end{align}

Note that, once again, this is a non-perturbative solution. Comparing the time evolution of the system with the two different initial states, it is clear that the difference in the dynamics is due to the factor $A_n$ which encapsulates the effects of initial correlations before the state preparation. If these correlations are included, every possible environment configuration occurs with the probability $A_nk_n /Z_{\text{tot}}$ instead of $k_n/Z_E$, thus leading to a possibly marked difference in the evolution of the Bloch vector components. To examine this difference in more detail, let us note that as long as the $\mathcal{SE}$ coupling strength is weak, we expect negligible evolution differences between the dynamics of the correlated and uncorrelated initial state. As we increase the coupling strength, the effect of the initial correlations should look more prominent. These forecasts are presented in Figs.~\ref{weakcouplingspin} and ~\ref{midcoupling}, where we have shown the evolution of $p^{\text{wc}}_x(t)$ and $p^{\text{woc}}_x(t)$ (the $x\text{-}$components of the Bloch vector) starting from the correlated initial state and the simply the product state respectively. Two points should be noted. First, the correlation effect is more pronounced in Fig.~\ref{midcoupling} (coupling strength $g=0.05$) as compared to Fig.~\ref{weakcouplingspin} where coupling strength is $ g = 0.01$. Second, as expected, with a stronger $\mathcal{SE}$ coupling, the oscillations in the Bloch vector dynamics die off more quickly. As the coupling strength is increased, the effect of the initial correlations becomes even more pronounced (see Fig.~\ref{comparison}). 

\begin{figure}[t]\begin{framed}
 		\includegraphics[scale = 1]{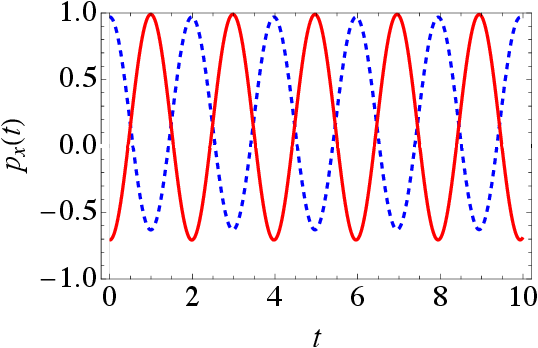}
 		\centering
		\caption{ Same as Fig.~\ref{weakcouplingspin}, but now $g = 1$ and $\beta = 10$.}
		\label{lowtemp}
\end{framed}\end{figure}

We can also investigate the effect of temperature. At higher temperatures, the total $\mathcal{SE}$ thermal equilibrium state (before applying the pulse) is almost a mixed state. Hence, at higher temperatures, there will be little difference as both initial states are effectively the same. We illustrate this in Figs.~\ref{hightemp} and \ref{lowtemp}. There are two points to be made regarding Fig.~\ref{hightemp}. First, at higher temperatures, the condition $\beta\varepsilon_0 \gg 1$ is not fulfilled. Therefore, the system Bloch vector, before the pulse operation, is not approximately along the negative $z\text{-}$axis. Consequently, the evolution of the Bloch vector component $p_x(t)$ does not start from $p_x \approx 1 $. Second, the correlation effect seen in Fig.~\ref{midcoupling} disappears at higher temperatures although the coupling strength is still $g=0.05$ in Fig.~\ref{hightemp}. The dynamics with the two different initial states at even lower temperatures is illustrated in Fig.~\ref{lowtemp} where $\beta = 10$. For the simple product initial state, since $\beta\varepsilon_0 \gg 1$, the initial system state just after the pulse is applied is approximately $p_x=1$. If we instead consider the joint $\mathcal{SE}$ thermal equilibrium state, the interaction Hamiltonian term $H_{\text{SE}}$ dominates; this leads to the system state being approximately `up' along the $z\text{-}$axis before the application of the pulse and `down' along the $x\text{-}$axis after the pulse operation.

\begin{figure}[t]\begin{framed}
		\includegraphics[scale = 1]{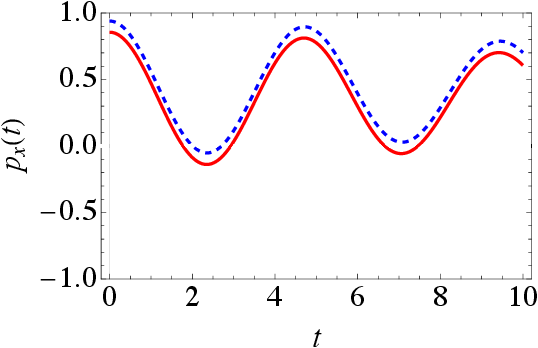}
 		\centering	
		\caption{ Same as Fig.~\ref{weakcouplingspin}, but now $N = 250$.}
		\label{environment}
\end{framed}\end{figure}

\begin{figure}[t]\begin{framed}
 		\includegraphics[scale = 1]{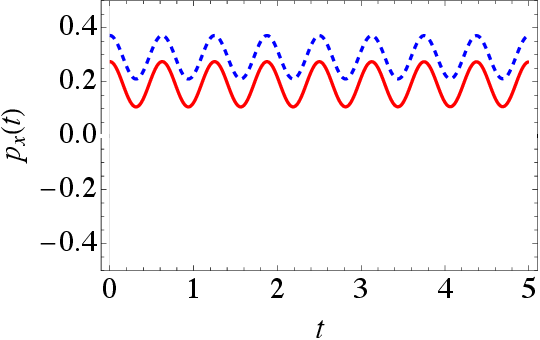}
 		\centering
		\caption{ Same as Fig.~\ref{weakcouplingspin}, but now $\Delta_0 = 10$ and $g = 0.05$.}
		\label{Delta}
\end{framed}\end{figure}

\begin{figure}[t]\begin{framed}
 		\includegraphics[scale = 1]{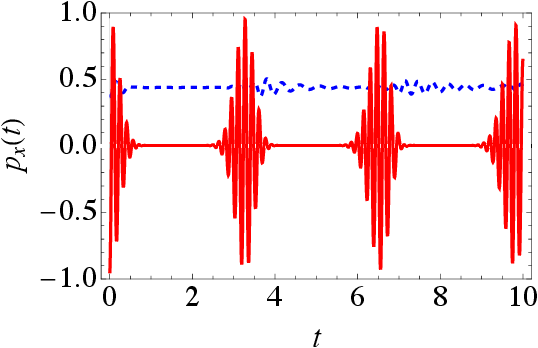}
 		\centering
		\caption{ Same as Fig.~\ref{Delta}, but now  $g = 1$.}
		\label{strongcoupling}
\end{framed}\end{figure}

\begin{figure}[t]\begin{framed}
 		\includegraphics[scale = 1]{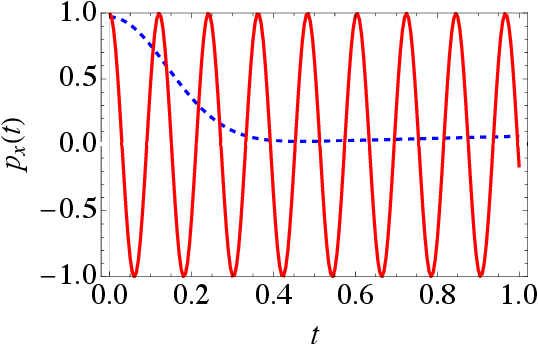}
 		\centering
		\caption{ Same as Fig.~\ref{weakcouplingspin}, but now $\beta=10$, $g = 1$ and $\varepsilon_i = 0.01$. }
		\label{weakenv}
\end{framed}\end{figure}

We should also note that with a larger spin environment, the effect of the initial correlations is more pronounced. This is illustrated in Fig.~\ref{environment}; one can compare Fig.~\ref{environment}, where $N = 250$, with Fig.~\ref{weakcouplingspin} ($N = 50$) to see the effect of the increased number of environmental spins.  
We can also investigate how the tunneling amplitude of the central spin system affects the system dynamics. As shown in Fig.~\ref{Delta}, where we have increased the tunneling amplitude to $\Delta_0=10$, with coupling $g=0.05$, the dynamical difference is still evident. With the same tunneling amplitude, if the $\mathcal{SE}$ coupling strength is made even stronger, there is an even more significant difference [see Fig.~\ref{strongcoupling}]. The difference in the dynamics persists with different values of the energy bias of the environment as well [see Fig.~\ref{weakenv}].
\begin{figure}[t]\begin{framed}
	\includegraphics[scale = 1]{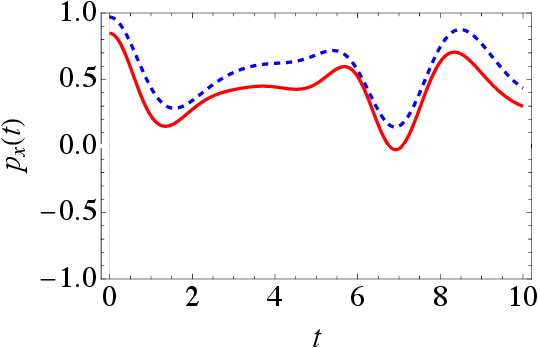}
	\centering	
	\caption{ Plot of $p_x(t)$ as a function of time $t$ for relatively strong coupling strength with initial correlations (solid, red line) and without initial correlations (dashed, blue line) and . Here we have also included the interactions between the environmental spins $\alpha=0.1$. The other parameters are $\Delta_0= 1$, $g_i = 0.5$, $\varepsilon_0=5, \varepsilon = 2$, $\varepsilon_i = 1$, $\beta = 1$, and $N = 10$.}
	\label{lowInterspin}
\end{framed}\end{figure}
Finally, let us consider the scenario where the environment spins are also interacting with each other. Once again, in general, we do see that the initial correlations play a significant role [see Fig.~\ref{lowInterspin}]. 

\begin{figure}[t]\begin{framed}
	\includegraphics[scale = 1]{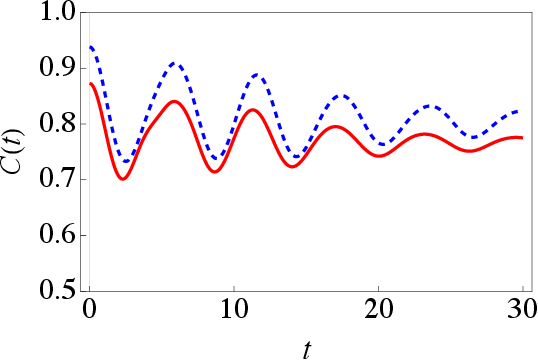}
	\centering
	\caption{ Plot of concurrence between the central spins as a function of time $t$ for relatively weak coupling strength $g_i = 0.05$ without initial correlations (dashed, blue line) and with initial correlations (solid, red line). We have also assumed that spins are not interacting with each other, that is, we have $\kappa = 0$. We have taken environment energy level spacing $\varepsilon_i=1$, and the other $\mathcal{SE}$ parameters are $\varepsilon^{(i)}_0=5, \varepsilon^{(i)} = 2$, $\Delta^{(i)}_{0} = 1$, $\beta =1$ and $N = 50$.}
	\label{2QweakCoupling}
\end{framed}\end{figure}

\begin{figure}[t]\begin{framed}
    \includegraphics[scale = 1]{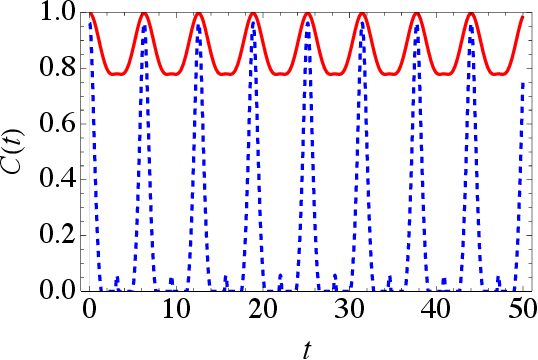}
 	\centering
 	\caption{ Same as Fig. \ref{2QweakCoupling}, but now $\beta =3$ and $g_i =0.5$.}
 	\label{2Qlowtemp}
\end{framed}\end{figure}
\begin{figure}[t]\begin{framed}
    \includegraphics[scale = 1]{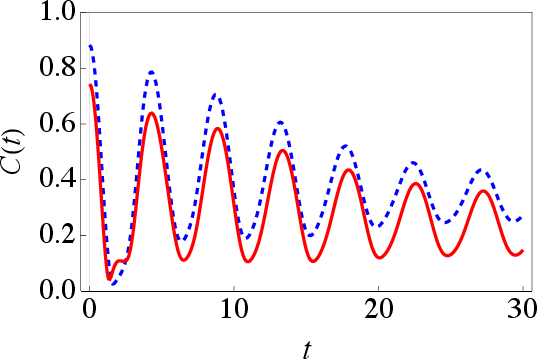}
 	\centering
 	\caption{ Same as Fig. \ref{2QweakCoupling}, but now $\kappa =0.5$.}
 	\label{qubitsinteraction}
\end{framed}\end{figure}

\section{Extension to two-qubit system}\label{2qubits}
We now consider the case of two qubits interacting with the common spin environment. Again, our goal is to investigate the difference in dynamics for correlated and uncorrelated initial states. This could possibly reveal aspects of dynamics that may be absent in the single qubit case. An example is entanglement sudden death (ESD \cite{EberlyPRL2004,EberlyScience2007,EberlyScience2009}, where the entanglement between the two qubits vanishes in a very short time.

The total Hamiltonian is now
\begin{align}
    H_{\text{tot}} =
    \begin{cases}
      H^{(1)}_{\text{S0}} + H^{(2)}_{\text{S0}} + H_{\text{12}} + H^{(1)}_{\text{SE}} + H^{(2)}_{\text{SE}} + H_E & t\le 0,\\
      H^{(1)}_S + H^{(2)}_S + H_{\text{12}} + H^{(1)}_{\text{SE}} + H^{(2)}_{\text{SE}} + H_E & t > 0,\\
    \end{cases}   
\end{align}
with 
\begin{align}
    H^{(i)}_{\text{S0}}
    &= \frac{\varepsilon^{(i)}_0}{2} \sigma^{(i)}_{z} + \frac{\Delta^{(i)}_0}{2} \sigma^{(i)}_{x},
\\
	H^{(i)}_{\text{S}}
	&= \frac{\varepsilon^{(i)}}{2} \sigma^{(i)}_{z} + \frac{\Delta^{(i)}_0}{2} \sigma^{(i)}_{x},
\\
    H_{\text{12}}
    &= \kappa \sigma^{(1)}_{z}\sigma^{(2)}_{z},
\\
    H^{(1)}_{\text{SE}}
    &= \frac{1}{2} \sigma^{(1)}_{z} \otimes \sum_{i=1}^{N} g_i\sigma^{(i)}_z , 
\\
    H^{(2)}_{\text{SE}}
    &= \frac{1}{2} \sigma^{(2)}_{z} \otimes \sum_{i=1}^{N} g_i\sigma^{(i)}_z ,
\\
    H_E
    &=\sum_{i=1}^{N} \frac{\varepsilon_i}{2} \sigma^{(i)}_z + \sum_{i=1}^{N} \alpha_i \sigma^{(i)}_z \sigma^{(i+1)}_z,
\end{align}
with $i=1,2$. Here the qubits are labeled as 1 and 2 with $\varepsilon^{(1)}_0$ and $\varepsilon^{(2)}_0$ the energy bias terms and $\Delta^{(1)}_0$ and $\Delta^{(2)}_0$ the tunneling amplitudes of the central qubit 1 and qubit 2 respectively. Both qubits are coupled with each other by $H_{\text{12}}$. We aim to look at the dynamics of entanglement between the two-qubit system, starting from correlated and uncorrelated initial states. To begin, let us comment on the initial state preparation. We prepare our initial state such that, starting from the thermal equilibrium state, the two qubits become entangled with each other. Note that with $\varepsilon^{(i)}_0 \gg \Delta^{(i)}_0$, our system initial state is (approximately) both spins `down' along the $z\text{-}$axis. We now apply the unitary operator (at $t = 0$)
\begin{align*}
    \text{CZ}
    =e^{i \frac{\pi}{4}\left(\sigma^{(1)}_x + \sigma^{(2)}_x -\sigma_x \otimes \sigma_x\right)},
\end{align*}
only on the system of two qubits to generate entanglement between them. We then have the two different initial states
\begin{align*}
    \varrho_{\text{entangled}}^{\text{woc}}
    &= \frac{1}{Z_{\text{woc}}} \text{CZ} e^{-\beta \left(H^{(1)}_{\text{S0}} + H^{(2)}_{\text{S0}} + H_{12} + H_E \right)}\text{CZ}^{\dagger},
\\    
    \varrho_{\text{entangled}}^{\text{wc}}
    &= \frac{1}{Z_{\text{wc}}} \text{CZ} e^{-\beta \left(H^{(1)}_{\text{S0}} + H^{(2)}_{\text{S0}} + H_{12}  + H^{(1)}_{\text{SE}} + H^{(2)}_{\text{SE}} + H_E \right)}\text{CZ}^{\dagger}.
\end{align*}
Here $Z_{\text{woc}}$ and $Z_{\text{wc}}$ are the partition functions for the corresponding states.

\begin{comment}The entanglement can be quantified by concurrence, defined as $\mathcal{C}(t) = \text{max}\left(0, \sqrt{\lambda_1}-\sqrt{\lambda_2}-\sqrt{\lambda_3}-\sqrt{\lambda_4} \right)$. Here $\lambda_i$ (with $i=1,2,3,4$) denotes the eigenvalues in decreasing order of the matrix defined as $\Lambda = \varrho_S(t)\left(\sigma^{(1)}_{y} \otimes \sigma^{(2)}_{y} \right) \varrho^*_S(t)\left(\sigma^{(1)}_{y} \otimes \sigma^{(2)}_{y} \right)$, with $\varrho_S(t)$ as the reduced density matrix of composite system of two qubits and $\varrho^*_S(t)$ being its complex conjugate. $ \mathcal{C}(t) = 1$ for maximally entangled and zero for unentangled states. 
\end{comment}

For simplicity, we first consider $\kappa=0$ (the direct qubit-qubit interaction is zero) and analytically calculate the reduced density matrices, starting from these two different initial states. To do so, we need to find the time evolution operator. The calculation is very similar to the single qubit case; therefore, we simply summarize the results. For the simple product initial state, we obtain 
\begin{align}
    \mathcolorbox{Apricot}{\varrho^{\text{woc}}_S(t)
    =\frac{1}{Z_{\text{woc}}}\sum_{n}k_n U^{(1)}_n(t) U^{(2)}_n(t)\varrho^{\text{woc}}_SU^{(2)\dagger}_n(t) U^{(1)\dagger}_n(t),}
\end{align}
where
\begin{align}
    &U^{(i)}_n(t)
	=e^{-i\frac{\epsilon_n}{4}t}e^{-i\frac{\lambda_n}{2}t}\left\{\mathds{1}\cos(\widetilde{\Delta}^{(i)}_nt)-\frac{i\sin(\widetilde{\Delta}^{(i)}_nt)}{\widetilde{\Delta}^{(i)}_n}H^{(i)}_{S,n}\right\},
\end{align}
with $\widetilde{\Delta}^{(i)}_n = (1/2)\sqrt{(\varepsilon^{(i)}_{n})^2 + (\Delta^{(i)})^2} $, $H_{S,n}^{(i)} = \frac{\varepsilon_n^{(i)}}{2} \sigma_z^{(i)} - \frac{\Delta_0^{(i)}}{2}\sigma_x^{(i)}$, $\varepsilon^{(i)}_n = e_n + \varepsilon^{(i)}$, and $Z_{\text{woc}} = \sum_n k_n$. On the other hand, with the correlated initial state, we get
\begin{align*}
    \varrho^{\text{wc}}_S(t)
    =\text{Tr}_E\left\{U(t)\varrho^{\text{wc}}_{\text{entangled}}U^\dagger(t)\right\},
\end{align*}
\begin{align}
    \mathcolorbox{Apricot}{\varrho^{\text{wc}}_S(t)
    =\frac{1}{Z_{\text{wc}}}\sum_{n}A_n k_n {U^{(1)}_n(t)U^{(2)}_n(t)\varrho_{S}^{\text{wc}}U^{(1)\dagger}_n(t)U^{(2)\dagger}_n(t)},}
\end{align}
where $Z_{\text{wc}} = \sum_n A_n k_n$ with $A_n = \text{Tr}\left\{e^{-\beta \left(H^{(1)}_{\text {S0},n} + H^{(2)}_{\text {S0},n}\right)}\right\}$ appearing due to the effect of initial correlations. Note that $H_{\text{S0},n}^{(i)}$ is the same as $H_{S,n}^{(i)}$ except the change of energy bias ($\varepsilon_{0,n}^{(i)}$ belongs to $H_{\text{S0},n}^{(i)}$ and $\varepsilon_n^{(i)}$ belongs to $H_{\text{S},n}^{(i)}$). Using our worked-out dynamics, we can look at the impact of the initial correlations on the entanglement dynamics. To quantify entanglement, we use the concurrence $C(t)$. The concurrence of a two-qubit state $\varrho(t)$ is defined as $C(t) = \text{max}\left(0, \sqrt{\lambda_1}-\sqrt{\lambda_2}-\sqrt{\lambda_3}-\sqrt{\lambda_4} \right)$, where the $\lambda_i$ (with $i=1,2,3,4$) denote the eigenvalues in decreasing order of $\varrho(t)\left(\sigma^{(1)}_{y} \otimes \sigma^{(2)}_{y} \right) \varrho^*(t)\left(\sigma^{(1)}_{y} \otimes \sigma^{(2)}_{y} \right)$. The concurrence is one for a maximally entangled state and zero for unentangled, separable states. We first see what examine the weak coupling regime. Fig.~\ref{2QweakCoupling} shows that even at weak $\mathcal{SE}$ coupling strength, we have considerable differences in the dynamics, and this difference is even more apparent at lower temperatures (see Fig.~\ref{2Qlowtemp}); in fact, the initial correlations can even be seen to significantly enhance the entanglement. 

For completeness, let us note that we can further investigate the dynamics by including the effect of the qubit-qubit interaction as well. That is, $\kappa$ is non-zero now. Following a similar formalism, the time evolution operator is now found to be
\begin{equation}
\label{lambdatotalunitarytimeoperator}
{U}(t)=\sum_{n=0}^{2^{N}-1}{U}_{n}^{(12)}(t)\ket{n}\bra{n},
\end{equation}
with
${U}_{n}^{(12)}(t)
={e^{-i\frac{\epsilon_{n}}{2}t}}{e^{-i\lambda_{n}t}}e^{-i(H_{S,n}^{(1)}+H_{S,n}^{(2)}+H_{12})t}$.
This operator helps us to write our final system state for both the correlated and uncorrelated cases as
\begin{align}
\mathcolorbox{Apricot}{{\varrho}^{\text{woc}}_S(t)
=\frac{1}{Z_E}\sum_{n=0}^{2^{N}-1} k_n U^{(12)}_{n}(t){\varrho}^{\text{woc}}_{\text{S0}}U_n^{(12)\dagger}(t),}
\\
\mathcolorbox{Apricot}{{\varrho}^{\text{wc}}_S(t)
=\frac{1}{Z_{\text{tot}}}\sum_{n=0}^{2^{N}-1}k_n A_n {U^{(12)}_{n}(t)}{\varrho}^{\text{wc}}_{\text{S0}}{U^{(12)}_{n}}^{\dagger}(t).}
\end{align}
The key difference is that now $A_n = \text{Tr}\left\{e^{-\beta \left(H^{(1)}_{\text {S0},n} + H^{(2)}_{\text {S0},n} + H_{12}\right)}\right\}$. We illustrate the entanglement dynamics with $\kappa = 0.5$ in Fig.~\ref{qubitsinteraction}. Once again, the effect of the initial correlations is quite apparent. 

\section{Summary}\label{ch3-sum}

In this chapter, we have explored the dynamics of a central spin system that is interacting with a spin environment, taking into account the $\mathcal{SE}$ correlations. In this particular model, both the diagonal and off-diagonal entries of the system's density matrix evolve. We found that the effects of the initial $\mathcal{SE}$ correlations generally cause a minimal difference in the regime of weak $\mathcal{SE}$ coupling and high temperatures. However, this difference becomes more appreciable when the $\mathcal{SE}$ coupling becomes stronger, the temperature is low, and the environment is really large. We also showed that kept at much lower temperatures, wherein the discrepancy due to state preparation vanished even with stronger coupling. A similar trend occurred when we considered inter-spin interaction. Next, we extended our study to two spins interacting with a common environment of spins, thereby showing that entanglement dynamics are also influenced by the initial correlations. Such results are promising, as they provide insights into the influence of initial correlations with spin environments.

\acresetall
\chapter*{Appendix}
\addcontentsline{toc}{chapter}{\tocEntry{Appendix}}
\section{Dynamics for uncorrelated case}\label{app:A}
\begin{align}
    \varrho^{\text{woc}}_S(t) 
    &=\text{Tr}_E\left\{U(t)\varrho^R_{\text{tot}}U^\dagger(t)\right\}\nonumber,
\\
    &=\text{Tr}_E\left\{\sum_{l,m}\ket{l}\bra{l}U_l(t)\left(\varrho^R_{\text{S0}}\otimes\frac{e^{-\beta H_E}}{Z_E}\right)\ket{m}\bra{m}U^\dagger_m(t)\right\},\nonumber
\\
    &=\frac{1}{Z_E}\sum_{l,m}\text{Tr}_E\left\{\ket{l}\bra{l}U_l(t)\varrho^R_{\text{S0}}e^{-\beta(\frac{\epsilon_m}{2}+\lambda_m)}U^\dagger_m(t)\ket{m}\bra{m}\right\},\nonumber
\\
    &=\frac{1}{Z_E}\sum_{l,m}k_m\sum_{n}\delta_{n,l}\bra{l}U_l(t)\varrho^R_{\text{S0}}U^\dagger_m(t)\ket{m}\delta_{m,n},\nonumber
\\
    &=\frac{1}{Z_E}k_n\sum_{n}\bra{n}U_n(t)\varrho^R_{\text{S0}}U^\dagger_n(t)\ket{n},\nonumber
\\
    &=\frac{1}{Z_E}\sum_{n}k_n{U_n(t)\varrho^R_{\text{S0}}U^\dagger_n(t)}.\nonumber
\end{align}

\acresetall
  
\chapter{Including the effect of initial correlations in the Non-Markovian master equation}\label{c:MasterEq}
  In this chapter, we present a general master equation, correct to second-order in the system-environment $(\mathcal{SE})$ coupling strength, that takes into account the initial $\mathcal{SE}$ correlations. We assume that the system and its environment are in a joint thermal equilibrium state, and thereafter, a unitary operation is performed to prepare the desired initial system state, with the system Hamiltonian possibly changing thereafter as well. We show that the effect of the initial correlations shows up in the second-order master equation as an additional term, similar in form to the usual second-order term describing relaxation and decoherence in quantum systems. We apply this master equation to a generalization of the paradigmatic Spin-Boson $(\mathcal{SB})$ model, namely, a collection of two-level systems interacting with a common environment of harmonic oscillators, as well as a collection of two-level systems interacting with a common spin environment. We demonstrate that, in general, the initial $\mathcal{SE}$ correlations need to be accounted for in order to accurately obtain the system dynamics.

Our problem is to derive such a master equation that is valid for weak $\mathcal{SE}$ coupling, and describes these system dynamics. We first write the total $\mathcal{SE}$ Hamiltonian as 
\begin{align}
    H_{\text{tot}} =
    \begin{cases}
      H_{\text{S0}} + H_E +\alpha H_{\text{SE}} & t\le 0,
\\
      H_{S} + H_E +\alpha H_{\text{SE}} & t > 0.\\
    \end{cases}   
\end{align}
Here $H_S$ is the system Hamiltonian corresponding to the coherent evolution of the system only after the initial time $t = 0$ at which the system state is prepared. $H_{\text{S0}}$ is similar to $H_S$ in the sense that both operators live in the same Hilbert space, but they may have different parameters. $H_E$ is the environment Hamiltonian, and $H_{\text{SE}}$ the interaction Hamiltonian that describes the $\mathcal{SE}$ coupling. $\alpha$ is simply a dimensionless parameter introduced to keep track of the perturbation order; later on, we will set $\alpha = 1$.

We organize this chapter as follows. In the first section \autoref{initial}, we derive a correlated initial state. In section \autoref{meqderv}, we derive our general time-local second-order master equation. Section \autoref{seclargespinapp} discusses the application of this master equation to the large $\mathcal{SB}$ model, while section \autoref{spinspinmodel} applies the master equation to the spin-spin model. We then present a summary in section \autoref{ch4sum}. The Appendices \autoref{relaxation}, \autoref{timedependent}, \autoref{bosonicbath} and \autoref{spinbathcorr} consists of some technical details regarding the initial system state preparation, the usual relaxation term in the master equation, the exactly solvable pure dephasing limit of the large $\mathcal{SB}$ model, the generalization of the master equation to a time-dependent system Hamiltonian, and the derivation of environment correlations functions respectively. Let us now briefly discuss the initial state preparation.

\section{The initial state}
\label{initial}

We first discuss the initial $\mathcal{SE}$ state. We let our system to come to a joint equilibrium state with the environment. What we mean by this is that the system's equilibrium state is not simply proportional to $e^{-\beta H_{\text{S0}}}$ - there are corrections as a consequence of finite $\mathcal{SE}$ coupling strength \cite{geva2000second}. We instead consider our system and the environment together in the Gibbs state proportional to  $e^{-\beta H_{\text{tot}}}$ with $H_{\text{tot}} =  H_{\text{S0}} + H_E +\alpha H_{\text{SE}}$; the system's reduced density matrix can be obtained by simply tracing out the environment degrees of freedom. A unitary operator $R$ is then applied to the system only. Consequently, the initial $\mathcal{SE}$ state becomes
\begin{align}
    \varrho_{\text{tot}}(0)
    &=\frac{R e^{-\beta H_{\text{tot}}}R^\dagger}{Z_{\text{tot}}}, \label{ref11}
\end{align}
with $Z_{\text{tot}}=\text{Tr}_{\text{S,E}}\left\{e^{-\beta H_{\text{tot}}}\right\}$ the total partition function whereas $\text{Tr}_{\text{S,E}}$ symbolizes the trace over the system and the environment. Now, assuming the interaction strength to be weak, we expand the initial $\mathcal{SE}$ state to second-order in the $\mathcal{SE}$ coupling strength perturbatively. In fact, we use the Kubo identity to expand the joint state given by Eq.~\eqref{ref11}. The Kubo identity tells us that for any two arbitrary operators $A$ and $B$, we have
\begin{align}
    e^{\beta(A+B)}&=e^{\beta A}\left[1+\int_{0}^{\beta} e^{-\lambda A}B e^{\lambda(A+B)}d\lambda\right]. \label{ref12}
\end{align}
By setting $A=-(H_{\text{S0}}+H_E)$  and  $B=-\alpha H_{\text{SE}}$, and using the Kubo identity twice, we obtain the required second-order expansion, given below
\begin{align}
    &e^{-\beta(H_{\text{S0}}+H_E+\alpha H_{\text{SE}})}
    =e^{-\beta(H_{\text{S0}}+H_E)} - \alpha e^{-\beta(H_{\text{S0}}+H_E)}\int_{0}^{\beta} e^{\lambda(H_{\text{S0}}+H_E)}H_{\text{SE}}  
    e^{-\lambda(H_{\text{S0}}+H_E)}d\lambda \nonumber 
\\ 
    &+\alpha^2e^{-\beta(H_{\text{S0}}+H_E)}\int_{0}^{\beta}d\lambda e^{\lambda(H_{\text{S0}}+H_E)}H_{\text{SE}}   e^{-\lambda(H_{\text{S0}}+H_E)}\int_{0}^{\lambda} e^{\lambda'(H_{\text{S0}}+H_E)} H_{\text{SE}}e^{-\lambda'(H_{\text{S0}}+H_E)}d\lambda'.\label{ref14}
\end{align}
We now write the system environment coupling $H_{\text{SE}}$ as $S \otimes E$, where $S$ and $E$ are the system and the environment operators living in their respective Hilbert spaces. The extension to the more general case where $H_{\text{SE}} = \sum_\alpha S_\alpha \otimes E_\alpha$ is straightforward. Eq.~\eqref{ref14} can then be simplified as
\begin{align} 
    e^{-\beta(H_{\text{S0}}+H_E+\alpha H_{\text{SE}})}
    &=e^{-\beta(H_{\text{S0}}+H_E)} -\alpha e^{-\beta(H_{\text{S0}}+H_E)}
    \int_{0}^{\beta} S(\lambda)\otimes E(\lambda)d\lambda + \alpha^2 e^{-\beta(H_{\text{S0}}+H_E)}\nonumber
\\
   &\times \int_{0}^{\beta}d\lambda S(\lambda)\otimes E(\lambda)\int_{0}^{\lambda} S(\lambda')\otimes E(\lambda')d\lambda',
\end{align}
where $S\left(\lambda\right)=e^{\lambda H_{\text{S0}}}Se^{-\lambda H_{\text{S0}}}$ and $E\left(\lambda\right)=e^{\lambda H_E} E e^{-\lambda H_E}$.
We now use this in Eq.~\eqref{ref11} in the main text and thereafter take the trace over the environment to find the initial system state correct to second-order in the interaction strength. This is important because our aim is to obtain a master equation valid up to second-order in the coupling strength. For consistency, the initial system state used to solve this master equation should also be exact up to second-order in the interaction strength. For ease of notation, we write the initial system state as
\begin{align}
    \varrho(0)=\varrho^{(0)}(0)+\varrho^{(1)}(0)+\varrho^{(2)}(0),
    \label{sysdminitial}
\end{align}
where
\begin{align}
    \varrho^{(0)}(0)
    &= \frac{1}{{Z_{\text{tot}}}} \text{Tr}_E\left\{R\left( e^{-\beta(H_{\text{S0}}+H_E)}\right)R^\dagger\right\},\nonumber
\\
    \varrho^{(1)}(0)
    &=\frac{1}{{Z_{\text{tot}}}} \text{Tr}_E\left\{-\alpha R\left(e^{-\beta(H_{\text{S0}}+H_E)}\int_{0}^{\beta} S(\lambda)\otimes E(\lambda)d\lambda\right)R^\dagger\right\},\nonumber
\\
    \varrho^{(2)}(0)
    &= \frac{1}{{Z_{\text{tot}}}} \text{Tr}_E\left\{\alpha^2 R\Big(e^{-\beta(H_{\text{S0}}+H_E)} \int_{0}^{\beta}d\lambda S(\lambda)\otimes E(\lambda) \int_{0}^{\lambda}S(\lambda')\otimes E(\lambda')d\lambda'\Big)R^\dagger\right\}. 
\end{align}
Let us simplify these relations one by one. $\varrho^{(0)}(0)$ can be simplified as
\begin{align*}
   \varrho^{(0)}(0)=\frac{e^{-\beta H^R_{\text{S0}}} Z_E}{Z_{\text{tot}}},
\end{align*}
where $Z_E= \text{Tr}_E\left\{ e^{-\beta H_E}\right\}$. As for $\varrho^{(1)}(0)$, we can write
\begin{align*}
    \varrho^{(1)}(0)= \frac{-\alpha Z_E\int_{0}^{\beta}R e^{-\beta H_{\text{S0}}}S(\lambda)R^{\dagger} \big\langle E(\lambda)\big\rangle_E d\lambda}{Z_{\text{tot}}},
\end{align*}
where $\langle \hdots \rangle_E = \text{Tr}_E \left\{e^{-\beta H_E} (\hdots)/Z_E\right\}$. Since $\langle E(\lambda)\rangle_E$ is zero for most $\mathcal{SE}$ models, we simply get that $\varrho^{(1)}(0)=0$. Carrying on, $ \varrho^{(2)}(0)$ can be simplified as
\begin{align*}
    \varrho^{(2)}(0) 
    &=\frac{1}{Z_{\text{tot}}}\alpha^2 Z_E R e^{-\beta H_{\text{S0}}}\int_{0}^{\beta}\int_{0}^{\lambda} S(\lambda) S(\lambda') R^{\dagger} \big\langle E(\lambda)  E(\lambda')\big\rangle_E d\lambda'd\lambda. 
\end{align*}
To proceed further, we evaluate the partition function $Z_{\text{tot}}$  that ensures that the trace of the system state $\varrho(0)$ in Eq.~\eqref{sysdminitial} is one. It is then clear that 
\begin{align}
    Z_{\text{tot}}
    &=Z_E\text{Tr}_{S}\left\{e^{-\beta H_{\text{S0}}} \right\} + \alpha^2Z_E \text{Tr}_S\left\{R e^{-\beta H_{\text{S0}}} \int_{0}^{\beta}\int_{0}^{\lambda} S(\lambda) S(\lambda') R^{\dagger} \big\langle E(\lambda)  E(\lambda')\big\rangle_E d\lambda'd\lambda\right\}\notag.
\end{align}
Putting these results together, we have finally
\begin{align}
    \mathcolorbox{Apricot}{\varrho(0)
    =\frac{e^{-\beta H^R_{\text{S0}}}}{Z_{\text{S0}}Z'}\left[\mathds{1}+\int_{0}^{\beta}\int_{0}^{\lambda} S^R(\lambda)S^R(\lambda')  \big\langle E(\lambda)  E(\lambda')\big\rangle_Ed\lambda'd\lambda\right],}
\end{align}
where $S^R(\lambda) = R S(\lambda) R^\dagger$, and $Z'=1 + \int_{0}^{\beta}\int_{0}^{\lambda} \langle S(\lambda) S(\lambda') \rangle_S 
   \langle E(\lambda)  E(\lambda')\rangle_E d\lambda'd\lambda$, $H^R_{\text{S0}} = R H_{\text{S0}}R^\dagger$ and $Z_{\text{S0}} = \text{Tr}_S\left\{e^{-\beta H_{\text{S0}}}\right\}$. With this initial system state in hand, we concentrate on deriving the second-order master equation.

\section{Derivation of master equation}\label{meqderv}

We now derive a master equation that details the system dynamics. The total $\mathcal{SE}$ Hamiltonian is 
\begin{align*}
    H_{\text{tot}} = H_S+H_E+\alpha H_{\text{SE}} \equiv H_0 + \alpha H_{\text{SE}}.
\end{align*}
Note that the system Hamiltonian $H_S$ can be different from the previous system Hamiltonian $H_{\text{S0}}$. In fact, $H_S$ can even be a time-dependent Hamiltonian without changing the subsequent derivation. Using perturbation theory, the unitary time evolution operator with such a Hamiltonian can be expressed as
\begin{align}
    U (t) \approx U_0\left(t\right) \left[ 1 - \alpha \int_0^t U_0^\dagger (s) H_{\text{SE}} U_0(s)\,ds \right], \label{ref15}
\end{align}
where $U_0(t) \equiv U_S(t) \otimes U_E(t)$ is known as the `free' unitary time evolution operator corresponding to $H_0$. $H_0$ is the `free' Hamiltonian representing the uncoupled system and its environment. The matrix elements of the system density matrix can be obtained via $\varrho_{\text{mn}}\left(t\right) = \text{Tr}_S\left\{\ket{n}\bra{m} \varrho \left(t\right) \right\}$, where $\ket{m}$ and $\ket{n}$ are some basis states of the system. Since $\varrho(t) = \text{Tr}_E \left\{ \varrho_{\text{tot}}(t)\right\}$, we can alternatively write
\begin{align*}
    \varrho_{\text{mn}}(t) = \text{Tr}_\text{S,E}\left\{X_{\text{nm}}^H\left(t\right) \varrho_\text{tot} (0) \right\},
\end{align*}
where $X_{\text{nm}}^H(t) = U^\dagger(t) (\ket{n}\bra{m} \otimes \mathds{1}_E)U(t)$. Our master equation can then be put in the general form 
\begin{align}
    \Dot{\varrho}_{\text{mn}}(t) 
    &=\text{Tr}_\text{S,E}\left\{\varrho_\text{tot}(0)\frac{d}{dt}X_{\text{nm}}^H(t)\right\}.
    \label{mastereqgen}
\end{align}
To make further progress, we note that $X_{\text{nm}}^H(t)$ is an operator written in the Heisenberg picture. Now using the Heisenberg equation of motion along with Eq.~\eqref{ref15}, it can be shown that 
\begin{align} \label{ref16}
    \Dot{X}_{\text{nm}}^H(t) &=i\left[H_0^H(t),X_{\text{nm}}^H(t)\right]+i\alpha\left[\widetilde{H}_{\text{SE}}(t),\widetilde{X}_{\text{nm}}(t)\right] + \alpha^2\int_{0}^{t}ds\left[\left[\widetilde{H}_{\text{SE}}(t),\widetilde{X}_{\text{nm}}(t)\right],\widetilde{H}_{\text{SE}}(s)\right],
\end{align}
where the `tildes' denote time evolution generated by the free unitary operator $U_0 \left(t\right)$ while the superscript `\textit{H}' stands for time evolution under the full-time evolution operator. Using Eq.~\eqref{ref16} and given the total initial state $\varrho_\text{tot}(0)$ in Eq.~\eqref{ref11}, we can derive the master equation by simplifying Eq.~\eqref{mastereqgen}. The result due to the first term in Eq.~\eqref{ref16} is very straightforward. We simply have that 
\begin{align}
    \text{Tr}_\text{S,E}\left\{\varrho_\text{tot}(0)i\left[H_0^H(t),X_{\text{nm}}^H(t)\right]\right\}
    &=i\text{Tr}_\text{S,E}\left\{\varrho_\text{tot}(t)\left[H_S + H_E,(\ket{n}\bra{m}\otimes \mathds{1}_E)\right]\right\},\nonumber
\\
    &=i\text{Tr}_S\left\{\varrho(t)\left[H_S,\ket{n}\bra{m}\right]\right\},\nonumber
\\
    &=i\bra{m}\left[\varrho(t),H_S\right]\ket{n}. \label{ref20}
\end{align}
This term simply tells us about free system evolution corresponding to $H_\text{S}$. To calculate the next term in our master equation, that is
 \begin{align*}
    i\alpha \text{Tr}_\text{S,E}\left\{\varrho_\text{tot}(0)\left[H_{\text{SE}}(t),X_{\text{nm}}(t)\right]\right\},
\end{align*}
we now expand the initial $\mathcal{SE}$ state perturbatively. It is useful to write $\varrho_{\text{tot}}(0) = \varrho_{\text{tot}}^{(0)} + \varrho_{\text{tot}}^{(1)}$, where [see Eq.~\eqref{ref14} in the section \autoref{initial}] 
\begin{align}
\varrho_\text{tot}^{(0)}(0)
    &=\frac{R e^{-\beta(H_{\text{S0}}+H_E)}R^\dagger}{Z_{\text{tot}}}  = \varrho^R_{\text{S0}} \otimes \varrho_E, \label{ref18}
\\
    \varrho_\text{tot}^{(1)}(0)
    &=\frac{-\alpha R e^{-\beta(H_{\text{S0}}+H_E)}Q_{\text{SE}}(\beta)R^\dagger}{Z_{\text{tot}}}. \label{ref19}
\end{align}
Here $\varrho^R_{\text{S0}} = e^{-\beta H^R_{\text{S0}}}/Z_{\text{S0}}$, $\varrho_E = e^{-\beta H_E}/Z_E$, the partition function $\tot{Z} = Z_{\text{S0}} Z_E$, and $Q_{\text{SE}}(\beta) = \int_0^\beta d\lambda S(\lambda)\otimes E(\lambda)$ with $S(\lambda) = e^{\lambda H_{\text{S0}}} S e^{-\lambda H_{\text{S0}}}$ and $E(\lambda) = e^{\lambda H_E} E e^{-\lambda H_E}$. We do not need the higher order terms since there is already a factor of $\alpha$ in $i\alpha\left[H_{\text{SE}}(t),X_{\text{nm}}(t)\right]$. Now, the contribution of $\varrho_\text{tot}^{(0)}$ is
\begin{align*}
    &i\alpha \text{Tr}_\text{S,E}\left \{\varrho_\text{tot}^{(0)}(0)\left[U_0^\dagger(t)H_{\text{SE}}U_0(t),U_0^\dagger(t)X_{\text{nm}}U_0(t)\right]\right\},
 \\   
    &=i\alpha \text{Tr}_\text{S,E}\left\{\varrho^R_{\text{S0}}\otimes\varrho_E U_0^\dagger(t) \left[S\otimes E,\ket{n}\bra{m}\otimes \mathds{1}_E\right] U_0(t)\right\},
\\
    &=i\alpha \text{Tr}_S\left \{\varrho^R_{\text{S0}}U_S^\dagger(t)\left[S,Y_{nm}\right]U_S(t)\right\}\times \langle E(it)\rangle_E.
\end{align*}
Since $\langle E(it)\rangle_E$ is usually zero for most $\mathcal{SE}$ models, this contribution turns out to be zero. The most interesting contribution is due to $\varrho_\text{tot}^{(1)}(0)$. Using this along with the second term in Eq.~\eqref{ref16}, we get 
\begin{align}
    &i\alpha \text{Tr}_\text{S,E}\left\{\varrho_\text{tot}^{(1)}(0)\left[U_0^\dagger(t)H_{\text{SE}}U_0(t),U_0^\dagger(t)X_{\text{nm}}U_0(t)\right]\right\},\nonumber
\\
    &=\frac{-i\alpha^2}{Z_{\text{S0}}}\int_{0}^{\beta}\text{Tr}_\text{S,E}\left\{\varrho_E{R e^{-\beta H_{\text{S0}}} S(\lambda)R^\dagger\otimes  E(\lambda)}U_S^\dagger(t) \left[S,\ket{n}\bra{m}\right]U_S(t)U_E^\dagger(t)EU_E(t)\right\}d\lambda,\nonumber
\\
    &=\frac{-i\alpha^2}{Z_{\text{S0}}}\int_{0}^{\beta}\text{Tr}_{S}\left\{{R e^{-\beta H_{\text{S0}}} S(\lambda)R^\dagger}U_S^\dagger(t)\left[S,\ket{n}\bra{m}\right] U_S(t)\right\} \text{Tr}_E\left\{\varrho_E E(\lambda)E(it)\right\}d\lambda,\nonumber
\\
    &=\frac{-i\alpha^2}{Z_{\text{S0}}}\int_{0}^{\beta}\bra{m}\left[U_S(t)R e^{-\beta H_{\text{S0}}} S(\lambda)R^\dagger U_S^\dagger(t),S\right]\ket{n}  E_{\text{corr}}(\lambda,t)\, d\lambda, \label{ref21}
\end{align}
here $E_{\text{corr}}(\lambda,t) = \text{Tr}_E \Big\{\varrho_E E(\lambda)E(it)\Big\}$. This is the additional term that appears in the master equation as a consequence of the initial correlations. In basis-independent form, we can write this term as 
\begin{align}
    -i\left[ \widetilde{\varrho}(t)J^R_\text{corr}(\beta,t),S\right],
\end{align}
where we have defined $\widetilde{\varrho}(t) = U_S(t) \varrho^R_{\text{S0}} U_S^\dagger (t)$ and
\begin{align} 
    J^R_\text{corr}(\beta,t)
    &=\int_{0}^{\beta}  \overleftarrow{S}^R(\lambda,t)E_\text{corr}(\lambda,t)d\lambda,
    \label{JR}
    \end{align}
    \begin{align}
    \overleftarrow{S}^R(\lambda,t)
    &= U_S(t)R e^{\lambda H_{\text{S0}}}Se^{-\lambda H_{\text{S0}}}R^\dagger U_S^\dagger(t).
    \label{FR}
\end{align}
We are allowed to replace $\widetilde{\varrho}(t)$ by $\varrho(t)$ because the corrections would be of order higher than second-order in the interaction strength. However, it can be checked that $-i\left[ \varrho(t)J^R_\text{corr}(\beta,t),S\right]$ is not guaranteed to be Hermitian. To proceed, we first write
\begin{align}
    -i\left[ \widetilde{\varrho}(t)J^R_\text{corr}(\beta,t),S\right] = -\frac{i}{2}\left(\left[ \widetilde{\varrho}(t)J^R_\text{corr}(\beta,t),S\right] - \text{H. c.}\right),
\end{align}
where $\text{H.~c.}$ denotes hermitian conjugate. This is permitted because $ -i\left[ \widetilde{\varrho}(t)J^R_\text{corr}(\beta,t),S\right]$ is Hermitian, so 
$\left[ \widetilde{\varrho}(t)J^R_\text{corr}(\beta,t),S\right]$ is anti-Hermitian. We now replace $\widetilde{\varrho}(t)$ by $\varrho(t)$. Consequently, the term in the master equation that incorporates the effect of initial correlations is $-\frac{i}{2}\left(\left[ \varrho(t)J^R_\text{corr}(\beta,t),S\right] - \text{H. c.}\right)$, and this is manifestly Hermitian.

We next simplify third term in Eq.~\eqref{ref16}. It is clear that now only $\varrho_{\text{tot}}^{(0)}(0)$ contributes. Similar manipulations to those performed above lead to (see the Appendix \autoref{relaxation} for details) 
$ \alpha^2\int_{0}^{t}\bra{m}\Big(\left[\Bar{S}(t,s)\widetilde{\varrho}(t),S\right]C_{\text{ts}}+\text{H.~c.}\Big)\ket{n}\,ds, $
where the environment correlation function is $C_{\text{ts}} = \langle E(it)E(is) \rangle_E$, $\Bar{S}(t,s)=U_S^\dagger(t,s) S U_S(t,s)$. We can further replace $\widetilde{\varrho}(t)$ by $\varrho(t)$ to get 
$$ \alpha^2\int_{0}^{t}\bra{m}\Big(\left[\Bar{S}(t,s)\varrho(t),S\right]C_{\text{ts}}+\text{H.~c.}\Big)\ket{n}\,ds. $$
Once again, this is permitted due to the same reason mentioned earlier.  We now put all the terms together to arrive at the general basis-independent structure of the master equation 

\begin{align}
    \mathcolorbox{Apricot}{\Dot{\varrho}(t)
    =i\left[\varrho(t),H_S\right]-\frac{i}{2}\left(\left[ \varrho(t)J^R_\text{corr}(\beta,t),S\right] - \text{H. c.}\right) +\int_{0}^{t}\left(\left[\Bar{S}(t,s)\varrho(t),S\right]C_{\text{ts}}+\text{H. c.}\right)ds.}
     \label{finalme}
\end{align}
Let us note that we have assumed implicitly that the timescale on which the unitary operator $R$ is implemented as well as the time taken to change the system Hamiltonian from $H_{\text{S0}}$ to $H_S$ are much smaller than the other timescales such as the environment correlation time, the relaxation time, and the free system evolution timescale. We also emphasize that the same master equation applies if the system Hamiltonian is time-dependent with the caveat that finding the free system time-evolution operator $U_S(t)$ will then be, in general, highly non-trivial. In fact, we use such a time-dependent Hamiltonian in Appendix \autoref{timedependent} to examine more carefully what happens when the system Hamiltonian parameters are not changed instantaneously.

\section{Application to the spin-boson model}
\label{seclargespinapp}

In this section, we apply our derived master equation to a variant of $\mathcal{SB}$ model \cite{BPbook} with numerous two-level systems interacting with their environment of harmonic oscillators \cite{KurizkiPRL20112,ChaudhryPRA2013a,ChaudhryPRA2013b}. Recall that the total $\mathcal{SE}$ Hamiltonian is $H_\text{tot} = H_{\text{S0}}+H_\text{E}+H_{\text{SE}}$ for $t < 0$, while $H_{\text{tot}} = H_S+H_E+H_{\text{SE}}$ for $t \geq 0$. Within the large $\mathcal{SB}$ model, we take the following set of Hamiltonians
\begin{align}
    H_{\text{S0}}
    &= \varepsilon_0 J_z + \Delta_0 J_x,\\
    H_S&= \varepsilon J_z + \Delta J_x,\\
    H_E&= \sum_{k} \omega_k b_{k}^{\dagger} b_k,\\
    H_{\text{SE}}&=J_z \sum_k \big( g_k^* b_k + g_k b_k^\dagger \big),
\end{align}
where $J_{x,y,y}$ are the collective spin angular momentum operators. $J^2 =J_x^2+J_y^2+J_z^2, \varepsilon$ is the energy bias, $\Delta$ is the tunneling amplitude, $H_E$ is the environment Hamiltonian. For convenience, we have ignored the zero-point energy. $H_{\text{SE}}$ represents the interaction between the common harmonic oscillator environment and the spin system. Note that the system operator $S=J_z,$ and the environment operator $E=\sum_k \big( g_k^* b_k + g_k b_k^\dagger \big). $ One imagines that the large-spin system has been interacting with the environment for a long time with a relatively large value of $\varepsilon_0$ and a small value of $\Delta_0$. In such a situation with $\beta \varepsilon_0 \gg 1$, realized, for example, by applying a suitably large static magnetic field, the system state will be approximately corresponding to the state with all spins down in the $z$-direction. At the time $t = 0$, we then apply a unitary operator to prepare the needful initial state. For example, if the desired initial state is one with all spins in the $x$-direction, then the unitary operator that should be applied is $R = e^{i\pi J_y/2}$. In other words, a $\frac{\pi}{2}$-pulse is applied to prepare the initial state, with the assumption that this pulse takes a very short time to be applied. In particular, we assume that the pulse duration is smaller than the inverse of the effective Rabi frequency $\widetilde{\Delta} = \sqrt{\varepsilon^2 + \Delta^2}$ as well as the inverse of the environment cutoff frequency. With the initial state approximately prepared, we can then change the parameters of the system Hamiltonian to whatever values we desire to generate any required system evolution - in our example, this entails changing the energy bias from $\varepsilon_0$ to $\varepsilon$ so that the effect of the tunneling term $\Delta J_x$ becomes more evident. Again, we assume that this change takes place over a very short time interval; this approximation is further critically examined in the Appendix \autoref{timedependent}. Let us then look at how the initial correlations appear in the system evolution using our general master equation. 

Our first objective is to calculate the operator $J^R_{\text{corr}}$. To do so, we first find [see Eq.~\eqref{FR}]
\begin{align*}
    \overleftarrow{S}^R(\lambda,t)
    &=U_S(t)\left[R\left(e^{\lambda H_{\text{S0}}} S e^{-\lambda H_{\text{S0}}}\right)R^{\dagger}\right]U_S^\dagger(t),
\\
    &=J_x\left[a_xd_x+a_yc_x-a_z b_x\right]+J_y\left[a_xd_y+a_yc_y-a_z b_y\right] + J_z\left[a_xd_z+a_yc_z-a_zb_z\right],
\end{align*}
with 
\begin{align*}
       a_x
    &=\frac{\varepsilon_0\Delta_0}{\Delta'^2} \left\{1-\cosh\left({\lambda\Delta'} \right)\right\},\\
    a_y
    &=\frac{-i\Delta_0}{\Delta'}\sinh \left(\lambda\Delta'\right),\\
    a_z
    &= \frac{\varepsilon_0^2 + \Delta^2_0 \cosh \left({\lambda\Delta'}\right)}{\Delta'^2},\\
     b_x
    &= \frac{\Delta^2 + \varepsilon^2 \cos \left({\widetilde{\Delta}t}\right)}{\widetilde{\Delta}^2},\\
    b_y
    &= \frac{ \varepsilon}{\widetilde{\Delta}}\sin \left( \widetilde{\Delta}t\right),\\
    b_z
    &= \frac{\varepsilon \Delta}{\widetilde{\Delta}^2} \left\{ 1 - \cos\left({\widetilde{\Delta}t} \right)\right\},\\
    c_x 
    &= -\frac{\varepsilon }{\widetilde{\Delta}}\sin{\left(\widetilde{\Delta}t\right)},\\
    c_y
    &= \cos{\left(\widetilde{\Delta}t\right)},\\
    c_z
    &= \frac{\Delta }{\widetilde{\Delta}}\sin{\left(\widetilde{\Delta}t\right)},\\
     d_x &= \frac{\varepsilon\Delta }{\widetilde{\Delta}^2}\left\{1-\cos{\left(\widetilde{\Delta}t\right)}\right\},\\
    d_y &=-\frac{\Delta }{\widetilde{\Delta}}\sin{\left(\widetilde{\Delta}t\right)},\\
    d_z &= 1+\frac{\Delta^2 }{\widetilde{\Delta}^2}\left\{\cos{\left(\widetilde{\Delta}t\right)}-1\right\}.
\end{align*}
Here $\Delta'^2 = \varepsilon_0^2 + \Delta_0^2$ and $\widetilde{\Delta}^2 = \varepsilon^2 + \Delta^2$. In short
\begin{align} \label{ref24}
    \overleftarrow{S}^R(\lambda,t)=\alpha_1(\lambda,t)J_x+\alpha_2(\lambda,t)J_y+\alpha_3(\lambda,t)J_z,
\end{align}
where
\begin{align*}
    \alpha_1(\lambda,t)
    &=a_xd_x+a_yc_x-a_z b_x,\\
    \alpha_2(\lambda,t)
    &=a_xd_y+a_yc_y-a_z b_y,\\
    \alpha_3(\lambda,t)
    &=a_xd_z+a_yc_z-a_zb_z.
\end{align*}

\begin{figure}[t!]\begin{framed} 
  \centering 
  \includegraphics[scale=1]{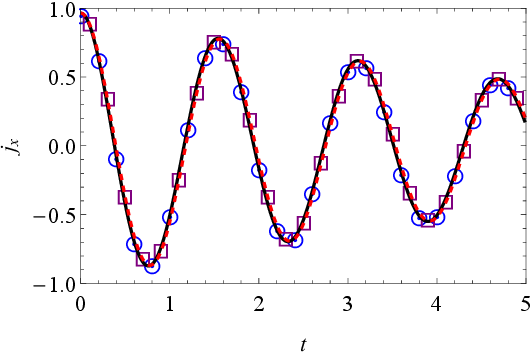}
\caption{Behavior of $j_x = 2\langle J_x \rangle/N$ as a function of} $t$ for $N = 1$ using the exact solution with (blue circled dot) and without (purple squares) initial correlations, as well as using the master
equation with (solid, black line) and without (dashed, red line) initial correlations in an Ohmic environment. We have used $\varepsilon=\varepsilon_0=4$, $G = 0.05$, $\beta=1$ and $\omega_c=5$. Here and in all other figures, the plotted variables are all in dimensionless units. \label{Puredephasing-N=1-unitary}
\end{framed}\end{figure} 
 \begin{figure}[t!]\begin{framed} 
  \centering 
  \includegraphics[scale=1]{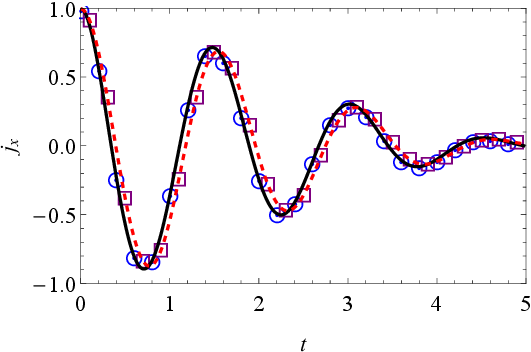}
\caption{Same as Fig. \ref{Puredephasing-N=1-unitary}, but now $N = 4$.}
\label{Puredephasing-N=4-unitary}
\end{framed}\end{figure}
\begin{figure}[t!]\begin{framed} 
  \centering 
  \includegraphics[scale=1]{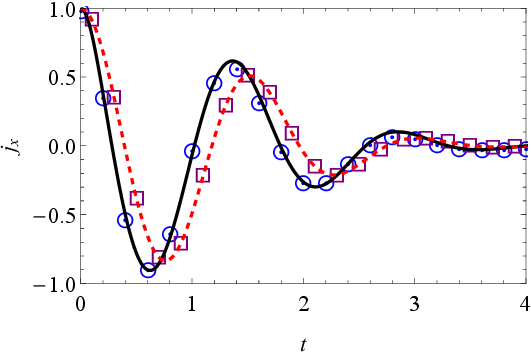}
\caption{Same as Fig. \ref{Puredephasing-N=1-unitary}, but now $N = 10$.} \label{Puredephasing-N=10-unitary}
\end{framed}\end{figure}

It then follows that [see Eq.~\eqref{JR}] 
\begin{align}
\label{initialcorrelations}
    J_\text{corr}^R(\beta,t)
    &=P(\beta,t)J_x+Q(\beta,t)J_y+R(\beta,t)J_z,
\end{align}
with
\begin{align*}
    P(\beta,t)
    &=\int_{0}^{\beta}  \alpha_1(\lambda,t)E_{\text{corr}}(\lambda,t)\,d\lambda,\\
    Q(\beta,t)
    &=\int_{0}^{\beta}  \alpha_2(\lambda,t)E_{\text{corr}}(\lambda,t)\,d\lambda,\\
    R(\beta,t)
    &=\int_{0}^{\beta}  \alpha_3(\lambda,t)E_{\text{corr}}(\lambda,t)\,d\lambda.
\end{align*}
We now calculate environment correlation term $E_{\text{corr}}(\lambda,t)$. First
\begin{align}
    E(\lambda)=\sum_k\left(g_k^*e^{-\lambda\omega_{k}}b_k+g_ke^{\lambda\omega_{k}}b_k^\dagger\right).
\end{align}
Since $E_{\text{corr}}(\lambda,t) =\text{Tr}\left\{\varrho_E E(\lambda) E(it)\right\}$, we find (see Appendix \autoref{bosonicbath} for details) 
\begin{align}
    \mathcolorbox{Apricot}{E_{\text{corr}}(\lambda,t)
    =\sum_k  |g_k|^2\Big\{e^{-\omega_{k}\left(\lambda-it\right)} +2n_k\cosh{\left(\lambda\omega_{k}-i\omega_{k}t\right)}\Big\},}\label{ref26}
\end{align}

\begin{figure}[t!]\begin{framed} 
  \centering 
  \includegraphics[scale=1]{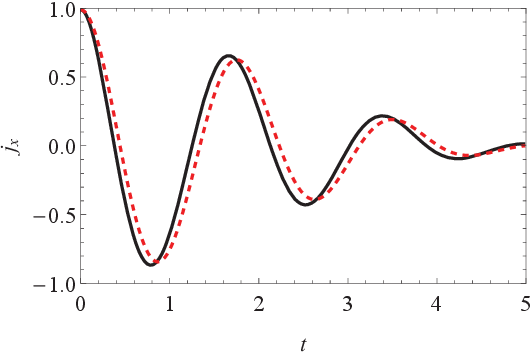}
\caption{Behavior of $j_x$ as a fuction of $t$ for $N=2$ with (black, solid) and without (dashed, red) including the effect of initial correlations. Here energy biases are $\varepsilon_0 = 4$, $\varepsilon=2.5$ and tunneling is $\Delta=\Delta_0=0.5$, while other parameters are same as Fig. \ref{Puredephasing-N=1-unitary}.} \label{Beyond-PD-N=2}
\end{framed}\end{figure}
\begin{figure}[t!]\begin{framed} 
  \centering 
  \includegraphics[scale=1]{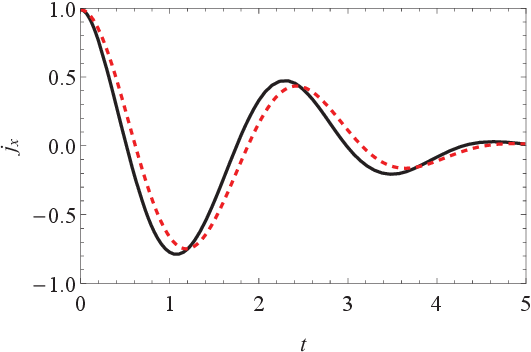}
\caption{Same as Fig. \ref{Beyond-PD-N=2}, but now $N = 4$.} \label{Beyond-PD-N=4}
\end{framed}\end{figure}
\begin{figure}[t!]\begin{framed} 
  \centering 
  \includegraphics[scale=1]{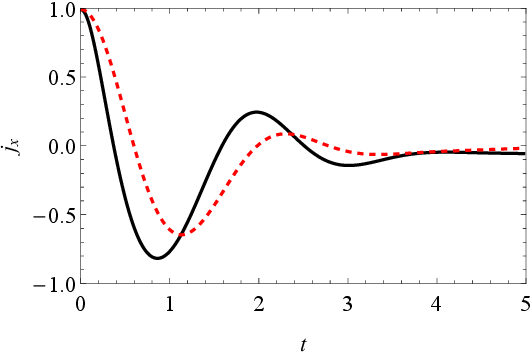}
\caption{Same as Fig. \ref{Beyond-PD-N=2}, but now $N = 10$.} \label{Beyond-PD-N=10}
\end{framed}\end{figure}

We imagine the environment harmonic oscillators are dense in frequency, allowing us to compute the sum over the environment modes via $\sum_k |g_k|^2 (\hdots) \rightarrow \int_0^\infty d\omega \, J(\omega) (\hdots)$. Here $J(\omega)$ is the standard spectral density function. We generally use an Ohmic spectral density written as $J(\omega) = G\omega e^{-\omega/\omega_c}$. The integrals are performed numerically to find $J_{\text{corr}}(\beta,t)$, and the results are incorporated in the numerical simulations of the master equation. We first examine the pure dephasing case where $\Delta=\Delta_0=0$, since this case can be solved exactly and serves as a useful benchmark (details of the exact solution are given in the section \autoref{pure}, chapter \ref{c:formulism}). We illustrate our results in Fig.~\ref{Puredephasing-N=1-unitary} for $N = 1$ by plotting $j_x = 2\langle J_x \rangle/N$. Two points should be noted. First, the role played by initial correlations is very small. Second, our master equation reproduces the exact results very well. Since the role of the initial correlations is expected to increase with increasing $N$, we next look at $N = 4$ and $N = 10$. Results are shown in Figs.~\ref{Puredephasing-N=4-unitary} and \ref{Puredephasing-N=10-unitary}. It is evident that as $N$ increases, the initial correlations play a larger and larger role. This is a manifestation of the fact that the environment harmonic oscillators can be understood to be displaced as a consequence of the $\mathcal{SE}$ interaction [see the displaced harmonic oscillator modes Eqs.~\eqref{ch2-1}, \eqref{ch2-2}], and as $N$ increases, the environment harmonic oscillator modes are displaced more. Moreover, the extra term in the master equation is able to take into account the effect of the initial correlations very well.

\begin{figure}[htp]\begin{framed} 
  \centering 
  \includegraphics[scale=1]{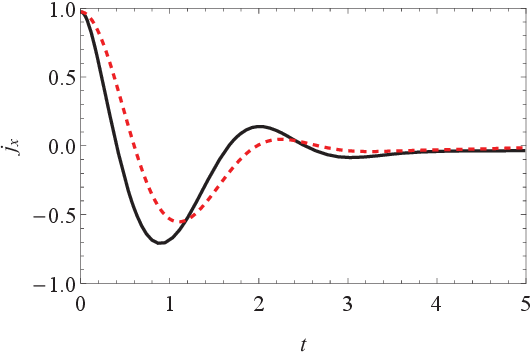}
\caption{Behavior of $j_x$ as a function of} $t$ for $N=10$ with (black, solid) and without (dashed, red) including the effect of initial correlations. The parameters used are same as Fig.~\ref{Beyond-PD-N=10}, except that $\beta = 0.5$. \label{Beta=0.5}
\end{framed}\end{figure}

\begin{figure}[htp]\begin{framed} 
  \centering 
  \includegraphics[scale=1]{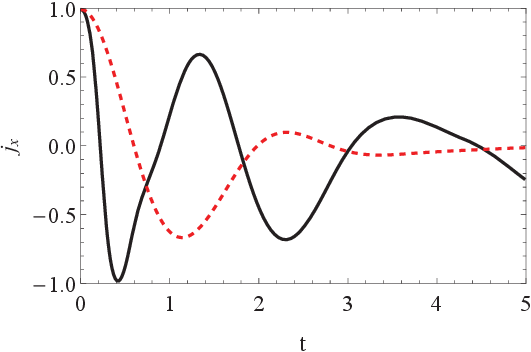}
\caption{Same as Figs.~\ref{Beyond-PD-N=10} and \ref{Beta=0.5}, but now $\beta=1.5$.} \label{Beta=1.5}
\end{framed}\end{figure}

Having shown that our master equation is able to reproduce results for the pure dephasing model, we are now in a position to go beyond the pure dephasing model and see the effects of the initial correlations. In Fig.~\ref{Beyond-PD-N=2}, we have shown the dynamics of $j_x$ with a non-zero value of the tunneling amplitude for $N = 2$. It is clear that the initial correlations do have a small influence on the dynamics. This effect becomes more pronounced as we increase $N$ (see Figs.~\ref{Beyond-PD-N=4} and \ref{Beyond-PD-N=10}), which signifies that the environment harmonic oscillators are more influenced by the system as $N$ increases. We have also looked at how the role played by the initial correlations changes as the temperature changes. To this end, we compare Fig.~\ref{Beyond-PD-N=10}, where the inverse temperature is $\beta = 1$, with Fig.~\ref{Beta=0.5} where $\beta = 0.5$ and Fig.~\ref{Beta=1.5} where $\beta = 1.5$. At higher temperatures, the effect of the initial correlations decreases, while at lower temperatures, the effect of the initial correlations increases. Mathematically, this can be seen in Eq.~\eqref{initialcorrelations}, where $P(\beta,t)$, $Q(\beta,t)$, and $R(\beta,t)$ become negligible as the temperature increases. This illustrates that our master equation produces sensible results since we do expect the role of the initial correlations to decrease as the temperature increases.\\

\begin{figure}[htp]\begin{framed} 
  \centering 
  \includegraphics[scale=1]{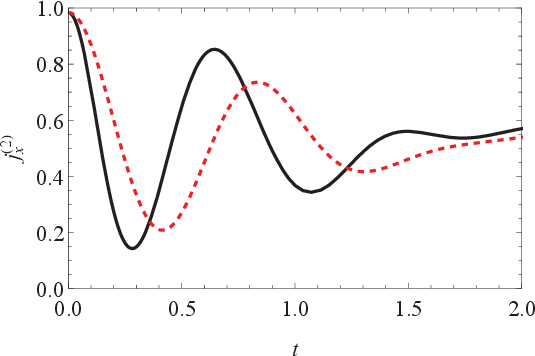}
\caption{Behavior of $j_x^{(2)}$ as a function of $t$ for $N = 10$
with (black, solid) and without (dashed, red) including the effect of initial correlations. The other parameters are same as in Fig.~\ref{Beyond-PD-N=2}.} \label{Jx^2-plot-for-N=10}
\end{framed}\end{figure}

\begin{figure}[htp]\begin{framed} 
  \centering 
  \includegraphics[scale=1]{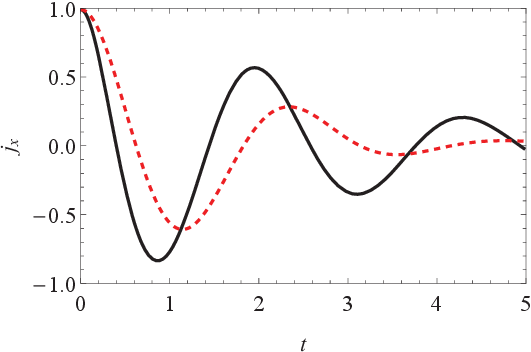}
\caption{Behavior of $j_x$ as a function of $t$ for $N=4$ with (black, solid) and without (dashed, red) including the effect of initial correlations. Here we have used a sub-Ohmic environment with $s = 0.5$. We also have $\varepsilon_0 = 4$, $\varepsilon=2.5$ and $\Delta=\Delta_0=0.5$, while other parameters are same as in Fig. \ref{Puredephasing-N=1-unitary}.} \label{subOhmic1}
\end{framed}\end{figure}

\begin{figure}[htp]\begin{framed} 
  \centering 
  \includegraphics[scale=1]{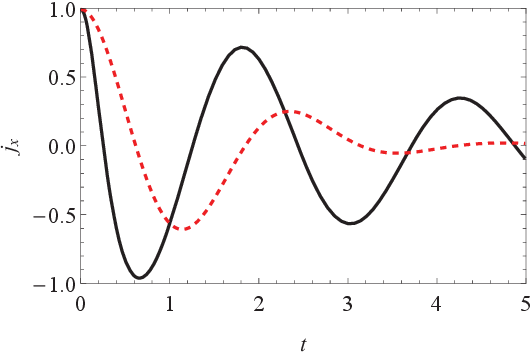}
\caption{Same as Fig.~\ref{subOhmic1}, but now $N = 10$.} \label{subOhmic2}
\end{framed}\end{figure}

Let us now demonstrate that the effect of the initial correlations is not evident in the dynamics of $j_x$ alone. We illustrate in Fig.~\ref{Jx^2-plot-for-N=10} the dynamics of $j_x^{(2)} = 4\av{J_x^2}/N^2$, which is not merely the sum of single-particle operators. Such an observable is relevant in the research on spin squeezing. It is clear from the figure that the effect of the initial correlations may also need to be accounted for when studying the dynamics of quantities beyond single-particle observables. 
Finally, in order to illustrate that we can equally well deal with other kinds of environments, we also demonstrate the effect of the initial correlations with a sub-Ohmic environment, that is, $J(\omega) = G\omega^s \omega_c^{1 - s} e^{-\omega/\omega_c}$ with $s < 1$. Since sub-Ohmic environments have longer correlation times, we expect that the effect of the initial correlations will be greater as well. This is indeed the case, as can be seen by comparing Figures \ref{subOhmic1} and \ref{subOhmic2} with Figs.~\ref{Beyond-PD-N=4} and Figs.~\ref{Beyond-PD-N=10} where an Ohmic environment had been used. 

\section{Application to the spin-spin model}
\label{spinspinmodel}

Now consider a collection of identical two-level systems interacting with an environment consisting of two-level systems \cite{cucchietti2005decoherence, Schlosshauerbook,camalet2007effect,villar2009spin,segal2014two}. We have
\begin{align*}
    H_{\text{S0}}
    &= \varepsilon_0 J_z + \Delta_0 J_x,\\
    H_S&= \varepsilon J_z + \Delta J_x,\\
    H_E&= \sum_{k} \frac{\omega_k}{2} \sigma_{x}^{(k)},\\
    H_{\text{SE}}&=J_z \otimes \sum_k g_k \sigma_{z}^{(k)}.
\end{align*}
where $\sigma_{z}^{(k)}$ and $ \sigma_{x}^{(k)}$ are the $z$ and $x$ components of Pauli spin operators of the $k^{\text{th}}$ environment spin respectively, $\omega_k$ symbolizes the tunneling matrix element for the $k^{\text{th}}$ environment spin, and $g_k$ quantifies the interaction strength. The different environment leads to a different correlation function $C_{\text{ts}}$ as well as a different factor $J^R_\text{corr}(\beta,t)$ that incorporates the effect of the initial correlations. The calculation of the environment correlation function is sketched out in Appendix \autoref{spinbathcorr}. A similar calculation leads to 

\begin{figure}[htp]\begin{framed} 
  \centering 
  \includegraphics[scale=1]{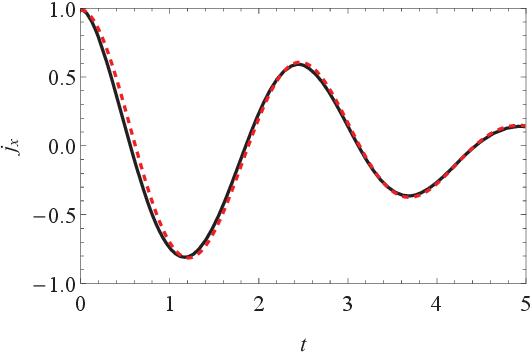}
\caption{Behavior of $j_x$ as a function of $t$ for $N = 4$
with (black, solid) and without (dashed, red) including the effect of initial correlations. The parameter we have used here are $\varepsilon_0 = 4$, $\varepsilon=2.5$, $\Delta=\Delta_0=0.5$, $G = 0.05$, $\beta=1$ and $\omega_c=5$.} \label{spinenv1}
\end{framed}\end{figure}
\begin{figure}[htp]\begin{framed} 
  \centering 
  \includegraphics[scale=1]{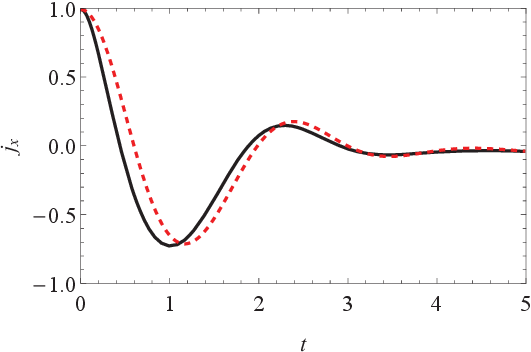}
\caption{Same as Fig.~\ref{spinenv1}, but now $N = 10$.} \label{spinenv2}
\end{framed}\end{figure}


\begin{align}
\mathcolorbox{Apricot}{
    E_{\text{corr}}(\lambda,t)
    =\sum_k  \abs{g_k}^2\left\{\text{tanh}{\left(\frac{\beta \omega_k}{2}\right)}e^{-\omega_{k}\left(\lambda-it\right)} + 2n_k\sinh\left(\lambda\omega_{k}-i\omega_{k}t\right)\right\},}
\end{align}
here the time-dependent coefficients $\alpha_1(\lambda,t)$, $\alpha_2(\lambda,t)$, and $\alpha_3(\lambda,t)$ remain same as before, this allows us to quantify the role of the initial correlations. Results are shown in Figs.~\ref{spinenv1} and \ref{spinenv2}. Once again, the role of the initial correlations is relatively small for a smaller value of $N$. However, as $N$ increases, it is clear that we need to encompass the role of the initial correlations to obtain an accurate picture of the system dynamics even in the spin environment.
\section{Summary}\label{ch4sum}

In this chapter, we have shown that if we start from the joint thermal equilibrium state of a quantum system and its environment and then apply a unitary operation to the system to prepare the system quantum state, the initial correlations that exist in the joint thermal equilibrium state influence the subsequent dynamics of the system. We have derived a time-local master equation, correct to second-order in the $\mathcal{SE}$ coupling strength, that takes into account the effect of these correlations, showing therefore that one need not necessarily be in the strong $\mathcal{SE}$ coupling regime to observe the effects of the initial correlations. The structure of this master equation is very interesting, as the form of the term that takes into account the initial correlations is the same as the relaxation and dephasing term. In this sense, one can say that the initial correlations affect the decoherence and dephasing rates, a fact that was already pointed out in studies of the role of initial correlations in pure dephasing models [24]. Finally, we actually applied our master equation to the large $\mathcal{SB}$ model as well as to a collection of two-level systems interacting with a spin environment to quantitatively investigate the role of the initial correlations. We found that when the number of spins is small, then the initial correlations do not play a significant role. However, for a larger number of spins, the initial correlations must be accounted for in order to explain the dynamics accurately.

\chapter*{Appendices}
\addcontentsline{toc}{chapter}{\tocEntry{Appendices}}
\section{The relaxation/dephasing term in the master equation}
\label{relaxation} 
We look at the contribution coming from the third term in Eq.~\eqref{ref16}. We need to consider only $\varrho_\text{tot}^{(0)}(0)$ since we are restricted to consider up to second-order terms in the master equation. Therefore
\begin{align}
    &\alpha^2 \text{Tr}_\text{S,E}\left \{\varrho_\text{tot}^{(0)}(0)\int_{0}^{t}\left[\left[\widetilde{H}_{\text{SE}}(t),\widetilde{X}_{\text{nm}}(t)\right],\widetilde{H}_{\text{SE}}(s)\right]ds\right \} ,\notag 
\\
    &=\alpha^2 \text{Tr}_\text{S,E}\left\{\varrho^R_{S0}\otimes\varrho_E \int_{0}^{t}\left[\left[\widetilde{H}_{\text{SE}}(t),\widetilde{X}_{\text{nm}}(t)\right],\widetilde{H}_{\text{SE}}(s)\right]ds\right \} ,\notag 
\\
    &=\alpha^2 \text{Tr}_\text{S,E} \Bigg\{\varrho^R_{S0}\otimes \varrho_E\Bigg(\int_{0}^{t}\widetilde{H}_{\text{SE}}(t)\widetilde{X}_{\text{nm}}(t)\widetilde{H}_{\text{SE}}(s)ds-\int_{0}^{t}\widetilde{H}_{\text{SE}}(s)\widetilde{H}_{\text{SE}}(t)\widetilde{X}_{\text{nm}}(t)ds ,\nonumber
\\
    &-\int_{0}^{t}\widetilde{X}_{\text{nm}}(t)\widetilde{H}_{\text{SE}}(t)\widetilde{H}_{\text{SE}}(s)ds+\int_{0}^{t}\widetilde{H}_{\text{SE}}(s)\widetilde{X}_{\text{nm}}(t)\widetilde{H}_{\text{SE}}(t)ds\Bigg) \Bigg\}.
\end{align}  
The first term is 
\begin{align*}    
    &\alpha^2 \text{Tr}_\text{S,E}\left \{\varrho^R_{S0}\otimes\varrho_E\int_{0}^{t}\widetilde{H}_{\text{SE}}(t)\widetilde{X}_{\text{nm}}(t)\widetilde{H}_{\text{SE}}(s)ds\right \} ,
\\
    &=\alpha^2\int_{0}^{t}\text{Tr}_\text{S,E}\Big \{\varrho^R_{S0}\otimes\varrho_EU_0^\dagger(t)H_{\text{SE}}U_0(t)U_0^\dagger(t)X_{\text{nm}}U_0(t) U_0^\dagger(s)H_{\text{SE}} U_0(s)\Big \}ds,
\\
    &=\alpha^2\int_{0}^{t}\text{Tr}_{S}\left \{\varrho^R_{S0}U_S^\dagger(t) S Y_{\text{nm}}U_S(t,s)FU_S(s)\right \} \text{Tr}_{E}\left \{\varrho_E E(t)E(s)\right \}ds,
\\
    &=\alpha^2\int_{0}^{t}\bra{m}\Bar{S}(t,s)\widetilde{\varrho}(t)S\ket{n} C_{\text{ts}} ds.
\end{align*}
In a similar fashion, we can simplify the other terms of the master equation. Putting them all back together, and shifting to the basis-independent representation, we obtain the third term in Eq.~\eqref{finalme}.

\section{Master equation with time-dependent system Hamiltonian} 
\label{timedependent}

\begin{figure}[t!] 
\begin{framed}
  \centering 
  \includegraphics[scale =0.85]{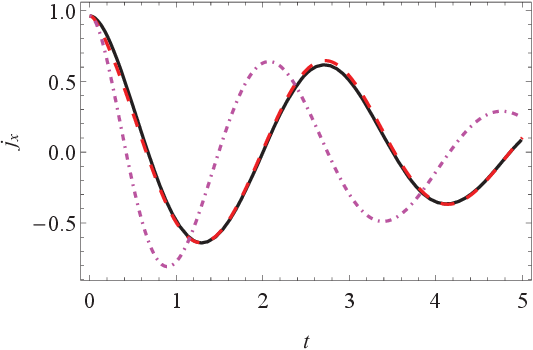}
\caption{Behavior of $j_x$ as a function of $t$ for $N = 2$, including the effect of the initial correlations, with the time-dependent system Hamiltonian. The solid, black curve is with $t_\varepsilon \rightarrow 0$, for the dashed red curve $t_\varepsilon = 0.1$, while $t_\varepsilon = 1$ for the dot-dashed magenta curve. Parameters being $\varepsilon_0 = 4$, $\varepsilon=2$, $\Delta=\Delta_0=1$, $G = 0.05$, $\beta=1$ and $\omega_c=5$.} \label{spinone}
\end{framed}
\end{figure}

In section \autoref{seclargespinapp}, we applied the master equation [see Eq.~\eqref{finalme}] to the large spin-boson model with the system Hamiltonian parameters changed suddenly. In particular, for the numerical results presented, the tunneling amplitude was not changed, that is, $\Delta_0 = \Delta$, while the energy level-splitting was changed from $\varepsilon_0$ to $\varepsilon$ instantaneously at $t = 0$. We examine in this appendix what happens if we do not change the energy level spacing instantaneously. In particular, we consider that for $t \geq 0$, the system Hamiltonian is $H_S(t) = \frac{\epsilon(t)}{2}J_z + \Delta J_x$, where $\epsilon(t) = \left(\varepsilon_0 - \varepsilon \right)e^{-t/t_\varepsilon} + \varepsilon$. $t_\varepsilon$  is a measure of how quickly we change the energy level-spacing with a smaller value of $t_\varepsilon$ indicating a quicker transition from $\varepsilon_0$ to $\varepsilon$. This time-dependent Hamiltonian can be used in the master equation Eq.~\eqref{finalme}, with the system unitary time-evolution operator calculated numerically via the split-operator method, thereby also entailing numerical evaluation of the operator $J_{\text{corr}}^R(\beta,t)$ as well as the third term in the master equation. Plots for different values of $t_\varepsilon$ are shown in Fig.~\ref{spinone}. As expected, for small values of $t_\varepsilon$, the results with the time-dependent Hamiltonian agree very closely with our previous results where we assumed that the system Hamiltonian is changed instantaneously. Such agreement is expected when $t_\varepsilon$ is smaller than the environment correlation time (which is related to the inverse of the cutoff frequency) as well as the timescale set by the system Hamiltonian (which is on the order of $1/\sqrt{\varepsilon^2 + \Delta^2}$).

\section{Environment correlation function with harmonic oscillator environment}
\label{bosonicbath}

To calculate $E_{\text{corr}}(\lambda,t)=\text{Tr}_E\{\varrho_E E(\lambda) E(it)\}=\big\langle  E(\lambda)E(it) \big\rangle_E$ for the harmonic oscillator environment, we first note that 
\begin{align}
    E(\lambda)
    &=\sum_k\left(g_k^*e^{-\lambda\omega_{k}}b_k+g_ke^{\lambda\omega_{k}}b_k^\dagger\right)\nonumber.
\end{align}
Using this relation, we find that 
\begin{align*}
    E_{\text{corr}}(\lambda,t)
    &=\sum_{k,k^\prime}\Big\langle  g_k^*g_{k^\prime}e^{-\lambda\omega_{k}}  e^{i\omega_{k^\prime}t}\left(\mathds{1}+b_k^\dagger b_k^\prime\right) + g_k g_{k^\prime}^*e^{\lambda\omega_{k}} e^{-i\omega_{k^\prime}t}b_k^\dagger b_k^\prime \Big\rangle_E ,\nonumber
\\
    &=\sum_k  |g_k|^2\left\{e^{-\omega_{k}\left(\lambda-it\right)}+\left(e^{-\omega_{k}\left(\lambda-it\right)}+e^{\omega_{k}\left(\lambda-it\right)} \right)n_k\right\}    ,
\end{align*}
finally
\begin{align*}  
\mathcolorbox{Apricot}{
    E_{\text{corr}}(\lambda,t)
    =\sum_k  |g_k|^2\left\{e^{-\omega_{k}\left(\lambda-it\right)}+2n_k\cosh{\left(\lambda\omega_{k}-i\omega_{k}t\right)}\right\},}
\end{align*}
with $n_k$ given by  
\begin{align}
    n_k
    &=\frac{1}{2}\left\{\coth{\left(\frac{\beta\omega_k}{2}\right)}-1\right\}\nonumber.
\end{align}

\section{Environment correlation function with spin environment}\label{spinbathcorr}

Consider the system-environment Hamiltonian given in section \autoref{spinspinmodel}. We evaluate
\begin{align*}
E_{\text{corr}}(\tau)
=\text{Tr}_{E}\left\{\varrho_E E(i\tau) E\right\}.
\end{align*}
Here $E(i\tau)=e^{i H_E\tau} E e^{-i H_E\tau}$ is the collective environment operator $E = \sum_k g_k \sigma_z^{(k)}$ but written in the interaction picture. The exponential $e^{i H_E\tau}$ factors into single-spin terms, leading to
\begin{align*}
    e^{i H_E \tau} \equiv e^{i \sum_k H_E^{(k)}\tau} = \prod_ke^{iH_E^{(k)}\tau},
\end{align*}
thus environment self-correlation function becomes
\begin{align*}
    E(\tau)=\sum_{k j} g_{k} g_{j}\text{Tr}_{E}\left\{\varrho_E e^{iH_E^{(k)}\tau} \sigma_{z}^{(k)}  e^{-iH_E^{(k)}\tau} \sigma_{z}^{(j)}\right\}.
\end{align*}
Since the environmental spins are uncorrelated, this can be simplified to \cite{Schlosshauerbook}
\begin{align*}
    E(\tau)= \sum_{k} |g_{k}|^{2} \text{Tr}_{E}\left\{\varrho_E \sigma_{z}^{(k)}(\tau) \sigma_{z}^{(k)}\right\},
\end{align*}
where $  \sigma_{z}^{(k)}(\tau)=e^{i H_E^{(k)} \tau} \sigma_{z}^{(k)} e^{-i H_E^{(k)} \tau}$ and $H_E^{(k)} = \frac{\omega_k}{2}\sigma_x^{(k)}$. This simplifies to a product of traces over the individual environment spins, that is
\begin{align}\label{spinbath2}
    E(\tau)
    =\sum_{k}  \frac{|g_{k}|^{2}}{Z_{k}} \text{Tr}_{E_{k}}\left\{e^{-\beta H_E^{(k)}} \sigma_{z}^{(k)}(\tau) \sigma_{z}^{(k)}\right\},
\end{align}
where, $Z_k=\text{Tr}_{E_{k}}\left\{e^{- \beta H_E^{(k)}}\right\}$. These traces are most easily evaluated by working in the eigenbasis of $\sigma_x^{(k)}$. We find that 
\begin{align*}
e^{-\beta H_E^{(k)}} \sigma_{z}^{(k)}(\tau) \sigma_{z}^{(k)}
&=e^{- \beta \omega_{k} / 2 + i \omega_{k} \tau}\ket{+}_k\bra{+}_k  + e^{\beta \omega_{k} / 2 - i \omega_{k} \tau}\ket{-}_k \bra{-}_k,  
\end{align*}
where $\ket{+}_k$ and $\ket{-}_k$ are the eigenstates of $\sigma_x^{(k)}$, and $Z_k=\text{Tr}_{E_{k}}\left\{e^{- \beta H_E^{(k)}}\right\}=e^{\beta \omega_{k} / 2}+e^{-\beta \omega_{k} / 2}$. Using these, we obtain the environment correlation function
\begin{align*}
\mathcolorbox{Apricot}{
  E_{\text{corr}}(\tau) =\sum_{k} |g_{k}|^{2}\left\{\cos \left(\omega_{k} \tau\right)- i \text{tanh} \left(\frac{\omega_{k}}{2  T}\right) \sin \left(\omega_{k} \tau\right)\right\}.} 
\end{align*}

\acresetall
  
\chapter{Estimating the environment parameters}\label{c:fisher}
Our central goal in this chapter is to improve the precision of estimating the parameters of a harmonic oscillator environment by using two two-level systems (or qubits) as a quantum probe. The two qubits coupled to the common environment, with the environment imprinting itself upon the dynamics of the two qubits. Within the pure dephasing model, we are able to explore the dynamics of the two qubits exactly, with and without invoking the effect of initial correlations. Next, by taking a partial trace, we determine the time evolution of the state of only one of these qubits. The environment shows up in the density matrix of this single qubit in various factors. First, there is a factor describing decoherence. This factor is the same as that of simply using a single qubit as a probe to estimate the environment parameters. The novelty of our scheme is the emergence, in the single qubit density matrix, of the factor taking into account the indirect interaction between the two qubits due to their interaction with the common environment. Moreover, if we take the initial correlations into account, then the initial correlations, which also contain information about the environment parameters, also show up in the dynamics of the single qubit. 

Our primary purpose in this chapter, then, is to simply show that the environment parameters can be estimated in general far more precisely with our scheme rather than simply using a single qubit as the probe. We do this by showing that the quantum Fisher information $(\mathcal{QFI})$ with our scheme is in general far greater than that obtained with just a single qubit interacting with the environment. We start by working out the dynamics of the two qubits interacting with the common harmonic oscillator environment exactly. We then find the density matrix for the single qubit exactly. Using this state, the $\mathcal{QFI}$ is calculated as a function of time. The maximum value of this $\mathcal{QFI}$ is then found for different values of the environment parameters, with the most emphasis given to the cutoff frequency. Finally, we quantify the measurements needed to be performed in order to obtain the $\mathcal{QFI}$.  

This chapter is ordered as follows. In section \autoref{model}, we show a explicit derivation of reduced dynamics for both correlated and uncorrelated cases. In the next section \autoref{QFI}, we obtain the formula for the $\mathcal{QFI}$ and present results for the estimation of the environment's cutoff frequency $\omega_c$, System-Environment $(\mathcal{SE})$ interaction strength $G$, and the environment's temperature $T$ respectively. Section \autoref{optM}, we derive the formula for Classical Fisher Information $(\mathcal{CFI})$ so that our findings can be compared with practically performed measurements. In the last section \autoref{ch5-sum} we summarize our work. 

\section{The model}
\label{model}

We consider two qubits interacting with a common harmonic oscillator environment. Rather than treating the two qubits collectively as we did in the section \autoref{pure}, Chapter \ref{c:formulism}, here we will find it useful to label the two qubits 1 and 2, since we intend to later take the partial trace over one of the two qubits. The dynamics of our two-qubit system can be expressed by the following time-independent Hamiltonian
\begin{align}
    H
    =H_S+H_E+H_{\text{SE}}, \nonumber
\end{align}
where
\begin{align}
    H_S
    &= \frac{\omega_0}{2}\left( \sigma_z ^{\left(1\right)} + \sigma_z ^{\left(2\right)}\right),
\\
    H_E
    &= \sum_r \omega_r b_{r}^{\dagger} b_r,
\\
    H_{\text{SE}}
    &=\left( \sigma_z ^{\left(1\right)} + \sigma_z ^{\left(2\right)}\right)\sum_{r} \left(g_{r}^{*}b_r + g_r b_r^\dagger \right).
\end{align}
Here, $\omega_0$ is the energy bias, $H_E$ is the environment Hamiltonian (we have ignored the zero point energy for convenience), while $H_{\text{SE}}$ corresponds to $\mathcal{SE}$ interactions. This is very similar to the collective spin model that we solved in the section \autoref{pure}, Chapter \ref{c:formulism}, and thus we proceed in a similar manner. We transform the interaction Hamiltonian into the interaction picture by using the unitary operator $U_0(t)=e^{-i \left(H_E+H_S\right) t}$, that is
\begin{align*}
    H_{\text{SE}}(t)
    &=U_0^\dagger(t) H_{\text{SE}}  U_0(t), \nonumber
\\
    &=\left( \sigma_z ^{\left(1\right)} + \sigma_z ^{\left(2\right)}\right) \sum_{r} \left(g_{r}^{*}b_r e^{-i\omega_r t} + g_r b_r^\dagger e^{i\omega_r t}\right).
\end{align*}
The use of Magnus expansion leads us to the total unitary time evolution operator being (see Chapter \ref{c:formulism} for the analogous calculation)
\begin{align}
    U(t)  
    &=\text{exp}\left\{-i \left(\frac{\omega_0}{2}\left( \sigma_z ^{\left(1\right)} + \sigma_z ^{\left(2\right)}\right) + \sum_r \omega_r b_{r}^{\dagger} b_r \right)t\right\}\nonumber
\\    
    &\times \text{exp}\Bigg\{\frac{1}{2}\left( \sigma_z ^{\left(1\right)} + \sigma_z ^{\left(2\right)}\right) \sum_r\left[ \alpha_r \left(t\right)b_r^\dagger -  \alpha_r^* \left(t\right)b_r\right]-\frac{i}{2}\left(\mathds{1}+\sigma_z ^{\left(1\right)} \sigma_z ^{\left(2\right)}\right) \Delta \left( t \right)\Bigg\},
\end{align}    
with $\alpha_r\left( t \right) = \frac{2g_r\left(1-e^{i\omega_r t} \right)}{\omega_r}$, and $\Delta \left( t \right) 
    =\sum_r  \frac{4\abs{g_r}^2}{\omega_r^2}\left[\sin(\omega_r t) - \omega_r t\right]$. The reduced density operator of the two-qubit system can be obtained via  $\varrho_S (t) 
        =\text{Tr}_E \left\{ U(t) \varrho(0)U^\dagger(t) \right\}$.
It is useful to express our reduced density operator in matrix form using the eigenbasis of $\sigma_z ^{\left(1\right)}$ and $\sigma_z ^{\left(2\right)},$ that is, $\ket{u,v}$, where $\sigma_z ^{\left(1\right)}\ket{u,v}= u\ket{u,v}$ and $\sigma_z ^{\left(2\right)}\ket{u,v}=v\ket{u,v}$. The two-qubit density matrix is then 
\begin{align}
    \left[\varrho_S \left(t\right)\right]_{u',v';u,v} 
    =e^{-i\frac{\omega_0}{2} \left(u'+v'-u-v\right)t} e^{-i\frac{\Delta\left(t\right)}{2} \left(u'v'-uv\right)}\text{Tr}_{\text{S,E}} \left\{  \varrho \left(0\right) e^{-R_{uv,u'v'}\left(t\right)}P_{uv,u'v'} \right\},\label{ch5-3} 
\end{align}
where $P_{uv,u'v'} \equiv \ket{u,v}\bra{u',v'}$, and 
\begin{align}
    R_{uv,u'v'}\left(t\right) 
    &=\sum_r \left[ \widetilde{\alpha}_r\left(t\right)b_r^\dagger -  \widetilde{\alpha}^*_r \left(t\right)b_r \right],
\\
    \widetilde{\alpha}_r\left(t\right) 
    &=\frac{1}{2}\left(u+v-u'-v'\right)\alpha_r\left(t\right).
\end{align}
\subsection{Factorized Initial State}
To make further progress, we now assume that the total state is a product state. In other words, denoting the initial state of the two qubits as $\varrho_S(0)$ and the total state as $\varrho(0)$, we have 
\begin{align}
    \varrho \left(0\right) 
    = \varrho_S \left(0\right) \otimes \varrho_E,
\end{align}
where $\varrho_E 
= \frac{e^{-\beta H_E}}{Z_E} \  \text{with}  \  Z_E = \text{Tr}_E \left\{ e^{-\beta H_E} \right\}.$ From Eq.~\eqref{ch5-3}, we then have  
\begin{align}
    \left[\varrho_S \left(t\right)\right]_{u',v';u,v} 
    &=\left[\varrho_S \left(0\right)\right]_{u',v';u,v}  \text{Tr}_{E} \left\{ \varrho_Ee^{-R_{uv,u'v'}\left(t\right)}\right\}  e^{-\frac{i \omega_0}{2} \left(u'+v'-u-v\right)t} e^{-\frac{i\Delta\left(t\right)}{2} \left(u'v'-uv\right)}.\label{ch5-4}
\end{align}
We simplify $\text{Tr}_{E} \left\{ \varrho_Ee^{-R_{uv,u'v'}\left(t\right)}\right\}$ in the same way as done in Chapter \ref{c:formulism}. Since the modes of the harmonic oscillator are independent of each other, by using the Bloch identity\footnote{If $C$ is a linear combination of the harmonic oscillator raising and lowering operators, then $\av{e^C} = e^{\av{C^2}/2}$}, we can write
\begin{align}
    \av{e^{-R_{uv,u'v'}\left(t\right)}}
    &= \prod_r  \text{exp} \left\{-\frac{1}{2}\abs{\widetilde{\alpha}_r\left(t\right)}^2 \av{2n_r + 1 }\right\},
\end{align}
where we have defined $n_r = \av{b_r^\dagger b_r}$. Since the environment is in thermal equilibrium, $n_r$ is simply the Bose-Einstein distribution, i.e., $n_r =  \frac{1}{e^{{\beta\omega_r }} -1}=\frac{1}{2}\left\{\coth\left(\frac{\beta \omega_r}{2}\right)-1\right\}$, therefore
\begin{align}
    \text{Tr}_{E} \left\{ \varrho_Ee^{-R_{uv,u'v'}\left(t\right)}\right\}
    &= \text{exp}\left\{ -\frac{1}{4}\left(u+v-u'-v'\right)^2\Gamma\left(t\right)\right\}\nonumber,
\end{align}
with
\begin{align}
    \Gamma\left(t\right)
    =&\sum_r  \frac{ 4\abs{g_r}^2}{\omega_r^2} \left[1-\cos\left(\omega_r t \right)\right]\coth{\left( \frac{\beta \omega_r}{2}\right)}.
\end{align}
The final state can be therefore be written as
\begin{align}
    \left[\varrho_S \left(t\right)\right]_{u',v';u,v} 
    =\left[\varrho_S \left(0\right)\right]_{u',v';u,v} e^{-i\frac{\omega_0}{2} \left(u'+v'-u-v\right)t} e^{-i\frac{\Delta\left(t\right)}{2} \left(u'v'-uv\right)}e^{ -\frac{1}{4}\left(u+v-u'-v'\right)^2\Gamma\left(t\right)}.
\end{align}
Note that $\Gamma(t)$ describes decoherence, while $\Delta(t)$ describes the indirect interaction between the qubits due to the interaction with the common environment. We take the initial state to be `pointing up' along the $x$-axis, that is, $\varrho_S\left(0\right)=\ket{+,+}\bra{+,+},$ where $  \sigma_x\ket{+}=\ket{+}$. Then, the density matrix representing the state of the first qubit (by taking a partial trace over the second qubit) is 
\begin{align}
    \varrho_{\text{S1}}\left(t\right)
    &=\frac{1}{2}
\begin{pmatrix}
    1 
    & e^{-i\omega_0 t   -\Gamma\left(t\right)}\cos{\left[\Delta\left(t\right)\right]}\\
    e^{i\omega_0 t -\Gamma\left(t\right)}\cos{\left[\Delta\left(t\right)\right]}
    & 1 \nonumber
\end{pmatrix}.
\end{align}
Now as the spectral density function $J(\omega)$ effectively converts a sum over the environment modes. This function usually assumed to be of the form  $J\left(\omega\right)
    = G\frac{\omega^s}{\omega_c^{s-1}}F(\omega,\omega_c)$, where $F(\omega,\omega_c)$ is a cutoff function containing the cutoff frequency $\omega_c$ \cite{BPbook}. Also, $G$ is the coupling strength, and $s$ is the Ohmicity parameter with $s<1$, $s = 1$ and $s > 1$ representing sub-Ohmic, Ohmic, and super-Ohmic spectral densities respectively. Here, we will only be considering an exponential cutoff function of the form $e^{-\omega/\omega_c}$. To sum up, the state of the first qubit, which will be our probe, is (without incorporating initial correlations)
\begin{align}\mathcolorbox{Apricot}{
\varrho_{\text{S1}}^\text{un}\left(t\right)
=
\frac{1}{2} \begin{pmatrix}
    1 
    &e^{-i\omega_0 t - \Gamma_{\text{un}}\left(t\right)}\cos\left[\Delta\left(t\right)\right]\\
    e^{i\omega_0 t - \Gamma_{\text{un}}\left(t\right)}\cos\left[\Delta\left(t\right)\right]
    & 1
\end{pmatrix},}
\end{align}
with
\begin{align*}
    \Gamma_\text{un}\left(t\right)
    &=\int_0^\infty J(\omega)   \left\{1-\cos\left(\omega t \right)\right\}\coth{\left( \frac{\beta \omega}{2}\right)d\omega},
\\
    \Delta\left(t\right)
    &=\int_0^\infty \frac{ J(\omega)}{\omega^2} \left\{\sin\left(\omega t \right)-\omega t\right\} d\omega.
\end{align*}
It is useful to split $\Gamma_{\text{un}}(t)$ into temperature-dependent and temperature-independent parts, that is, $\Gamma_{\text{un}}(t) = \Gamma_{\text{vac}}(t) + \Gamma_{\text{th}}(t)$ \cite{BPbook}. At zero temperature, $\Gamma_{\text{th}}(t) = 0$. On the other hand
\begin{align*}
    \Gamma_{\text{vac}}(t)
    &=\begin{cases}
    \frac{G}{2}\ln\left(1+ \omega^2_c t^2\right)  &  s=1,
\\    
    G \Bar{\Gamma}[s-1] -  \frac{1}{2}\left(\frac{G \Bar{\Gamma}[s-1]}{\left(1 + i\omega_c t\right)^{s-1}} + \frac{G \Bar{\Gamma}[s-1]}{\left(1+i\omega_c t\right)^{s-1}}\right)  &  s\neq 1,
    \end{cases}
\end{align*}
where $\bar{\Gamma}$ is the usual gamma function defined as $\bar{\Gamma}=\int_{0}^{\infty} t^{z-1} e^{-t} dt$.
\subsection{Correlated Initial State}

We now consider preparing the initial state of the two qubits in the initial state $\ket{\psi} = \ket{+,+}$ via a projective measurement. Following the treatment in the section \autoref{corrStatPrep}, Chapter \ref{c:formulism}, the initial state is then 
\begin{align}
    \varrho \left(0\right) 
    =\ket{\psi}\bra{\psi} \otimes \frac{\bra{\psi} e^{-\beta H} \ket{\psi}}{Z}, \label{ch5-7}
\end{align}
where 
    $Z =  \text{Tr}_{\text{S,E}}\left\{ e^{-\beta H} \right\}$ is total partition function. Inserting the completeness relation $\sum_{p,q} \ket{p,q}\bra{p,q} $ such that $\sigma^{(1)}_z \ket{p,q} = p \ket{p,q}$ and $\sigma^{(2)}_z \ket{p,q} = q \ket{p,q}$, and using the displaced harmonic oscillator modes (see section \autoref{corrStatPrep}, Chapter \ref{c:formulism})
    $B_{r,p,q}
    = b_r + \frac{\left( p+q\right)g_r}{\omega_r}$, it is straightforward to prove
\begin{align}
    Z
    &=\sum_{p,q}e^{-\frac{\beta\omega_0}{2}\left( p+q\right)}e^{\beta \sum_r(p+q)^2\frac{\abs{g_r}^2}{\omega_r}
    }Z_E.
\end{align}
After some algebraic manipulations in the spirit of what has already been done in Chapter \ref{c:formulism}, we arrive at the final expression of $\varrho_S \left(t\right)$, namely
\begin{align}
    \left[\varrho_S (t)\right]_{u',v';u,v} 
    =\left[\varrho_S \left(0\right)\right]_{u',v';u,v} e^{-i\frac{\omega_0}{2} \left(u'+v'-u-v\right)t} e^{-i\frac{\Delta\left(t\right)}{2} \left(u'v'-uv\right)}e^{ -\frac{1}{4}\left(u+v-u'-v'\right)^2\Gamma\left(t\right)} X\left(t\right), \nonumber
\end{align}
where
\begin{align}
    \left[\varrho_S (0)\right]_{u',v';u,v} 
    &=\ip{\psi}{u,v}\ip{u',v'}{\psi},\nonumber
\\
    X\left(t\right)
    &=\frac{\sum_{p,q}e^{-\frac{\beta\omega_0}{2}\left(p+q\right)}\abs{\ip{pq}{\psi}}^2e^{\beta(p+q)^2 \frac{\mathcal{C}}{4}}e^{-i(p+q)\widetilde{\Phi}_{u',v';u,v}\left(t\right)} }{\sum_{p,q}e^{-\frac{\beta\omega_0}{2}\left( p+q\right)}\abs{\ip{pq}{\psi}}^2e^{\beta(p+q)^2 \frac{\mathcal{C}}{4}}},
\end{align}
$\mathcal{C}
    =\sum_r \frac{4\abs{g_r}^2}{\omega_r}$ and  $\widetilde{\Phi}_{u',v';u,v}\left(t\right) 
    =\frac{1}{2}\left(u+v-u'-v'\right)\phi\left(t\right)$
with
\begin{align}
    \phi\left(t\right)
    =\int_0^\infty G\frac{\omega^s}{\omega_c^{s-1}}e^{-\frac{\omega}{\omega_c}} \frac{\sin\left(\omega t \right)}{\omega^2}d\omega.
\end{align} 
From this state, we get the state describing the dynamics of the first spin system by taking a partial trace over the second spin system, as we did in the uncorrelated case. We write the final result as 
\begin{align}\mathcolorbox{Apricot}{
    \varrho_{\text{S1}}^\text{corr} \left(t\right)
    =\frac{1}{2} \begin{pmatrix}
    1 
    & e^{-i\xi\left(t\right)-\Gamma\left(t\right)}\cos\left[\Delta\left(t\right)\right] 
\\
    e^{i\xi\left(t\right)-\Gamma\left(t\right)}\cos\left[\Delta\left(t\right)\right]
    & 1
\end{pmatrix},} \label{ch5-1}
\end{align}
where $\xi\left(t\right)=\omega_0 t + \chi\left(t\right)$, and $ \omega_0 $ is the natural frequency of probe. Again, $\Gamma\left(t\right)$ incorporates the decoherence effect of the environment, while $\Delta(t)$ captures the indirect interaction. Moreover, as result of the initial correlations, the effect of the environment is also encoded in $\chi\left(t\right)$. In particular, 
\begin{align}
    \Gamma\left(t\right)
    &=\Gamma_\text{un}\left(t\right)+\Gamma_\text{corr}\left(t\right),\nonumber
\\
    \Gamma_\text{corr}\left(t\right)
    &=\text{ln}\left[\frac{1+e^{\beta \mathcal{C}}\cosh\left(\beta \omega_0\right)}{\sqrt{a^2 \left(t\right) + b^2 \left(t\right) }}\right],\label{ch5-12}
\\
    \chi\left(t\right)
    &=\text{tan}^{-1}
    \left[\frac{b \left(t\right)}{a \left(t\right)}\right],\nonumber
\end{align}
here we have defined time-dependent coefficients $a (t) = 1+e^{\beta \mathcal{C}}\cosh\left(\beta \omega_0\right)\cos[2\phi(t)]$ and $b(t) = e^{\beta \mathcal{C}}\sinh\left(\beta \omega_0\right)\sin[2\phi(t)]$ with
\begin{align}
    \phi(t)=
    \begin{cases}
    G \tan^{-1}\left(\omega_c t\right) & \quad s=1,
\\
    \frac{G}{2i} \left(\frac{1}{\left(1 - i\omega_c t\right)^{s-1}} - \frac{1}{\left(1+i\omega_c t\right)^{s-1}}\right)\Bar{\Gamma}[s-1] & \quad s\neq 1,
    \end{cases}\nonumber
\end{align}

\section{The Quantum Fisher Information}
\label{QFI}

To quantify the precision with which a general environment parameter $\textit{x}$ can be estimated, we use the $\mathcal{QFI}$ \cite{benedetti2018quantum}. It can be shown that the $\mathcal{QFI}$ is related to the Cramer-Rao bound - the greater the $\mathcal{QFI}$, the greater our precision of the estimate. The general expression for the $\mathcal{QFI}$ is given by \cite{benedetti2018quantum}
\begin{align}
    \mathds{F}_{Q}\left(x\right)
    &=\sum_{n=1}^2 \frac{(\partial_x\varrho_n)^2}{\varrho_n}+2\sum_{n\neq m }\frac{(\varrho_n-\varrho_m)^2}{\varrho_n+\varrho_m}\abs{\ip{\varepsilon_m}{\partial_x\varepsilon_n}}^2,\label{ch5-2}
\end{align}
where $\ket{\varepsilon_n}$ is the $n^{\text{th}}$ eigenstate of our probe state and $\varrho_n$ is the corresponding eigenvalue. For our probe state, which is a $2 \times 2$ matrix, it is straightforward to calculate the eigenvalues and eigenstates. We find that $\varrho_1=\frac{1}{2}[1-\mathcal{F}\left(t\right)] $ and $\varrho_2=\frac{1}{2}[1+\mathcal{F}\left(t\right)] $ with $\mathcal{F}\left(t\right)= \cos\left[\Delta(t)\right]e^{-\Gamma\left(t\right)}$. The corresponding eigenstates are
\begin{align*}
    \ket{\varepsilon_1\left(t\right)}
    &=\frac{1}{\sqrt{2}}\left\{\ket{0} + e^{i\xi \left(t\right)}\ket{1}\right\},
\\
    \ket{\varepsilon_2\left(t\right)}
    &=\frac{1}{\sqrt{2}}\left\{\ket{0} - e^{i\xi \left(t\right)}\ket{1}\right\},
\end{align*}
where $\ket{0}$ and $\ket{1}$ being the eigenstates of $\sigma_z$ and following the eigenvalue equation $\sigma_z\ket{n}=(-1)^n\ket{n}$. Now,  
\begin{align}
    \left(\partial_x \varrho_1 \right)^2
    =\left(\partial_x \varrho_2 \right)^2
    &=\frac{1}{4}e^{-2\Gamma} \left(\sin \Delta \partial_x \Delta + \cos \Delta \partial_x \Gamma \right)^{2}.\nonumber 
\end{align}
Calculating also the derivatives of the eigenstates, and substituting in Eq.~\eqref{ch5-2}, the $\mathcal{QFI}$ comes out to be
\begin{align}\mathcolorbox{Apricot}{
    \mathds{F}_{Q}\left(x\right)
    =\frac{\left( \partial_x \Delta \sin{\Delta}   + \partial_x\Gamma \cos{\Delta}  \right)^2}{e^{2\Gamma}-\cos^2{\Delta}} + \frac{\left(\partial_x \chi\right)^2 \cos^2{\Delta}}{e^{2\Gamma}}.}\label{ch5-5}
\end{align}
This expression reduces to the expression presented in Ref. \cite{ather2021improving} for a single qubit case by setting $\Delta = 0$ in Eq. \eqref{ch5-5} gives the $\mathcal{QFI}$ for the case where we take the initial correlations into account. If we start with the simple product, then we can obtain the $\mathcal{QFI}$ by setting $\chi = 0$ and replacing $\Gamma$ by $\Gamma_{\text{un}}$. 

\subsection{Estimation of the cutoff frequency of the environment}

We now look in detail at the estimation of the cutoff frequency of the environment using our two-qubit scheme. To use Eq.~\eqref{ch5-5}, we note that 

\begin{figure}[t]
\begin{framed}
 		\includegraphics[scale = 1]{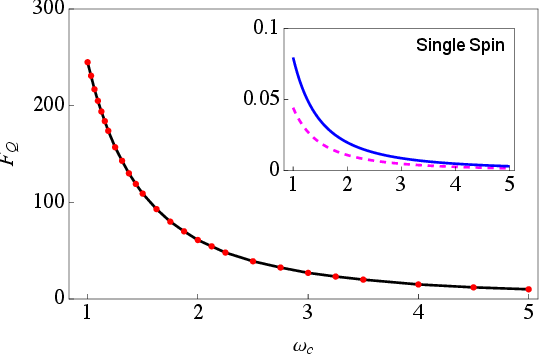}
 		\centering
		\caption{ The main figure shows the behavior of the optimized $\mathcal{QFI}$ for the estimation of the cutoff frequency as a function of the cutoff frequency. The black, solid curve is obtained by including the effects of the initial correlations, while the dotted, red curve ignores these effects. We have taken $\omega_0 = 1$ and the rest of the parameters are  $G= 0.01$, $s=0.5$, and the temperature $T = 0$. The inset shows the optimized $\mathcal{QFI}$ if we simply use a single qubit, both with (solid, blue curve) and without correlations (dashed, magenta curve). The parameters used are the same as the main figure.}
		\label{weakcoupling}
\end{framed}
\end{figure}
\begin{figure}[t]
\begin{framed}
 		\includegraphics[scale = 1]{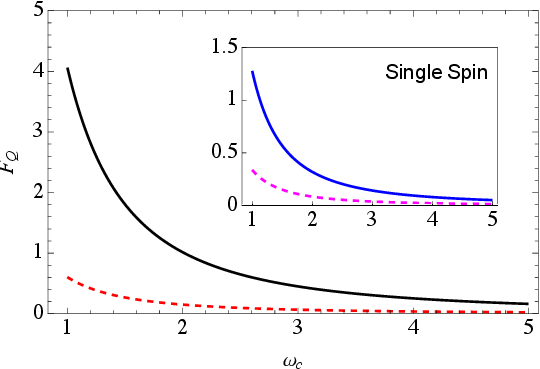}
 		\centering
		\caption{ Same as Fig.~\ref{weakcoupling}, except that we now have $G=1$.}
		\label{StrongCoupling}
\end{framed}
\end{figure}
\begin{figure}[t]
\begin{framed}
 		\includegraphics[scale = 1]{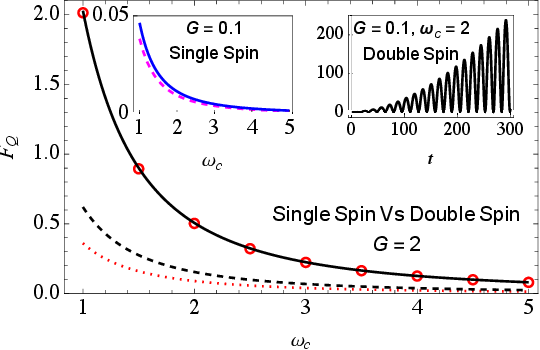}
 		\centering
		\caption{ Comparison of optimized $\mathcal{QFI}$ while estimating $\omega_c$ for single-qubit probe versus two-qubit probe case in Ohmic environment $(s=1)$. In the main plot, solid black (with correlations) and dashed black (without correlations) show the optimized $\mathcal{QFI}$ for the two-qubit case while red circles (with correlations) and dotted red (without correlations) show the optimized $\mathcal{QFI}$ for the single-qubit case. In the top-left inset, optimized $\mathcal{QFI}$ is plotted with (solid blue) and without (magenta dashed) correlations at $G=0.1$ for the single-qubit case while in the top-right inset, $\mathcal{QFI}$ is plotted with (solid black) and without (red dashed) correlations at $G=0.1$ for the two-qubit case. Other parameters are the same as Fig. \ref{weakcoupling}.}
		\label{ohmic}
\end{framed}
\end{figure}
\begin{figure}[t]
\begin{framed}
\includegraphics[scale = 1.2]{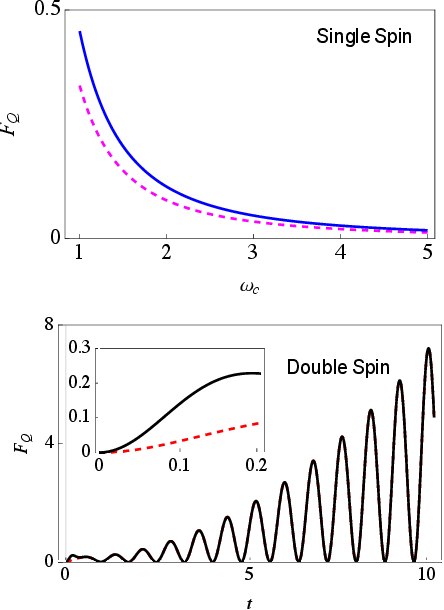}
 		\centering
		\caption{ The top plot shows the optimized $\mathcal{QFI}$ for estimating $\omega_c$ with (solid blue) and without (dashed magenta) correlations using a single qubit as the probe. The bottom plot shows the $\mathcal{QFI}$ with (solid black) and without (dashed red) correlations within the two-qubit scheme. The inset simply zooms in on the $\mathcal{QFI}$ for small values of time to illustrate the $\mathcal{QFI}$ with and without correlations. Here, we are considering a super-Ohmic $(s=2)$ environment. Other parameters used are the same as Fig.~\ref{weakcoupling} except that now we have $G=2$.}
		\label{Super1}
\end{framed}
\end{figure}

\begin{align*}
\frac{\partial\Gamma}{\partial \omega_c}
&=\begin{cases}
    \frac{G \omega_c t^2}{1 + \omega^{2}_c t^2} & \quad s = 1,
\\    
    iG \Bar{\Gamma}[s]t\left(\frac{1}{2\left(1+i\omega_c t\right)^s} - \frac{1}{2\left(1-i\omega_c t\right)^s}\right) & \quad s \neq 1,
\end{cases}
\\
    \frac{\partial\Delta}{\partial \omega_c}
    &=\begin{cases}
    \frac{G \omega_c^2 t^3}{1 + \omega^{2}_c t^2} & \hspace{0.18cm} s = 1,
\\    
    G \Bar{\Gamma}[s]t\left(\frac{1/2}{\left(1+i\omega_c t\right)^s} + \frac{1/2}{\left(1-i\omega_c t\right)^s} - 1 \right) & \hspace{0.18cm} s \neq 1,
\end{cases}    
\\
\frac{\partial\chi}{\partial \omega_c}
&=\begin{cases}
    \frac{2G t}{1 + \omega^{2}_c t^2} & \hspace{0.5cm} s = 1,
\\    
    -G \Bar{\Gamma}[s]t\left(\frac{1}{\left(1+i\omega_c t\right)^s} + \frac{1}{\left(1-i\omega_c t\right)^s}\right) & \hspace{0.5cm} s \neq 1,
\end{cases}
\end{align*}
using these in Eq.~\eqref{ch5-5}, we obtain the $\mathcal{QFI}$ for the estimation of the cutoff frequency as a function of time. We then optimize this $\mathcal{QFI}$ over the interaction time to find the maximum possible $\mathcal{QFI}$. For example, one could plot the $\mathcal{QFI}$ as a function of time for different values of $\omega_c$, and thereby note the maximum value of $\mathcal{QFI}$ for each value of $\omega_c$. We can then investigate the behavior of this optimal $\mathcal{QFI}$ as a function of the cutoff frequency, as has been shown in Fig.~\ref{weakcoupling}. The main figure shows the typical behavior of the $\mathcal{QFI}$ for estimating the cutoff frequency for a sub-Ohmic environment using our two-qubit scheme, both with and without considering initial correlations. It is clear that in this weak coupling strength regime, the effect of the initial correlations is insignificant, as expected since the black, solid curve overlaps with the red dotted curve. The inset shows the optimized $\mathcal{QFI}$ if we simply use a single qubit interacting with the environment with the same set of parameters. What is most notable in this figure is the drastic increase of the $\mathcal{QFI}$ with our two-qubit scheme as compared to using a single qubit - it is a three orders of magnitude increase, which demonstrates in a remarkable manner the advantage of using our two-qubit scheme. The increase is simply because of the indirect qubit-qubit interaction (the $\Delta$ term). Interestingly, if we increase the coupling strength $G$, our two-qubit scheme improves the $\mathcal{QFI}$, although the increase is not as drastic as the in the case of weak coupling (see Fig.~\ref{StrongCoupling}) - the increased decoherence leads to the smaller values of the $\mathcal{QFI}$. We also investigated an Ohmic environment in Fig.~\ref{ohmic}. For strong coupling, we notice the overlap of red circles (using the simple single qubit probe with correlations included) and the solid black curve (using our two-qubit scheme with the effect of the correlations included), thereby indicating that the two schemes perform similarly for strong coupling with an Ohmic environment. However, the situation drastically changes for weaker coupling. As one can see from the inset, the $\mathcal{QFI}$ with our two-qubit scheme keeps on increasing as the qubits interact with their environment - the decoherence is now smaller, and the indirect interaction leads to a buildup of the information gained about the environment. On the other hand, the $\mathcal{QFI}$ obtained using a single qubit probe is bounded. Similar behavior is seen in super-Ohmic environments (see Fig.~\ref{Super1}) where again the inter-qubit interaction (the $\Delta$ term) plays a vital role in the $\mathcal{QFI}$. In fact, now the buildup of $\mathcal{QFI}$ with the two-qubit scheme persists even in the strong coupling regime. 

\subsection{Estimation of system-environment coupling strength}

We now consider estimating the coupling strength $G$. We again use the expression given in Eq.~\eqref{ch5-5} and optimize it over the interaction time to get optimized $\mathcal{QFI}$. We need now the derivatives 
\begin{align*}
\frac{\partial\Gamma}{\partial G}
&=\begin{cases}
    \frac{1}{2}\ln \left(1 + \omega^{2}_c t^2\right) &  s = 1,
\\    
    \Bar{\Gamma}[s-1] - \left(\frac{\Bar{\Gamma}[s-1]/2}{\left(1 + i\omega_c t\right)^{s-1}} + \frac{\Bar{\Gamma}[s-1]/2}{\left(1 + i\omega_c t\right)^{s-1}} \right) &  s \neq 1,
\end{cases}
\\
\frac{\partial\Delta}{\partial G}
    &=\begin{cases}
    \tan^{-1}\left(\omega_c t\right) - \omega_c t &  \hspace{0.1cm} s=1,
\\
    \Bar{\Gamma}[s] \omega_c t - \left(\frac{i\Bar{\Gamma}[s-1]/2}{\left(1-i\omega_c t\right)^{s-1}} -  \frac{i\Bar{\Gamma}[s-1]/2}{\left(1 + i\omega_c t\right)^{s-1}} \right)  & \hspace{0.1cm} s\neq 1,
    \end{cases}
\\
\frac{\partial\chi}{\partial G}
    &=\begin{cases}
    2\tan^{-1}\left(\omega_c t\right) & \hspace{0.2cm} s=1,
\\
    i\Bar{\Gamma}[s-1] \left( \frac{1}{\left(1 - i\omega_c t\right)^{s-1}} - \frac{1}{\left(1+i\omega_c t\right)^{s-1}}\right) & \hspace{0.2cm} s\neq 1.
    \end{cases}
\end{align*}
\begin{figure}[t]
\begin{framed}
 		\includegraphics[scale = 1]{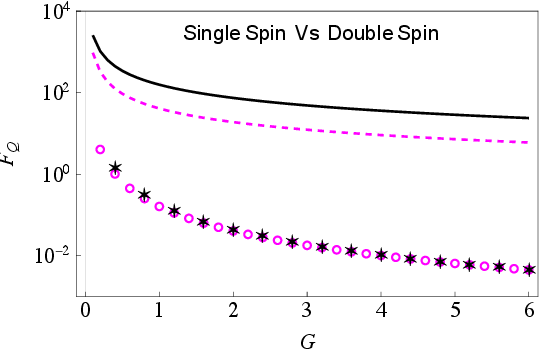}
 		\centering
		\caption{ Behavior of optimized $\mathcal{QFI}$ versus coupling strength $G$ obtained using a single qubit probe [magenta with (dashed) and without (circles) initial correlations], and our two-qubit scheme [black with (solid) and without (asterisks) correlations]. Here we have considered a sub-Ohmic ($s=0.1$) environment. Also, $\omega_{c}=5$, with the rest of the parameters the same as in Fig. \ref{weakcoupling}.}
		\label{HvsA}
\end{framed}
\end{figure}
\begin{figure}[t]
\begin{framed}
   \includegraphics[scale = 1]{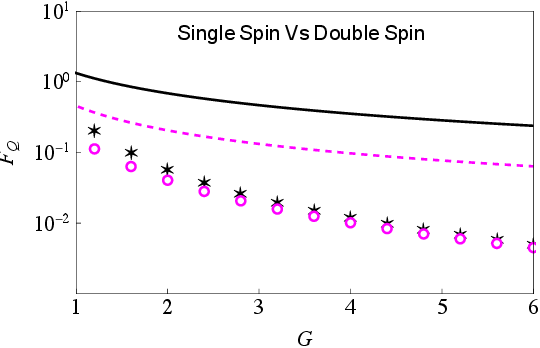}
 		\centering
		\caption{ Same as Fig.~\ref{HvsA}, except that now we considering an Ohmic environment ($s=1$). }
		\label{GOhmic}
\end{framed}
\end{figure}

\begin{figure}[t]
\begin{framed}
 		\includegraphics[scale = 1.2]{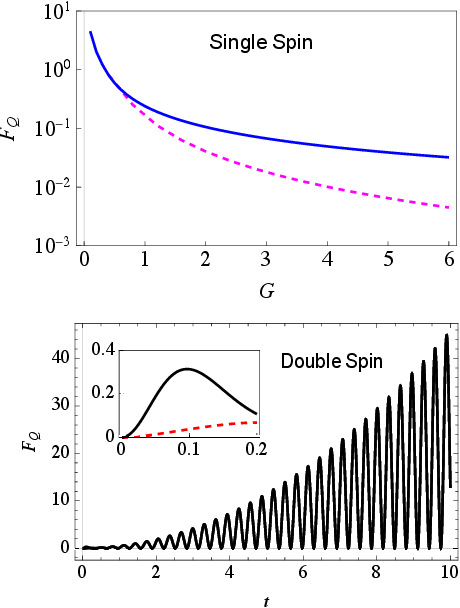}
 		\centering
		\caption{ Same as Fig.~\ref{Super1}, but here we are estimating the coupling strength. Also, we are using $\omega_c = 5$.}
		\label{GSuper}
\end{framed}
\end{figure}
We first compare the optimized $\mathcal{QFI}$ for estimating the coupling strength $G$ obtained using our two-qubit scheme with the $\mathcal{QFI}$ obtained using a single-qubit probe for a sub-Ohmic environment. Results are illustrated in Fig.~\ref{HvsA}, where we have shown the behavior of the optimized $\mathcal{QFI}$ versus the coupling strength $G$ using a single qubit probe both with and without incorporating the effect of the initial correlations - these are shown with the dashed, magenta curve and the circular markers respectively. We have also shown the $\mathcal{QFI}$ with our two-qubit scheme, both with  (solid, black curve) and without (the asterisk markers) including the initial correlations. At least three points should be noted here. First, if we ignore the initial correlations, then there is little difference between the two schemes. Second, the role of the initial correlation is, in general, very important. Third, with both indirect interactions and the initial correlations accounted for, there is a drastic increase in the $\mathcal{QFI}$. Following the same color scheme and parameters used in Fig. \ref{HvsA}, we demonstrate the optimized $\mathcal{QFI}$ in an Ohmic environment $s=1$ as well (see Fig. \ref{GOhmic}). In this environment, while the $\mathcal{QFI}$ is lower as compared to the sub-Ohmic environment, the benefit of using our two-qubit scheme is still evident.

The advantage of our two-qubit scheme becomes even more evident, as before, with super-Ohmic environments as shown in Fig.~\ref{Super1}. Once again, the $\mathcal{QFI}$ generally keeps on increasing as we increase the interaction time (see the bottom figure in Fig.~\ref{GSuper}) for the case of a two-qubit probe. If we compare this with the results obtained using a single qubit probe with (solid blue curve) or without (magenta dashed curve) initial correlations (see the top plot), we see that the $\mathcal{QFI}$ for the single qubit probe is far smaller.

\subsection{Estimation of Temperature}

\begin{figure}[t]
\begin{framed}
 		\includegraphics[scale = 1]{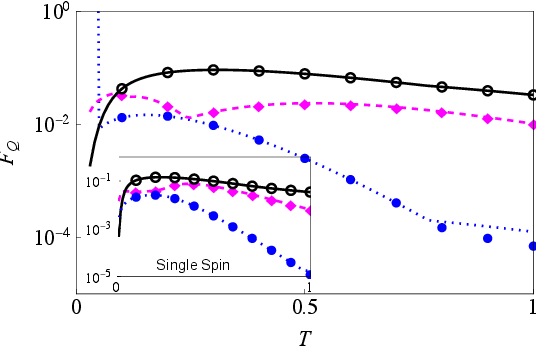}
 		\centering
		\caption{The $\mathcal{QFI}$ for the estimation of temperature. The black curves [with (dashed) and without  (circles) correlations], magenta curves [with (solid) and without  (squares) correlations], and blue curves [with (dotted) and without  (solid circles) correlations] denote the optimized $\mathcal{QFI}$ with a super-Ohmic $(s=2)$, Ohmic $(s=1)$ and sub-Ohmic $(s=0.5)$ environment respectively. Here we have $\omega_c=5$ and $G=1$. The inset follows the same parameters and color scheme but for the single qubit probe.}
		\label{TempAll}
\end{framed}
\end{figure}

Now we consider the estimation of temperature using a single qubit probe, as well as using our two-qubit scheme, for sub-Ohmic, Ohmic, and super-Ohmic environments. Since temperature is not zero here, therefore $\Gamma_{\text{corr}}\left(t\right)$ and $\Gamma_{\text{th}}$ are no longer zero. $\Gamma_{\text{corr}}\left(t\right)$ can be found analytically -  its expression is given in Eq.~\eqref{ch5-12} - while $\Gamma_{\text{th}}$ and its temperature derivative are found numerically. We illustrate our results in Fig.~\ref{TempAll}. The key point to note here is that the higher temperatures mean that the decoherence factor is greatly enhanced. This enhancement effectively washes out the advantage of using our two-qubit scheme, so that the $\mathcal{QFI}$ with a single-qubit probe and our two-qubit scheme are quantitatively similar.  
\begin{figure}[t]
\begin{framed}
 		\includegraphics[scale = 1]{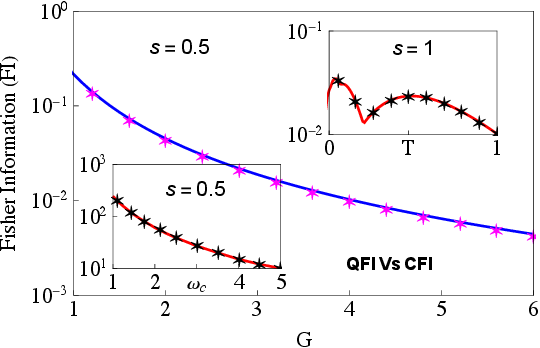}
 		\centering
		\caption{ Plot of $\mathcal{QFI}$ (solid curves) versus $\mathcal{CFI}$ (asterisk markers). The main plot shows the estimation of the coupling strength $G$ with $\omega_c=5$. In the insets, we have plotted the optimized Fisher information (quantum and classical) on the top right, we are estimating temperature $T$ with coupling strength $G=1$. At the bottom left, we estimate the cutoff frequency $\omega_c$ with coupling strength $G=0.01$.} \label{QC-OTC}
\end{framed}
\end{figure}
\section{Optimal measurement}
\label{optM}
Until now, we have found that by using our two-qubit scheme, the $\mathcal{QFI}$ is substantially increased. The question remains regarding which measurements need to be performed in order to obtain this maximum $\mathcal{QFI}$. This can be answered by calculating the $\mathcal{CFI}$ for a particular measurement scheme; if the $\mathcal{CFI}$ comes out to be equal to the $\mathcal{QFI}$, then we have found the optimal measurement to be performed. We guess that the optimal measurements are projective measurements described by the projection operators $\mathbf{P}_1 = \ket{\Psi_1}\bra{\Psi_1}$ and $\mathbf{P}_2 = \ket{\Psi_2}\bra{\Psi_2}$, with
\begin{align}
    \ket{\Psi_1} 
    &=\frac{1}{\sqrt{2}}\left\{
    \ket{\uparrow}_{z}
+ e^{i\varphi}\ket{\downarrow}_{z}\right\},
\\
\ket{\Psi_2} 
    &=\frac{1}{\sqrt{2}}\left\{
    \ket{\uparrow}_{z}
- e^{i\varphi}\ket{\downarrow}_{z}\right\}.
\end{align}
Here $\varphi$ is an equatorial angle in the Bloch sphere. The effect of this measurement is encapsulated by the probability distribution $\mathds{P}(k|x)$ with $k = 1, 2$ and $x$ is the parameter we intend to estimate. For the discrete case, the $\mathcal{CFI}$ is simply \cite{hall2000quantum}
\begin{align}
    \mathds{F}_{c}(x)
    = \sum^{2}_{k=1} \left(\partial^{2}_{x} \ln\left[\mathds{P}(k|x)\right]\right)\mathds{P}(k|x),
\end{align}
where $\mathds{P}(k|x)$ is the conditional probability of getting measurement result $k$, and $\partial^{2}_{x}$ denotes the double derivative with respect to the parameter \emph{x} that is to be estimated. Using the projection operators along with the final state \eqref{ch5-1}, we find that (we have set $\Theta = \chi + \omega_0 t -\varphi $ to show a more compact form)
\begin{align}
    \mathcolorbox{Apricot}{\mathds{F}_{c}(x)
    =\frac{\left[\left(\partial_x\Delta \sin\Delta + \partial_x\Gamma \cos\Delta \right) \cos\Theta - \partial_x\chi \cos\Delta  \sin\Theta\right]^{2}}
    {e^{2\Gamma} - \cos^{2}\Delta \cos^{2}\Theta}.}\label{ch5-9}
\end{align}
If we disregard the effect of the initial correlations, then this expression reduces to
\begin{align}\mathcolorbox{Apricot}{
    \mathds{F}^{\text{woc}}_{c}(x)
    =\frac{\left(\partial_x\Delta \sin\Delta + \partial_x\Gamma \cos\Delta \right)^{2}}
    {\sec^{2}\left(\omega_0 t -\varphi\right) e^{2\Gamma} - \cos^{2}\Delta}.} \label{ch5-10}
\end{align}
We aim to maximize the $\mathcal{CFI}$ [Eq.~\eqref{ch5-9}] over the angle $\varphi$. If this maximized $\mathcal{CFI}$ is equal to the $\mathcal{QFI}$, then we have found the optimal measurement. If the effect of initial correlations is not included, then it is clear that $\varphi=\omega_0 t$ is the optimal value. In this case, the $\mathcal{CFI}$ reduces to the $\mathcal{QFI}$, and so we have found the optimal measurement. On the other hand, if $\chi \neq 0$, we can show that for
\begin{align}\mathcolorbox{Apricot}{
   \varphi
    = \omega_0 t + \chi- \tan^{-1} \left[\frac{\partial_{x}{\chi} \cos\Delta \left( e^{2\Gamma} - \cos^{2} \Delta \right)}
    {e^{2\Gamma} \left(\partial_{x}{\Delta} \sin\Delta + \partial_{x}{\Gamma} \cos\Delta \right)} \right],} \label{ch5-11}
\end{align}
the $\mathcal{CFI}$ reduces to the $\mathcal{QFI}$. Again, this means that we have managed to find the optimal measurement. We further support these claims by plotting both the $\mathcal{QFI}$ and $\mathcal{CFI}$ while estimating environment's cutoff frequency $\omega_c$, $\mathcal{SE}$ coupling strength $G$ and the environment's temperature $T$ [see Fig.~\ref{QC-OTC}], where the overlap between the $\mathcal{CFI}$ and the $\mathcal{QFI}$ shows that we have successfully found the optimal measurements to be performed. 

\section{Summary}
\label{ch5-sum}

In this chapter, we have explored the dynamics of a  two-qubit system that is interacting with an environment of harmonic oscillators, with and without including the effect of initial correlations. In association with these dynamics, we minimize the error in the environment parameter estimation by maximizing the $\mathcal{QFI}$ over the interaction time. By comparison with the single-qubit probe results, we demonstrated that it is generally beneficial to consider a two-qubit scheme in order to improve the precision of the estimates, especially for Ohmic and super-Ohmic environments. 
 
\chapter{Counting Statistics for \emph{WORK}}\label{c:workstat}
Development in experimental methods has enabled the exploration of the dissipative dynamics of mesoscopic as well as quantum systems \cite{collin2005verification,liphardt2001reversible,douarche2005experimental,pekola2013calorimetric}. Unlike for classical systems, the work statistics are still relatively less established in the quantum regime. Much attention has been given towards deriving the quantum versions of fluctuation relations for open quantum systems \cite{campisi2009fluctuation,campisi2011erratum,esposito2009nonequilibrium,crooks1999entropy,crooks2008jarzynski,mukamel2003quantum,de2004quantum}. Sub-Poissonian statistics for photon counts have been investigated indicating the nonclassical states of an electromagnetic field in quantum optics \cite{mandel1995optical}. Counting statistics of charge and heat transfer have also been scrutinized mainly in nonequilibrium mesoscopic systems previously \cite{bagrets2003full,kindermann2004statistics,esposito2009nonequilibrium}. Of particular interest, full work statistics via the Lindblad master equation approach have been presented, where the environment is supposed to be Markovian \cite{silaev2014lindblad}. In this chapter, a generating function is derived that determines the counting statistics of work and heat exchange. The fluctuations of work done by the driving field are also calculated. However, these findings hold only in the weak coupling regime. Namely, we formulate a Lindblad master equation to investigate the counting statistics for work in driven quantum systems. We use the Spin-Boson $(\mathcal{SB})$ model where a single two-level system is coupled to its environment composed of harmonic oscillators. We follow a two-point measurement scheme to construct the characteristic function as the Fourier transform of the probability distribution. Our goal is to study the exchange of energy between the system and the environment in terms of bosons under the action of the driving field. We also aim to differentiate between the work and heat statistics, which becomes crucial for small energy exchanges.

\section {Formalism}\label{form}
\begin{figure}[t]
 		\includegraphics[scale = 0.5]{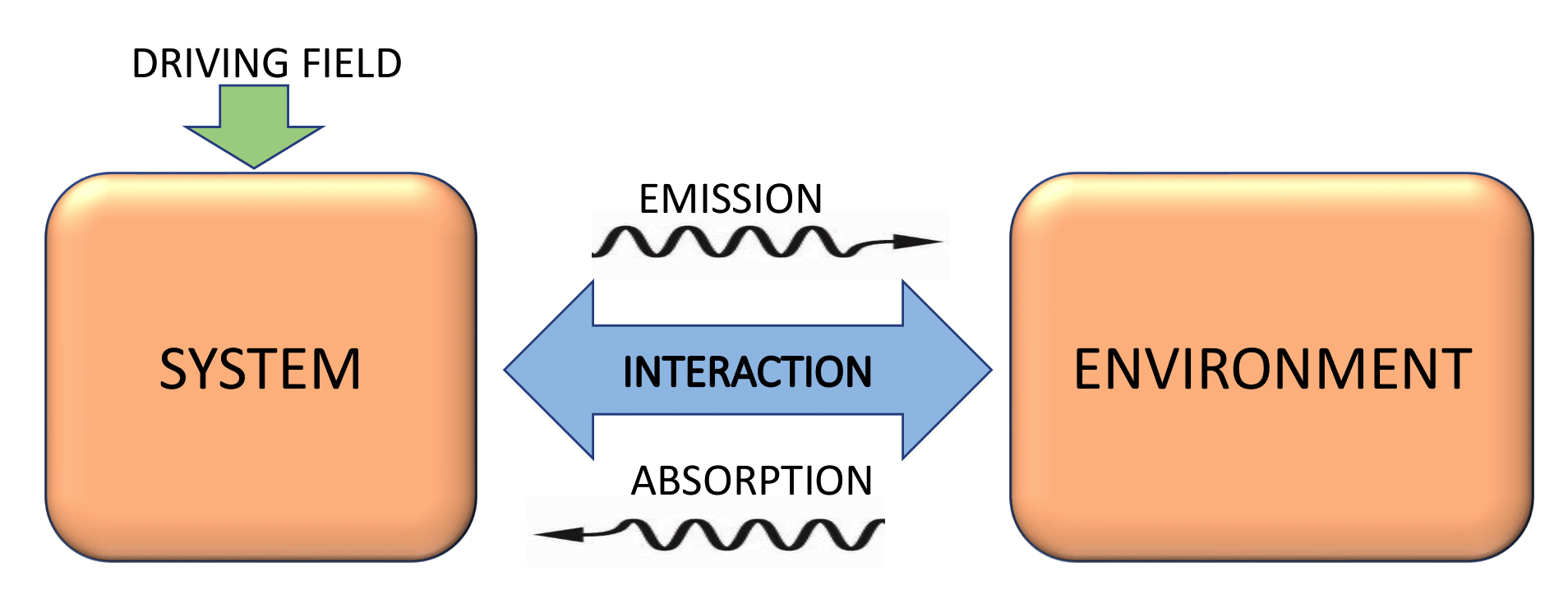}
 		\centering
		\caption{A driven spin system coupled to the harmonic oscillator environment.}
		\label{c:c0}
\end{figure}

We consider a driven quantum spin system coupled to the environment of harmonic oscillators. Our system can exchange its energy in terms of emission and absorption of bosons as shown in Fig.~\ref{c:c0}. The usual driven spin-boson system-environment $(\mathcal{SE})$ Hamiltonian is 
\begin{align}
    H= 
\begin{cases}
       H_{\text{S0}}+H_E &\quad t\leq 0, 
\\
      H_{S}(t)+H_E+ H_{\text{SE}}  & \quad t> 0,
\end{cases}
\end{align} \label{ch6-10}
with
\begin{align}
    H_{\text{S0}}
    &=\frac{\epsilon }{2}\sigma_z,
\\
    H_E
    &= \sum_k \omega_k b_k^{\dagger}b_k,
\\
    H_{\text{SE}}
    &= \sigma_z \otimes \sum_k g_k \left(b_k^{\dagger} + b_k\right).
\end{align}
It is convenient to define $H(0)=H_{\text{S0}}+H_E$ at $t=0$ (and before) and $H(t) = H_{S}(t)+H_E+ H_{\text{SE}}$ later on. With the driving field switched on for $t > 0$, the system Hamiltonian is
\begin{align}
    H_S(t)
    &= \frac{\epsilon }{2}\sigma_z +\Delta\cos{(\omega_l t)} \sigma_x,
\end{align}
where $\epsilon$ is the two-level system energy bias and $\sigma_z$ and $\sigma_x$ are the usual Pauli spin operators, $g_k$ denotes $\mathcal{SE}$ coupling strength, and $\Delta$ and $\omega_l$ are the amplitude and frequency of the applied field respectively. Within a two-point measurement scheme, work is defined as the difference in the measurements of the energy of a closed system at the beginning and end of a process. Since there is no heat dissipated in a closed quantum system, we can associate changes in energy with work. As such, we perform two projective measurements of the full $\mathcal{SE}$ energy at $t=0$ and at some later time $t$ with outcomes $h_0$ and $h_t$ respectively. The probability of getting outcome $h_0$ is  
\begin{align}
    \mathds{P}_{h_0}
    &= \Tr\left\{\prod_{h_0} \varrho _0\right\},
\end{align}
where $\prod_{h_0}= \ket{h_0}\bra{h_0}$  with $H(0) \ket{h_0}=h_0 \ket{h_0}$, and $\varrho_0$ is the joint $\mathcal{SE}$ state at $t=0$ . After performing the first measurement, this state collapses to  
\begin{align}
    \varrho_0'= \frac{\prod_{h_0}\varrho_0\prod_{h_0}}{Z_0},
\end{align}
with $Z_0=\Tr{\prod_{h_0}\varrho_0}$. Now we perform another measurement at time $t>0$ and obtain the outcome $h_t$. Note that $ H(t) \ket{h_t}=h_t \ket{h_t}$. The corresponding probability distribution function can be written as
\begin{align}
    \mathds{P}_{h_t|h_0}
    &= \Tr\left\{\prod _{h_t}  U(t,0) \varrho_0' U^{\dagger}(t,0) \right\}.
\end{align}
It is useful to define the characteristic function as the Fourier transform of the probability distribution
\begin{align}
    \Phi(\zeta)
    = \sum _ {h_0, h_t} \mathds{P}_{h_0} \mathds{P}_{h_t|h_0} e^{i\zeta (h_t-h_0)}.
\end{align}
Using the expressions for $\mathds{P}_{h_0} $ and $\mathds{P}_{h_t|h_0}$, we put an identity inside trace and using the cyclic invariance feature of the trace operation, we arrive at
\begin{align}
    \Phi(\zeta)
    &= \Tr\left\{e^{i\zeta \frac{H(t)}{2}}  U(t,0) e^{-i\zeta \frac{H(0)}{2}}\Bar{\varrho}_0 e^{-i\zeta \frac{H(0)}{2}}U^{\dagger}(t,0)  e^{i\zeta \frac{H(t)}{2}}\right\},\notag
\\
    &= \Tr\left\{e^{i\zeta \frac{H(t)}{2}}\varrho (\zeta, t)e^{i\zeta \frac{H(t)}{2}}\right\},\label{ch6-1}
\end{align}
where $\Bar{\varrho}_0 = \sum_{h_0} \prod_{h_0}\varrho_0\prod_{h_0} $, and
\begin{align}
    \varrho (\zeta, t)
    = U(t,0) e^{-i\zeta \frac{H(0)}{2}}\Bar{\varrho}_0 e^{-i\zeta \frac{H(0)}{2}}U^{\dagger}(t,0).    
\end{align}
Now we make a transformation with the unitary operator $R(t) = e^{i \omega_l \sigma_z t/2}$. We then get
\begin{align}
    \Phi(\zeta)
    = \Tr\left\{e^{i\zeta \frac{H^R(t)}{2}}R(t)\varrho(\zeta, t)R^{\dagger}(t) e^{i\zeta \frac{H^R(t)}{2}}\right\}, \label{ch6-2}
\end{align}
where we have defined $H^R(t) =  R(t) H(t) R^{\dagger} (t)$. We then simplify $H^R(t)$ using the rotating-wave approximation\footnote{The terms oscillating much faster than the system frequency can be ignored.}. Doing so, we get
\begin{align}
    H^R(t)
    &= \frac{\epsilon }{2}\sigma_z +\Delta\cos{(\omega_l t)} R(t) \sigma_x R^{\dagger}(t) + H_E+H_{\text{SE}}, \nonumber
\\
    & \approx \frac{\epsilon }{2}\sigma_z +\frac{\Delta}{2} \left(\sigma_+ +\sigma_- \right) + H_E+H_{\text{SE}}, \nonumber
\\
    &= H_{S}^{\text{rwa}}  + H_E+H_{\text{SE}} \nonumber,
\\
    &= H_0^{\text{rwa}} + H_{\text{SE}} \equiv H^{\text{rwa}},
\end{align}
where we have defined the free Hamiltonian $H_0^{\text{rwa}}=H_{S}^{\text{rwa}} + H_E$, and the system Hamiltonian $H_{S}^{\text{rwa}} = \frac{\epsilon }{2}\sigma_z +\frac{\Delta}{2} \sigma_x $. The corresponding unitary operator is $U_{\text{rwa}}(t)=e^{-i H^{\text{rwa}} t }$. Using the cyclic invariance feature of the trace here again, the characteristic function becomes
\begin{align}
    \Phi(\zeta) 
    = \Tr\left\{{e^{i\zeta H^{\text{rwa}}}R(t)\varrho(\zeta, t)R^{\dagger}(t)}\right\}.
\end{align}    
We have made an approximation $e^{i\zeta H^{\text{rwa}}}\approx e^{i\zeta H_S^{\text{rwa}}}e^{i\zeta H_E}$, that is, we have ignored the interaction part in the second measurement. Therefore, we can write 
\begin{align}
    \Phi(\zeta)
    &= \text{Tr}_S\left\{e^{i\zeta H_S^{\text{rwa}}}\text{Tr}_E\left\{e^{i\zeta \frac{H_E}{2}}R(t)\varrho(\zeta, t)R^{\dagger}(t)e^{i\zeta \frac{H_E}{2}}\right\}\right\},\nonumber
\\
    &= \text{Tr}_S\left\{e^{i\zeta H_S^{\text{rwa}}}\text{Tr}_E\left\{ \varrho^R (\zeta,t)\right\}\right\},\label{ch6-4}
\end{align}
where
\begin{align}
    \varrho^R (\zeta,t)
    =e^{i\zeta \frac{H_E}{2}}R(t)\varrho(\zeta, t)R^{\dagger}(t)e^{i\zeta \frac{H_E}{2}}.  \label{ch6-3} 
\end{align}

\section{Derivation of the LINDBLAD master equation}\label{weakEq}

Now we focus our attention towards deriving the master equation since we need the density matrix to find the characteristic function. Before proceeding, let us clarify our notation. Given a Hamiltonian $H$, we define the transformed Hamiltonian $H(\pm \zeta) =  e^{\pm i\zeta \frac{H_E}{2}} H e^{\mp i\zeta \frac{H_E}{2}} $, and a `primed' Hamiltonian follows $H' = H - \frac{\omega_l \sigma_z}{2}$. We then have $H_S^{'\text{rwa}}=\frac{\varepsilon}{2}\sigma_z +\frac{\Delta}{2} \sigma_x$ with $\varepsilon= \epsilon - \omega_l $. Now, taking the time derivative of Eq.~\eqref{ch6-3}, we get the time-evolution equation
\begin{align}
    \Dot{\varrho}^R(\zeta, t)
    =-i \Big(H^{'\text{rwa}}(\zeta)\varrho^R(\zeta, t)-\varrho^R(\zeta, t)H^{'\text{rwa}}(-\zeta)\Big).
\end{align}
Now, decomposing $H^{' rwa}(\pm\zeta)$ into the free and interaction Hamiltonian as $H^{'\text{rwa}}(\pm\zeta) = H_0^{'\text{rwa}} + H_{\text{SE}}(\pm\zeta)$, we can write
\begin{align}
\Dot{\varrho}^R(\zeta, t)
    = -i \left[H_0^{'\text{rwa}},\varrho^R(\zeta, t)\right]-i \Big(H_{\text{SE}}^{}(\zeta)\varrho^R(\zeta, t)-\varrho^R(\zeta, t)H_{\text{SE}}(-\zeta)\Big).
\end{align}
It is more advantageous to switch in the interaction picture via the unitary operator $U_0^{'\text{rwa}}(t) = e^{-i H_0^{'\text{rwa}}t}$. The density matrix written in the interaction picture looks like $\widetilde{\varrho}^R(\zeta, t)
    = U_0^{'\dagger \text{rwa}}(t) \varrho^R(\zeta, t) U_0^{'\text{rwa}}(t)$. Taking its time derivative, we get
\begin{align}
    \Dot{\widetilde{\varrho}}^R(\zeta, t)
    =-i\Big( \widetilde{H}_{\text{SE}}(\zeta,t)\widetilde{\varrho}^R(\zeta, t)-\widetilde{\varrho}^R(\zeta, t)\widetilde{H}_{\text{SE}}(-\zeta,t)\Big), \label{ch6-5}
\end{align}
here $\widetilde{H}_{\text{SE}}(\zeta,t)=U_0^{'\dagger \text{rwa}}(t) H_{\text{SE}}(\zeta)U_0^{'\text{rwa}}(t)$. Now integrating Eq.~\eqref{ch6-5} with respect to time, and putting the outcome back in the same equation, we get 
\begin{align}
    \Dot{\widetilde{\varrho}}^R(\zeta, t)
    =&-i\left( \widetilde{H}_{\text{SE}}(\zeta,t)\varrho^R(\zeta, 0)-\varrho^R(\zeta, 0)\widetilde{H}_{\text{SE}}(-\zeta,t)\right),\nonumber
\\
    &-\int_{0}^{t} ds \widetilde{H}_{\text{SE}}(\zeta,t) \widetilde{H}_{\text{SE}}(\zeta,s)\widetilde{\varrho}^R(\zeta, s)+ \int_{0}^{t} ds \widetilde{H}_{\text{SE}}(\zeta,t)\widetilde{\varrho}^R(\zeta, s)\widetilde{H}_{\text{SE}}(-\zeta,s),\nonumber
\\
    &+\int_{0}^{t} ds \widetilde{H}_{\text{SE}}(\zeta,s)\widetilde{\varrho}^R(\zeta, s) \widetilde{H}_{\text{SE}}(-\zeta,t)-\int_{0}^{t} ds\widetilde{\varrho}^R(\zeta, s) \widetilde{H}_{\text{SE}}(-\zeta,s) \widetilde{H}_{\text{SE}}(-\zeta,t).\label{ch6-6}
\end{align}
We decompose the interaction Hamiltonian into system and environment parts as $H_{\text{SE}}(\pm\zeta) =  \mathcal{S} \otimes \mathcal{E}(\pm\zeta) $. An equivalent relation holds in the interaction picture as well, that is $\widetilde{H}_{\text{SE}}(\pm\zeta,t)
    =  \widetilde{\mathcal{S}}(t) \otimes \widetilde{\mathcal{E}}(\pm\zeta,t)$, where $\widetilde{\mathcal{S}}(t) = U_S^{'\dagger \text{rwa}}(t) {\mathcal{S}}U_S^{'\text{rwa}}(t)$, and $\widetilde{\mathcal{E}}(\pm\zeta,t) = U_E^{\dagger }(t) e^{\pm i\zeta \frac{H_E}{2}}\mathcal{E} e^{\mp i\zeta \frac{H_E}{2}}U_E(t)$, with $U_E(t) = e^{- i H_E t}$. Hence, Eq.~\eqref{ch6-6} becomes
\begin{align}
    \Dot{\widetilde{\varrho}}^R(\zeta, t)
    =&-i\left( \widetilde{\mathcal{S}}( t)\widetilde{\mathcal{E}}(\zeta,t)\varrho^R(\zeta, 0)-\varrho^R(\zeta, 0)\widetilde{\mathcal{S}}( t)\widetilde{\mathcal{E}}(-\zeta,t)\right),\nonumber
\\
    &-\int_{0}^{t} ds \Bigg[ \widetilde{\mathcal{S}}( t)\widetilde{\mathcal{S}}( s)\widetilde{\mathcal{E}}(\zeta,t)\widetilde{\mathcal{E}}(\zeta,s)\widetilde{\varrho}^R(\zeta, s)
    -  \widetilde{\mathcal{S}}( t)\widetilde{\mathcal{E}}(\zeta,t)\widetilde{\varrho}^R(\zeta, s)\widetilde{\mathcal{S}}(s)\widetilde{\mathcal{E}}(-\zeta,s),\nonumber
\\
    &- \widetilde{\mathcal{S}}( s)\widetilde{\mathcal{E}}(\zeta,s)\widetilde{\varrho}^R(\zeta, s)\widetilde{\mathcal{S}}( t)\widetilde{\mathcal{E}}(-\zeta,t)
    +\widetilde{\varrho}^R(\zeta, s)\widetilde{\mathcal{S}}( s)\widetilde{\mathcal{S}}( t)\widetilde{\mathcal{E}}(-\zeta,s)\widetilde{\mathcal{E}}(-\zeta,t)\Bigg]. \label{ch6-7}
\end{align}
Now we make the approximation $\widetilde{\varrho}^R(\zeta, t) \approx \widetilde{\varrho}^R_S(\zeta, t)\varrho_\mathcal{E}(\zeta) $. Also, we use the Markov approximation to replace $\widetilde{\varrho} (\zeta, s) \rightarrow  \widetilde{\varrho} (\zeta, t)$ and extending the lower limit of integration to $-\infty$. Taking the trace over the environment, the first term in Eq.~\eqref{ch6-7} reduces to zero since
\begin{align}
    &-i \widetilde{\mathcal{S}}( t)\varrho^R_S(\zeta, 0)\text{Tr}_E\left\{\widetilde{\mathcal{E}}(\zeta,t)\varrho_E(\zeta)\right\}+i\varrho^R_S(\zeta, 0)\widetilde{\mathcal{S}}( t)\text{Tr}_E\left\{\widetilde{\mathcal{E}}(-\zeta,t)\varrho_E(\zeta)\right\}, \nonumber
\\    
    &=-i \widetilde{\mathcal{S}}( t)\varrho^R_S(\zeta, 0)\langle \widetilde{\mathcal{E}}(\zeta,t)\rangle-i\varrho^R_S(\zeta, 0)\widetilde{\mathcal{S}}( t)\langle \widetilde{\mathcal{E}}(-\zeta,t)\rangle,\nonumber
\\    
    &=0.\nonumber
\end{align}
The last four terms become 
\begin{align}
    &-\int_{-\infty}^{t} ds \widetilde{\mathcal{S}}( t)\widetilde{\mathcal{S}}( s)\widetilde{\varrho}^R_S(\zeta, t)\text{Tr}_E\left\{\widetilde{\mathcal{E}}(\zeta,t)\widetilde{\mathcal{E}}(\zeta,s)\varrho_E(\zeta)\right\},\nonumber
\\    
    &+ \int_{-\infty}^{t} ds \widetilde{\mathcal{S}}(t)\widetilde{\varrho}^R_S(\zeta, t)\widetilde{\mathcal{S}}( s)\text{Tr}_E\left\{\widetilde{\mathcal{E}}(\zeta,t)\varrho_E(\zeta)\widetilde{\mathcal{E}}(-\zeta,s)\right\},\nonumber
\\
    &+\int_{-\infty}^{t} ds \widetilde{\mathcal{S}}(s)\widetilde{\varrho}^R_S(\zeta, t)\widetilde{\mathcal{S}}( t)\text{Tr}_E\left\{\widetilde{\mathcal{E}}(\zeta,s)\varrho_E(\zeta)\widetilde{\mathcal{E}}(-\zeta,t)\right\},\nonumber
\\
    &-\int_{-\infty}^{t} ds\widetilde{\varrho}^R_S(\zeta, t)\widetilde{\mathcal{S}}( s)\widetilde{\mathcal{S}}( t)\text{Tr}_E\left\{\varrho_E(\zeta)\widetilde{\mathcal{E}}(-\zeta,s)\widetilde{\mathcal{E}}(-\zeta,t)\right\}.\nonumber
\end{align}
Now we set $s=t-\tau$ which means $\int_{-\infty}^{t}ds\rightarrow \int_{0}^{\infty}d\tau$. Using the cyclic invariance feature of the trace operation, it is straight forward to show that $\av{\widetilde{\mathcal{E}}(t)\widetilde{\mathcal{E}}(t')}
    =\av{\widetilde{\mathcal{E}}(t-t')\mathcal{E}}$. Hence our master equation takes the following form
\begin{align}
    \Dot{\widetilde{\varrho}}_S^R(\zeta, t)
    =&-\int_{0}^{\infty} d\tau \widetilde{\mathcal{S}}( t)\widetilde{\mathcal{S}}( t-\tau)\widetilde{\varrho}^R_S(\zeta, t)\av{\widetilde{\mathcal{E}}(\zeta,\tau){\mathcal{E}}(\zeta)}_E,\nonumber
\\    
    &+ \int_{0}^{\infty} d\tau \widetilde{\mathcal{S}}( t)\widetilde{\varrho}^R_S(\zeta, t)\widetilde{\mathcal{S}}( t-\tau)\av{\widetilde{\mathcal{E}}(-\zeta,-\tau){\mathcal{E}}(\zeta)}_E,\nonumber
\\
    &+\int_{0}^{\infty} d\tau \widetilde{\mathcal{S}}(t-\tau)\widetilde{\varrho}^R_S(\zeta, t)\widetilde{\mathcal{S}}( t)\av{\widetilde{\mathcal{E}}(-\zeta,\tau){\mathcal{E}}(\zeta)}_E,\nonumber
\\
    &-\int_{0}^{\infty} d\tau\widetilde{\varrho}^R_S(\zeta, t)\widetilde{\mathcal{S}}( t-\tau)\widetilde{\mathcal{S}}( t)\av{\widetilde{\mathcal{E}}(-\zeta,-\tau){\mathcal{E}}(-\zeta)}_E.\label{ch6-9}
\end{align}
In order to write our master equation in the Lindblad form, we need to diagonalize the interaction Hamiltonian. To do so, we first determine the eigenstates of our system Hamiltonian 
\begin{align}
    H_S^{'\text{rwa}}
    =\frac{\varepsilon}{2}\sigma_z +\frac{\Delta}{2} \sigma_x.
\end{align}
We rotate this Hamiltonian by an angle $\theta/2$ about the $y$-axis to get
\begin{align}
    H_S^{r}
    &=e^{i\frac{\theta}{2}\sigma_y}\left(\frac{\varepsilon}{2}\sigma_z +\frac{\Delta}{2}\sigma_x\right)e^{i\frac{\theta}{2}\sigma_y},\nonumber
\\
    &=\left(\frac{\varepsilon}{2}\cos{\theta} + \frac{\Delta}{2}\sin{\theta}\right)\sigma_z + \left(\frac{\Delta}{2}\cos{\theta} - \frac{\varepsilon}{2}\sin{\theta}\right)\sigma_x.\nonumber
\end{align}
Setting $\frac{\Delta}{2}\cos{\theta} - \frac{\varepsilon}{2}\sin{\theta}=0$ to get $\theta= \arctan(\frac{\Delta}{\varepsilon})$, we find that $\sin{\theta}= \frac{\Delta}{\eta}$ and $\cos{\theta}= \frac{\varepsilon}{\eta}$ with $\eta = \sqrt{\varepsilon^2 + \Delta^2}$. Therefore our rotated Hamiltonian becomes
\begin{align}
    H_S^{r}
    &=\left(\frac{\varepsilon^2}{2\eta}+ \frac{\Delta^2}{2\eta}\right)\sigma_z,\nonumber
\\
    &=\frac{\eta}{2}\Big(\ket{e}\bra{e}-\ket{g\bra{g}}\Big).\nonumber
\end{align}
Now we switch back to find
\begin{align}
    H_S^{'\text{rwa}}
    &=\frac{\eta}{2}\left(e^{-i\frac{\theta}{2}\sigma_y}\ket{e}\bra{e}e^{i\frac{\theta}{2}\sigma_y}-e^{-i\frac{\theta}{2}\sigma_y}\ket{g\bra{g}}e^{i\frac{\theta}{2}\sigma_y}\right),\nonumber
\\
    &=\frac{\eta}{2}\Big(\ket{+}\bra{+}-\ket{-}\bra{-}\Big).\nonumber
\end{align}
We thus have $H_S^{'\text{rwa}}\ket{\pm}=\pm\frac{\eta}{2}\ket{\pm}$, where we have defined new set of basis states $\ket{\pm}$ in which our Hamiltonian is diagonal, namely
\begin{align}
    \ket{+}
    &=e^{-i\frac{\theta}{2}\sigma_y}\ket{e} = \cos{\frac{\theta}{2}}\ket{e}+\sin{\frac{\theta}{2}}\ket{g},
\\
    \ket{-}
    &=e^{-i\frac{\theta}{2}\sigma_y}\ket{e} = \cos{\frac{\theta}{2}}\ket{g}-\sin{\frac{\theta}{2}}\ket{e}.
\end{align}
We also have
\begin{align}
    \ket{g}
    &=\cos{\frac{\theta}{2}}\ket{-}+\sin{\frac{\theta}{2}}\ket{+},
\\
    \ket{e}
    &=\cos{\frac{\theta}{2}}\ket{+}-\sin{\frac{\theta}{2}}\ket{-}.
\end{align}
The interaction Hamiltonian written in the interaction picture $\widetilde{H}_{\text{SE}}(\pm\zeta, t)$ can be decomposed into the system and the environment part as 
\begin{align}
    \widetilde{H}_{\text{SE}}(\pm\zeta, t) 
    &= \widetilde{\mathcal{S}}(t) \otimes \widetilde{\mathcal{E}}(\pm\zeta,t),\nonumber
\\
    &= \widetilde{\sigma}_z(t) \otimes \sum_k g_k\left( b_k^{\dagger} e^{i\zeta\frac{\omega_k}{2}}e^{i\omega_k t}  + b_k e^{-i\zeta\frac{\omega_k}{2}}e^{-i\omega_k t}\right),
\end{align}
where $\widetilde{\mathcal{S}}(t) = \widetilde{\sigma}_z(t)$ is a system operator written in the interaction picture with respect to $H^{'\text{rwa}}_S$. It is convenient to work in the eigenbasis of $H^{'\text{rwa}}_S$, that is, $H_S^{'\text{rwa}}\ket{\pm}=\pm\eta\ket{\pm}$. To determine $\widetilde{\sigma}_z(t)$, we first write ${\sigma}_z$ in terms of the eigenbasis of the system Hamiltonian:  
\begin{align}
    \sigma_z
    &= \ket{e}\bra{e}-\ket{g}\bra{g}= \cos{\theta}\Big(\ket{+}\bra{+}-\ket{-}\bra{-}\Big) - \sin{\theta}\Big(\ket{+}\bra{-}+\ket{-}\bra{+}\Big),\nonumber
\end{align}
we can thus write $\widetilde{\sigma}_z(t) = e^{i H^{'\text{rwa}}_S t} \sigma_z e^{-i H^{'\text{rwa}}_S t}$ as 
\begin{align}
    \widetilde{\sigma}_z(t)
    &=\cos{\theta}\Big(\ket{+}\bra{+}-\ket{-}\bra{-}\Big) - \sin{\theta}\Big(\ket{+}\bra{-}e^{i\eta t}+\ket{-}\bra{+}e^{-i\eta t}\Big),\nonumber
\\
    &=S_0 - \Big(S_{\eta}^{\dagger}e^{i\eta t}+S_{\eta}e^{-i\eta t}\Big),
\end{align}
where $S_0 = \frac{\varepsilon}{\eta}\left(\ket{+}\bra{+}-\ket{-}\bra{-}\right)$ and $S_{\eta} = \frac{\Delta}{\eta}\ket{-}\bra{+}$.
\begin{figure}[t] \begin{framed}
 		\includegraphics[scale = 0.85]{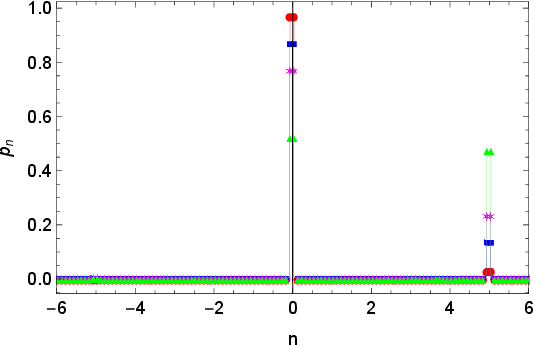}
 		\centering
		\caption{ Work probability distribution at various times, corresponding to $\Delta t=0.1$ (red), $\Delta t=0.5$ (blue), $\Delta t=1$ (magenta) and $\Delta t=5$ (green) We have set coupling strength $G=0.1$ and the other parameters are $\varepsilon=5, \beta=1 $ and $\Delta=0.01$.}
		\label{c:weak}
\end{framed}\end{figure}
Having diagonalized the interaction Hamiltonian, we make the secular approximation, that is, the terms oscillating with frequencies $\pm \eta t$ or $\pm 2\eta t$ are neglected since their contributions are small on the system relaxation timescale. The right hand side terms of Eq.~\eqref{ch6-9} transform under secular approximation as
\begin{align}
    &\widetilde{\mathcal{S}}(t)\widetilde{\mathcal{S}} (t-\tau)\widetilde{\varrho}^R_S(\zeta,t) \longrightarrow
    \left\{S_0^2+ S_{\eta}^{\dagger }S_{\eta}e^{i\eta \tau} + S_{\eta}S_{\eta}^{\dagger }e^{-i\eta \tau} \right\}\widetilde{\varrho}^R_S(\zeta,t),\nonumber
\\
    &\widetilde{\mathcal{S}}(t)\widetilde{\varrho}^R_S(\zeta,t)\widetilde{\mathcal{S}} (t-\tau) \longrightarrow
    S_0\widetilde{\varrho}^R_S(\zeta,t)S_0
    +S_{\eta}^{\dagger}\widetilde{\varrho}^R_S(\zeta,t)S_{\eta}e^{i\eta \tau}
    +S_{\eta}\widetilde{\varrho}^R_S(\zeta,t)S_{\eta}^{\dagger}e^{-i\eta \tau}
    ,\nonumber
\\
    &\widetilde{\mathcal{S}} (t-\tau)\widetilde{\varrho}^R_S(\zeta,t) \widetilde{\mathcal{S}}(t) \longrightarrow
    S_0\widetilde{\varrho}^R_S(\zeta,t)S_0
    +S_{\eta}^{\dagger}\widetilde{\varrho}^R_S(\zeta,t)S_{\eta}e^{-i\eta \tau}
    +S_{\eta}\widetilde{\varrho}^R_S(\zeta,t)S_{\eta}^{\dagger}e^{i\eta \tau}
    ,\nonumber
\\
    &\widetilde{\varrho}^R_S(\zeta,t)\widetilde{\mathcal{S}} (t-\tau)\widetilde{\mathcal{S}}(t) \longrightarrow
    \widetilde{\varrho}^R_S(\zeta,t)\left\{S_0^2+ S_{\eta}^{\dagger }S_{\eta}e^{-i\eta \tau} + S_{\eta}S_{\eta}^{\dagger }e^{i\eta \tau} \right\}.\nonumber
\end{align}
\begin{figure}[t] \begin{framed}
 		\includegraphics[scale = 0.85]{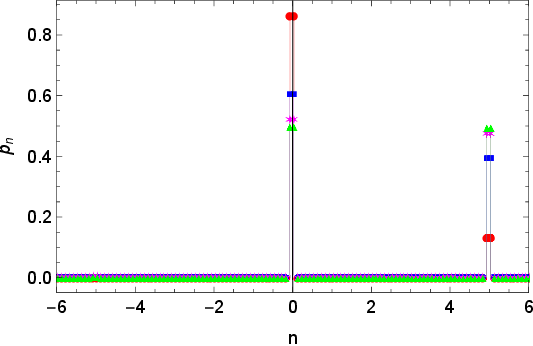}
 		\centering
		\caption{ Same as Fig.~\ref{c:weak} except that now we have set $G=0.5$.}
		\label{c:strong}
\end{framed}\end{figure}
\begin{figure}[t] \begin{framed}
 		\includegraphics[scale = 0.85]{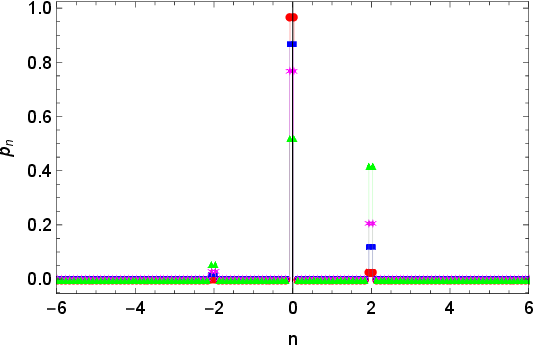}
 		\centering
		\caption{ Same as Fig.~\ref{c:weak} except that now we have set $\varepsilon=2$.}
		\label{c:bias}
\end{framed}\end{figure}
\begin{figure}[t] \begin{framed}
 		\includegraphics[scale = 0.85]{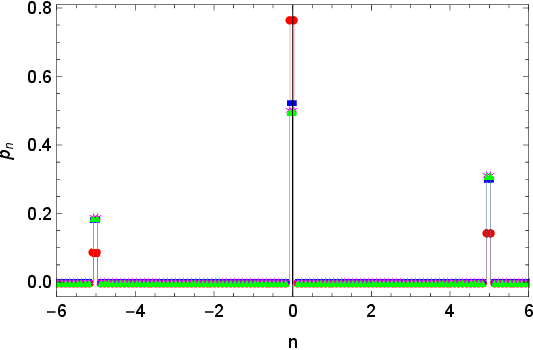}
 		\centering
		\caption{ Same as Fig.~\ref{c:weak} except that now we have set $\beta=0.1$.}
		\label{c:temp}
\end{framed}\end{figure}
Using these results along with the environment correlations functions, and performing the integral $\int_0^{\infty} d\tau e^{\pm i \varepsilon \tau}= \pi \delta(\varepsilon)\pm i \mathcal{P}(1/\varepsilon)$, we  can arrive at the final form of Lindblad master equation
\begin{align}
    \Dot{\varrho}^R_S (\zeta, t)
    =&-i \left[H'_S,\varrho^R_S(\zeta,t)\right] + \Gamma_0 \Big(S_0 \varrho^R_S(\zeta,t)S_0 - \frac{1}{2}\left\{S^2_0 , \varrho^R_S(\zeta,t)\right\}\Big),\nonumber
\\
    &+ \Gamma (\eta)\left(1+N(\eta)\right)\Big(S_{\eta}\varrho^R_S(\zeta,t)S^{\dagger}_{\eta}e^{i\eta\zeta} -\frac{1}{2}\left\{S^{\dagger}_{\eta}S_{\eta}, \varrho^R_S(\zeta,t)\right\}\Big),\nonumber
\\
    &+ \Gamma (\eta)N(\eta)\Big(S^{\dagger}_{\eta}\varrho^R_S(\zeta,t)S_{\eta}e^{-i\eta\zeta} -\frac{1}{2}\left\{S_{\eta}S^{\dagger}_{\eta}, \varrho^R_S(\zeta,t)\right\}\Big),\label{cfmeq}
\end{align}
where we have switched back to the Schrodinger picture and defined the following decay rates
\begin{align}
    \Gamma_0 
    &= 2\pi \lim_{\omega\rightarrow 0}J(\omega)\left(1+2n(\omega)\right),
\\
    \Gamma(\eta) 
    &= 2\pi J(\eta).
\end{align}
$J(\omega)$ is the usual spectral density that encapsulates the effect of the environment, and the system's shifted Hamiltonian is
\begin{align}\mathcolorbox{Apricot}{
    H'_S
    = H_S^{'\text{rwa}} - \Lambda_0 S^2_0 +\Lambda_1 \left(S^{\dagger}_{\eta}S_{\eta}-S_{\eta}S^{\dagger}_{\eta}\right)+\Lambda_2 \left(S^{\dagger}_{\eta}S_{\eta}+S_{\eta}S^{\dagger}_{\eta}\right),}
\end{align}
with
\begin{align}
    \Lambda_0
    &=\mathcal{P} \int_0^{\infty} d\omega \frac{J(\omega)}{\omega},
\\
    \Lambda_1
    &=\mathcal{P} \int_0^{\infty} d\omega \frac{\eta J(\omega)(1+2n(\omega))}{\eta^2-\omega^2},
\\
    \Lambda_2
    &=\mathcal{P} \int_0^{\infty} d\omega \frac{ \omega J(\omega)}{\eta^2-\omega^2}.
\end{align}
We solve our master equation \eqref{cfmeq} numerically to obtain $\varrho^R_S(\zeta, t),$ and hence the characteristic function by using $\Phi(\zeta)= \text{Tr}_S\left\{e^{i\zeta H_S^{\text{rwa}}}\varrho^R_S(\zeta, t)\right\}.$ The inverse Fourier transform of the characteristic function gives us the probability that the difference in the measurement outcomes $h_t - h_0$ corresponds to $n$ bosons. Positive values of boson number $n$ correspond to energy emission from the system towards the environment. Similarly, negative values of $n$ means system is taking energy from its environment. We demonstrate the system dynamics in Fig.~\ref{c:weak} where we have shown the work probability distribution at different times, corresponding to $\Delta t=0.1$ (red), $\Delta t=0.5$ (blue), $\Delta t=1$ (magenta) and $\Delta t=5$ (green). We have set the coupling strength $G=0.1$ and the other parameters are $\varepsilon=5, \beta=1 $ and $\Delta=0.01$. The plot markers on the vertical line $n=0$ correspond to a zero-emission or absorption event taking place. We notice here, for weak coupling strength ($G=0.1$), there is only a very small emission probability at small times, which increases later on. However, as we increase the coupling strength, the emission probability increases as illustrated in Fig.~\ref{c:strong}. Further interpretation is a work in progress. 

\begin{comment}
Some simple examples of the resulting probability distributions are shown in Fig. 1 for various times, where we take the bath spectral density to be $J(ω) = γω$ with dimensionless coupling γ. We have chosen an initial system state corresponding to $ρSn (0) = δn,0 |g⟩ ⟨g|.$ For this choice, the structure of Eq. (54) allows only a single emission or absorption event [1], hence the probability distributions show non-zero values for only n = 0, 1 in the case of a zero temperature bath (left plot), and $n = 0, ±1$ for a non-zero temperature bath (right plot) since both absorption and emission are now possible.
More interesting plots can be obtained if we include another dissipative process [1]. By way of illustration, let us add a phenomenological decay term to Eq. (54), which now becomes
\end{comment}

\chapter*{Appendix}
\addcontentsline{toc}{chapter}{\tocEntry{Appendix}}
\section{Environment correlation function}\label{app:C}

Here we will evaluate $\av{\widetilde{\mathcal{E}} (\zeta,\tau){\mathcal{E}} (\zeta)}_\mathcal{E} $ which is a so-called counting field-dependent environment correlation function. First, consider
\begin{align}
    \mathcal{E}(\zeta) 
    &= e^{ i\zeta \frac{H_E}{2}}\mathcal{E} e^{- i\zeta \frac{H_E}{2}},\nonumber
\\
    &= \sum_k g_ke^{\frac{i\zeta}{2}\sum_k \omega_k b_k^{\dagger}b_k} \left(b_k^{\dagger} + b_k\right) e^{-\frac{i\zeta}{2}\sum_k \omega_k b_k^{\dagger}b_k}.\nonumber
\end{align}
Using the Baker–Campbell–Hausdorff (BCH) identity and the commutation relation $[b_k, b_k^{\dagger}]=1$, we have
\begin{align}
    \mathcal{E}(\zeta) 
    =\sum_k g_k\left( b_k^{\dagger} e^{i\zeta\frac{\omega_k}{2}}  + b_k e^{-i\zeta\frac{\omega_k}{2}}\right).\nonumber
\end{align}
Once again making the unitary transformation with time evolution operator $U_E(t)$, we have
\begin{align}
    \widetilde{\mathcal{E}}(\zeta, t)
    &=U_E(t) \mathcal{E}(\zeta) U_E^{\dagger}(t),\nonumber
\\
    &=\sum_k g_k\left( b_k^{\dagger} e^{i\zeta\frac{\omega_k}{2}}e^{i\omega_k t}  + b_k e^{-i\zeta\frac{\omega_k}{2}}e^{-i\omega_k t}\right). \label{eq10}
\end{align}
Now we are equipped to derive an environment correlation function
\begin{align}
    &\av{\widetilde{\mathcal{E}}(\zeta,\tau){\mathcal{E}}(\zeta)}_\mathcal{E} \\
    &= \sum_{k,k'} g_k g_k'\Big(\av{b_k^{\dagger}b_k'^{\dagger}} e^{i\zeta} e^{i\omega_k \tau}+\av{b_k^{\dagger}b_k'} e^{i\omega_k \tau}+\av{b_kb_{k'}^{\dagger}} e^{-i\omega_k \tau}+\av{b_kb_k'} e^{-i\zeta} e^{-i\omega_k \tau}
    \Big).\nonumber
\end{align}
Using $\av{b_k^{\dagger}b_k'^{\dagger}}=0=\av{b_kb_k'}$, $\av{b_kb_k'^{\dagger}}=\delta_{k,k'}\left(1+ n(\omega_{k'})\right)$, and $\av{b_k'^{\dagger}b_k}=\delta_{k,k'}n(\omega_{k'})$, where $n_k=\left(e^{\beta \omega}-1\right)^{-1}$ is the environment occupation number,
\begin{align}
    \av{\widetilde{\mathcal{E}}(\zeta,\tau){\mathcal{E}}(\zeta)}_\mathcal{E}
    &=\sum_{k} \abs{g_k}^{2}\Big(n_ke^{i\omega_k \tau}+(1+n_k)e^{-i\omega_k \tau}\Big).
\end{align}

We now switch into the continuum regime via environment spectral density that characterizes the environment and $G$ is the coupling strength between system and the environment. To sum up
\begin{align}
    \av{\widetilde{\mathcal{E}}(\zeta,\tau){\mathcal{E}}(\zeta)}_\mathcal{E}
    &= \int_0^{\infty} d\omega  J(\omega)\Big(n_ke^{i\omega_k \tau}+(1+n_k)e^{-i\omega_k \tau}\Big),\nonumber
\\    
    \av{\widetilde{\mathcal{E}}(-\zeta,-\tau){\mathcal{E}}(-\zeta)}_\mathcal{E}
    &= \int_0^{\infty} d\omega  J(\omega)\Big(n_ke^{-i\omega_k \tau}+(1+n_k)e^{-i\omega_k \tau}\Big),\nonumber
\\    
    \av{\widetilde{\mathcal{E}}(-\zeta,-\tau){\mathcal{E}}(\zeta)}_\mathcal{E}
    &= \int_0^{\infty} d\omega  J(\omega)\Big(n_ke^{-i\omega_k(\zeta + \tau)}+(1+n_k)e^{i\omega_k(\zeta + \tau)}\Big),\nonumber
\\    
    \av{\widetilde{\mathcal{E}}(-\zeta,\tau){\mathcal{E}}(\zeta)}_\mathcal{E}
    &= \int_0^{\infty} d\omega  J(\omega)\Big(n_ke^{-i\omega_k(\zeta- \tau)}+(1+n_k)e^{i\omega_k(\zeta- \tau)}\Big).\nonumber
\end{align}

  \acresetall
  
  \chapter{Conclusions}\label{c:Conclusions}
  In this thesis, we have first explored the importance of initial $\mathcal{SE}$ correlations. In particular, we have shown that if we start from the joint thermal equilibrium state of a quantum system and its environment, and then apply a unitary operation to the system to prepare the required initial system state, the correlations in the joint thermal equilibrium state influence the subsequent dynamics of the system. In chapter \ref{c:SpinEnv}, we considered an exactly solvable spin-spin model to exactly put together the effect of the initial correlations. Thereafter, in chapter \ref{c:MasterEq}, we derived a time-local master equation that is correct to second-order in the system-environment $(\mathcal{SE})$ coupling strength and also takes into account the effect of these correlations. The structure of this master equation is very interesting, as the form of the term that takes into account the initial correlations is the same as the relaxation and dephasing term. In this sense, one can say that the initial correlations affect the decoherence and dephasing rates, a fact which has already been pointed out in studies of the role of initial correlations in pure dephasing models. Finally, we applied our master equation to the large spin-boson model as well as to a collection of two-level systems interacting with a spin environment to quantitatively investigate the role of the initial correlations. We found that when the number of spins is small, the initial correlations do not play a significant role. However, for a larger number of spins, the initial correlations must be accounted for in order to explain the dynamics accurately. We come to find that the initial $\mathcal{SE}$ have a minimal effect in the regimes of weak $\mathcal{SE}$ coupling and high temperatures. However, this difference becomes more appreciable when the $\mathcal{SE}$ coupling becomes stronger and the temperature is low. Such results are promising, as they provide insights into the effect of the initial correlations.

We next looked at estimating the environment parameters by using a two-qubit probe. We first worked out the exact dynamics of two qubits interacting with a common harmonic oscillator environment via pure dephasing. Thereafter, we minimized the error in the environment parameter estimation by maximizing the quantum Fisher information for the various environment parameters. By comparison with the single-qubit probe results, we have demonstrated that it is beneficial to consider a two-qubit system in order to improve estimates. 

\section*{Future Work}
Let us now briefly highlight what we plan to do in the near future as offshoots of the work done in this thesis. 

\begin{itemize}

    \item While working on the interpretation of the work counting statistics in the weak coupling regime, we are also extending our treatment to the strong $\mathcal{SE}$ coupling regime.

    \item The effect of the initial correlations can be masked by the effects of decoherence and dissipation. To clearly understand the effect of the initial correlations, we must find a way to eliminate the effect of decoherence and dissipation while retaining the effect of the initial correlations. It may be possible to apply suitable external control fields to the system that effectively reduce the effect of decoherence while amplifying the effect of the initial correlations. 
  
    \item Since the effect of the $\mathcal{SE}$ correlations is expected to be greatly significant in the strong coupling regime, we can attempt to derive a non-Markovian master equation (as we did in chapter \ref{c:MasterEq}) that also works if the $\mathcal{SE}$ coupling is strong, and also includes the effect of initial correlations. In this regard, the polaron transformation can be helpful.
    
    \item Selective measurements (like projective measurements) are ideal and difficult to realize experimentally, which is why measurements other than ideal projective measurements are of particular interest. We can then look at the effect of using non-ideal projective measurements to initialize the system state. 
    
    \item Dynamical decoupling technique has been used to maximize the quantum of Fisher information by considering a single qubit quantum probe \cite{ather2021improving}. We can try applying the same technique as an extension of our work done in chapter \ref{c:fisher}, hence precision in the estimation of environment parameters can be further improved. 
    
    \item With the solutions of the two-qubit system interacting with the harmonic oscillator environment, and various recently proposed non-Markovianity measures, the role played by any non-Markovian effects can also be quantified. We can also investigate the interplay between initial  correlations and non-Markovianity.

\end{itemize}
  \cleardoublepage

  %\singlespacing
  % Bibliography
\label{app:Bibliography} % Reference the bibliography elsewhere with \autoref{app:Bibliography}

\manualmark % Work-around to have small caps also here in the headline
\markboth{\spacedlowsmallcaps{\bibname}}{\spacedlowsmallcaps{\bibname}} % Work-around to have small caps also
%\phantomsection
\refstepcounter{dummy}

\addtocontents{toc}{\protect\vspace{\beforebibskip}} % Place the bibliography slightly below the rest of the document content in the table of contents
\addcontentsline{toc}{chapter}{\tocEntry{\bibname}}
\printbibliography
  \cleardoublepage
  
  %\pagestyle{empty}
  %\onehalfspacing
  %\input{text/app-colophon}
\end{document}